\documentclass[prd,showpacs,preprintnumbers,amsmath,amssymb,superscriptaddress,preprintstyle,preprint,nofootinbib]{revtex4-1}
\usepackage{graphicx}
\usepackage{dcolumn}
\usepackage{bm}
\usepackage{color}
\usepackage{soul,xcolor}
\usepackage{epsfig}
\usepackage{dcolumn}
\def\be{\begin{equation}}
\def\ee{\end{equation}}
\def\bea{\begin{eqnarray}}
\def\eea{\end{eqnarray}}
\def\gsim{\ \rlap{\raise 2pt\hbox{$>$}}{\lower 2pt \hbox{$\sim$}}\ }
\def\lsim{\ \rlap{\raise 2pt\hbox{$<$}}{\lower 2pt \hbox{$\sim$}}\ }
\def\dslash{\kern-4pt \not{\hbox{\kern-2pt $\partial$}}}
\def\pslash{\not{\hbox{\kern-2pt p}}}

\newcommand{\epssme}{\varepsilon^s_{\mu e}}
\newcommand{\epssmm}{\varepsilon^s_{\mu\mu}}
\newcommand{\epssmt}{\varepsilon^s_{\mu\tau}}

\newcommand{\epsdee}{\varepsilon^d_{ee}}
\newcommand{\epsdem}{\varepsilon^d_{e\mu}}

\newcommand{\epsdme}{\varepsilon^d_{\mu e}}
\newcommand{\epsdmm}{\varepsilon^d_{\mu\mu}}
\newcommand{\epsdmt}{\varepsilon^d_{\mu\tau}}
\newcommand{\epsdte}{\varepsilon^d_{\tau e}}
\newcommand{\epsdtm}{\varepsilon^d_{\tau\mu}}

\newcommand{\dcp}{\delta}

\newcommand{\dm}[1]{\Delta m^2_{#1}}
\newcommand{\essnusb}{ESS$\nu$SB}

\setstcolor{blue}

\begin{document}
\DeclareGraphicsExtensions{.eps,.ps}


\title{Exploring Source and Detector Non-Standard Neutrino Interactions at \essnusb}



\author{Mattias Blennow}
\email[Email Address: ]{emb@kth.se}
\affiliation{
Department of Theoretical Physics,
School of Engineering Sciences, KTH Royal Institute of Technology,
AlbaNova University Center, 106 91 Stockholm, Sweden
}
 
\author{Sandhya Choubey}
\email[Email Address: ]{sandhya@hri.res.in}
\affiliation{
Department of Theoretical Physics,
School of Engineering Sciences, KTH Royal Institute of Technology,
AlbaNova University Center, 106 91 Stockholm, Sweden
}
\affiliation{
Harish-Chandra Research Institute,
Chhatnag Road, Jhunsi, Allahabad 211 019, India
}
 
\author{Tommy Ohlsson}
\email[Email Address: ]{tohlsson@kth.se}
\affiliation{
Department of Theoretical Physics,
School of Engineering Sciences, KTH Royal Institute of Technology,
AlbaNova University Center, 106 91 Stockholm, Sweden
}

\author{Sushant K. Raut}
\email[Email Address: ]{raut@kth.se}
\affiliation{
Department of Theoretical Physics,
School of Engineering Sciences, KTH Royal Institute of Technology,
AlbaNova University Center, 106 91 Stockholm, Sweden
}

\begin{abstract}

We investigate source and detector non-standard neutrino interactions 
at the proposed \essnusb\ experiment. We analyze the effect of 
non-standard physics at the probability level, the event-rate 
level and by a full computation of the \essnusb\ setup. We find 
that the precision measurement of the leptonic mixing angle 
$\theta_{23}$ at \essnusb\ is robust in the presence of non-standard 
interactions, whereas that of the leptonic CP-violating phase $\dcp$ 
is worsened at most by a factor of two. We compute 
sensitivities to all the relevant source and decector non-standard 
interaction parameters and find that the sensitivities to the parameters 
$\epssme$ and $\epsdme$ are comparable to the existing 
limits in a realistic scenario, while they improve by a factor of 
two in an optimistic scenario. Finally, we show that the absence 
of a near detector compromises the sensitivity of \essnusb\ to non-standard interactions.

\end{abstract}
\maketitle

\section{Introduction}

Without comparison, the Standard Model of particle physics (SM) is the most successful physics model to date, accurately predicting an enormous number of observables with high precision from only a handful of fitted parameters. The success of the SM may have culminated in 2012 when the ATLAS and CMS experiments announced the discovery of the Higgs boson \cite{Aad:2012tfa,Chatrchyan:2012ufa}, predicted by the SM as a direct result of the electroweak symmetry breaking which was introduced to provide masses into the theory. Still, there are a number of observations which may not be explained within the SM itself. Most notable among these are the existence of dark matter, the exclusion of gravity and the observation of neutrino oscillations. In addition, there are conceptual theoretical problems with the SM, such as the hierarchy problem, indicating that the SM may only be a low-energy approximation of a more general theory. As such, the SM should be viewed as an effective theory and a priori higher-dimensional operators, suppressed by powers of a new mass scale $\Lambda$ should be added to the SM Lagrangian. At lower energies, the additional effective operators will generally produce very small corrections due to this suppression. This concept is further supported by the fact that the only gauge invariant operator allowed at dimension five, and therefore suppressed only by one power of $\Lambda$, is the so-called Weinberg operator~\cite{Weinberg:1979sa}, which results in a Majorana mass term for the left-handed neutrinos of the SM. It is therefore not unreasonable to imagine that the effect of neutrino masses would be among the first observations of physics beyond the SM, which indeed is the case due to neutrino oscillations requiring neutrino mass-squared differences to be non-zero.

Neutrino flavour conversion, although at that time not confirmed as such, was first observed in solar neutrino experiments where a discrepancy between the observed flux and the flux predicted by solar models was found~\cite{Davis:1968cp}. Since the first robust evidence of neutrino oscillations by the Super-Kamiokande experiment's observation of atmospheric neutrinos in 1998 \cite{Fukuda:1998mi}, they have been extensively studied experimentally in a variety of atmospheric, solar, reactor, and accelerator experiments, which have helped to constrain the neutrino mass and mixing parameters to very high precision (see Refs.~\cite{Capozzi:2013csa,Forero:2014bxa,Gonzalez-Garcia:2014bfa} for recent global fits). The remaining questions in neutrino oscillation physics today are the neutrino mass ordering, the existence or non-existence of CP violation in the lepton sector and the octant of the leptonic mixing angle $\theta_{23}$. Answering these three questions is the main aim of the next generation of neutrino oscillation experiments, such as the European Spallation Source Neutrino Super-Beam (\essnusb) experiment \cite{Baussan:2013zcy}, which is a proposed accelerator neutrino experiment based on the European Spallation Source (ESS) currently under construction in Lund, Sweden. The sensitivity of \essnusb\ to the CP-violating phase $\dcp$ was studied in Ref.~\cite{Baussan:2013zcy}, while the sensitivity to other standard oscillation parameters was discussed in Ref.~\cite{Agarwalla:2014tpa} and the sensitivity to light sterile neutrinos in Ref.~\cite{Blennow:2014fqa}.

While the Weinberg operator provides the neutrino masses necessary for neutrino oscillations to occur and neutrino oscillations have been firmly established as the leading mechanism behind neutrino flavour conversion, higher-order operators may give rise to sub-leading contributions to the neutrino conversion probabilities and their observation would allow us to gain additional insight into the high-energy completion of the SM and the generation of neutrino masses. In addition, it may be necessary to consider the robustness of the usual neutrino oscillation parameters when higher-order operators are also considered. One of the more common types of operators to be investigated in this respect is non-standard neutrino interactions (NSIs), which are effective four-fermion operators involving at least one neutrino field. For recent reviews on NSIs, see Refs.~\cite{Ohlsson:2012kf,Miranda:2015dra}.

In this work, we will consider the possible impact of NSIs at the \essnusb\ experiment. We will study both the influence of NSIs on the determination of the standard neutrino oscillation parameters and the bounds which \essnusb\ could place on the NSI parameters. In particular, we will focus on correlations in the determination of the leptonic CP violation and the NSI parameters, which is of large importance for \essnusb\ as the discovery of leptonic CP violation is the major scientific target of this experiment.

The rest of this work is organized as follows. In Sec.~\ref{sec:nsis},  we will briefly review non-standard neutrino interactions and present the current upper bounds on the source and detector NSI parameters. Next, in Sec.~\ref{sec:essnusb}, the setup of the proposed \essnusb\ experiment will be discussed. Then, in Sec.~\ref{sec:neuosc}, we will investigate the phenomenology of source and detector NSIs at probability and event-rate levels. In Sec.~\ref{sec:results}, the main results of our full computation on source and detector NSIs at \essnusb\ will be presented. Finally, in Sec.~\ref{sec:sc}, we will summarize and draw our conclusions.

\section{Non-standard neutrino interactions}
\label{sec:nsis}

When considering NSIs, we will be confronted with effective four-fermion operators of the type
\begin{equation}
\mathcal O = (\bar f_1 \gamma^\mu P_{L,R} f_2) (\bar f_3 \gamma_\mu P_{L,R} f_4) + h.c. ~,
\end{equation}
where $f_i$ ($i=1,2,3,4$) are SM fermion fields and $P_{L,R}$ are left- and right-handed projections. These operators are of dimension six and they will therefore appear together with an effective coupling constant of dimension minus two in the effective Lagrangian. Since we are interested in the NSIs of neutrinos, we require that at least one of the fermion fields in the operators is a neutrino field, which implies that the corresponding projection operator must be $P_L$. Furthermore, in order to keep the electromagnetic and strong interactions unbroken, we require that all operators are scalars under transformations of the corresponding gauge groups. Due to the weak interaction being broken, we do not impose any constraints on the transformation of the operators under ${\rm SU(2)}_{\rm L}$. It should be mentioned that imposing ${\rm SU(2)}_{\rm L}$ gauge symmetry on the dimension-six operators would lead to flavour constraints on these operators~\cite{Antusch:2008tz,Biggio:2009kv}, leaving only a few possible operators without significant constraints due to the non-observation of effective four-charged-fermion processes such as $\mu \to 3e$. The dimension-six operators which break ${\rm SU(2)}_{\rm L}$ may generally be induced from higher-dimensional operators such as $(\phi \phi^\dagger) \mathcal O$, where $\phi$ is the Higgs field, which are invariant under ${\rm SU(2)}_{\rm L}$, but generate ${\rm SU(2)}_{\rm L}$-breaking terms once the Higgs field takes on a vacuum expectation value $v$. Depending on  the dimension at which the NSIs are generated above the electroweak scale, we may expect the NSI coefficients to scale as $v^{n-6}/\Lambda^{n-4}$, where $n$ is the dimension and $\Lambda$ is the energy scale at which the NSIs are generated .

The different possible neutrino NSIs are generally divided into two categories of effective four-fermion operators. The neutral-current NSIs \cite{Wolfenstein:1977ue,Barger:1991ae}
\begin{equation}
\mathcal O_{\alpha \beta}^{f(L,R)} = (\overline{\nu_\alpha} \gamma^\mu P_L \nu_\beta)(\overline{f} \gamma_\mu P_{L,R} f) + h.c. ~,
\end{equation}
where $f$ is a charged fermion field, affects the neutrino flavour propagation in matter for $f = u, d, e$ by providing an effective potential analogous to the Mikheyev--Smirnov--Wolfenstein (MSW) potential \cite{Wolfenstein:1977ue,Mikheev:1986gs,Mikheev:1986wj}. For the neutral-current NSIs to be of importance, relatively large matter potentials and/or high neutrino energies are required. As this is not the case for the \essnusb\ experiment, we will not focus on such NSIs in this work. On the other hand, the charged-current NSIs \cite{Grossman:1995wx}
\begin{equation}
\mathcal O_{\alpha\beta}^{ff'(L,R)} = (\overline{\ell_\alpha}\gamma^\mu P_L \nu_\beta)(\overline{f}\gamma_\mu P_{L,R} f') ~,
\end{equation}
where $f$ and $f'$ are different fermion fields such that the operator is invariant under ${\rm U(1)}_{\rm EM}$ and ${\rm SU(3)}_c$, will instead affect the production and detection processes of neutrinos and this effect will not depend on the neutrino energy or the presence of matter along the neutrino propagation.

In the remainder of this work, we will focus on the charged-current NSI Lagrangian
\begin{equation}
\mathcal L_{\rm NSI} = -2\sqrt 2 G_F \sum_{X \in \{L,R\}} \sum_{\alpha,\beta} \varepsilon^{A}_{\alpha\beta} (\overline{\nu_\beta}\gamma^\mu P_L \ell_\alpha)(\overline d \gamma_\mu P_{X} u) + h.c. ~,
\end{equation}
which includes the operators that will appear in neutrino production by pion decays $\pi \to \ell_\alpha \nu$ and charged-current neutrino detection processes. Here, we have normalised the strength of the NSIs to that of the weak interaction by the introduction of the Fermi coupling constant $G_F$. The NSI parameters $\varepsilon_{\alpha\beta}$ are therefore dimensionless numbers expected to be of the order $(v/\Lambda)^{n-4}$. With the introduction of charged-current NSIs, the production amplitude of the neutrino mass eigenstate $|\nu_i\rangle$ in the $\pi^+$, which in the SM is proportional to $U_{\mu i}^*$, where $U$ is the leptonic mixing matrix, is now instead proportional to $\sum_\alpha (\delta_{\mu\alpha} + \varepsilon^s_{\mu\alpha}) U_{\alpha i}^*$, where the NSI parameters relevant for the source process are
\begin{equation}
\varepsilon^s_{\alpha\beta} = \varepsilon_{\alpha\beta}^R - \varepsilon_{\alpha\beta}^L ~.
\end{equation}
Unlike the source process, the detection process does not necessarily involve a pseudoscalar current in the quark sector. We instead define the NSI parameters relevant for the detection process as
\begin{equation}
\varepsilon^d_{\alpha\beta} = (\varepsilon_{\beta\alpha}^P)^* ~,
\end{equation}
where $P$ represents the quark current in the detection process. Due to the nature of the inverse beta decay involved in the detection process, this definition oversimplifies the neutrino oscillation probabilities that we will discuss in Sec.~\ref{sec:neuosc}.\footnote{In fact, the neutrino oscillation probabilities should be computed along the lines
$$
P_{\alpha\beta} \simeq \frac{1}{5.5}\left[P_{\alpha\beta}(\varepsilon^A,\varepsilon^V) + 4.5P_{\alpha\beta}(\varepsilon^A,\varepsilon^A)\right] ~,
$$
where $P_{\alpha\beta}(\varepsilon^s,\varepsilon^d)$ is the probability for a given source/detector NSI. Note that the largest prefactor comes from the contribution with the source and detector effects both dependent on the axial quark current. This would therefore indicate a relation between the source and detector NSIs.} However, we will use this as a simplified model for how NSIs may affect \essnusb. The complex conjugate and change of indices has been introduced to adhere to the usual convention in the field when considering detector NSI effects.
 The production rates of charged leptons at the detector in any neutrino oscillation experiment will be affected by this change in the production and detection amplitudes and we may ask the question whether or not the presence of such NSIs could be measured or have a negative impact on the experimental precision to the standard oscillation parameters. The experimental bounds (at the 90\% C.L.) on the NSI parameters relevant for the \essnusb\ experiment from non-oscillation experiments are given by \cite{Biggio:2009nt}
\begin{equation}
\begin{split}
|\varepsilon^s_{\mu e}| < 0.026 ~, \quad |\varepsilon^s_{\mu\mu}| < 0.078 ~, \quad |\varepsilon^s_{\mu\tau}| < 0.013 ~, \\
|\varepsilon^d_{ee}| < 0.041~, \quad |\varepsilon^d_{\mu e}| < 0.025 ~, \quad |\varepsilon^d_{\tau e}| < 0.041 ~,
\\
|\varepsilon^d_{e\mu}| < 0.026 ~, \quad |\varepsilon^d_{\mu\mu}| < 0.078 ~, \quad |\varepsilon^d_{\tau \mu}| < 0.013 ~.
\end{split}
\label{eq:bounds}
\end{equation}
Although these bounds are quite stringent, it should be kept in mind that the next generation of neutrino experiments is aiming for highly sensitive measurements of the neutrino oscillation parameters. As such, even sub-leading effects may be of interest and it is worth the effort to examine the possible impact of these effects. It is also worth noting that new oscillation experiments, such as those performed with nuclear reactors, may be sensitive to some of these NSI parameters as well~\cite{Ohlsson:2008gx}. However, the current bounds from these experiments are somewhat weaker than the bounds quoted above~\cite{Agarwalla:2014bsa}.

\section{The \essnusb\ experimental setup}
\label{sec:essnusb}

In this section, we describe the experimental setup for the proposed \essnusb\ experiment. 
We have used the standard flux (with 2 GeV protons) and cross-sections 
from the \essnusb\ collaboration~\cite{Baussan:2013zcy}. The source provides a  
neutrino beam for two years and an antineutrino beam for eight years.
We have assumed that a 500 kiloton 
water Cherenkov detector is placed at a distance of 540 km 
from the source, which corresponds to the location of the mine in
Garpenberg, Sweden. The detector specifications have been taken from 
the performance study of the MEMPHYS detector~\cite{Agostino:2012fd}. 
The energy range of interest is up to 2 GeV, which is divided 
into 20 energy bins. We have used 9\% (18\%) systematic errors 
on the signal (background) events.
Unless specified otherwise, we have also assumed the existence of 
a near detector with mass 1 kiloton, 1 km from the source and the same flux as at the 
detector at 540 km scaled by the distance-squared. As  a crude approximation, 
we assume the same characteristics for both these detectors. 

To this end, we have written our 
own probability engine to calculate the neutrino oscillation 
probability in the presence of source and detector NSIs. This 
probability engine interfaces with 
GLoBES~\cite{Huber:2004ka,Huber:2007ji} for calculating 
the neutrino event rates at \essnusb. The large parameter 
space is handled with the help of the GLoBES plugin MonteCUBES~\cite{Blennow:2009pk}.

\section{Neutrino oscillations with NSIs}
\label{sec:neuosc}

Standard three-flavour neutrino oscillations depend on six fundamental 
parameters -- two mass-squared differences, $\dm{21}$ 
and $\dm{31}$, three mixing angles, $\theta_{12}$, $\theta_{13}$ 
and $\theta_{23}$ and one CP-violating phase $\dcp$. In addition, 
if the neutrinos are propagating through matter, the 
charged-current interactions of the neutrinos with electrons 
modify the oscillations. This effect can be incorporated 
into the probability formalism by using the MSW potential term 
$A = 2\sqrt{2} G_F n_e E$ \cite{Wolfenstein:1977ue,Mikheev:1986gs,Mikheev:1986wj}, where $n_e$ is the number density 
of electrons in the matter and $E$ is the  
neutrino energy. For an experiment like \essnusb\ with a 
short baseline length as well as low neutrino energy, we can ignore the matter 
effects for the sake of this discussion. (The 
numerical results presented in this work do not make any 
such assumption.)

Non-standard neutrino interactions can affect the 
production and detection of neutrinos at the source 
and detector, respectively. In the SM, interactions 
of charged leptons with neutrinos are strictly flavour-diagonal. 
However, charged-current NSIs can introduce a non-zero 
overlap between charged leptons and neutrinos of different flavours. 
Thus, a neutrino produced at a source in association 
with a charged lepton $\ell_\alpha$ is not simply $\nu_\alpha$, but 
is given by \cite{Grossman:1995wx,GonzalezGarcia:2001mp,Bilenky:1992wv,Meloni:2009cg}
\begin{equation}
 | \nu_\alpha^s \rangle  =  | \nu_\alpha \rangle  +  \sum_{\gamma=e,\mu,\tau} \varepsilon^s_{\alpha\gamma}  | \nu_\gamma \rangle  ~.
\end{equation}
Similarly, a neutrino that produces a charged lepton $\ell_\beta$ 
in a detector is 
\begin{equation}
 \langle \nu_\beta^d |  =  \langle \nu_\beta |  +  \sum_{\gamma=e,\mu,\tau} \varepsilon^d_{\gamma\beta}  \langle \nu_\gamma |  ~.
\end{equation}
The matrices $\varepsilon^s$ and $\varepsilon^d$ are in general complex,
giving 36 new parameters. These 
are 9 amplitudes and 9 phases of each NSI parameter 
in the source and detector NSI matrices. 

Not all of the 36 NSI parameters are relevant for the 
experiment under consideration.
Since we are only interested in the oscillation channels 
$\nu_\mu \to \nu_e$ and $\nu_\mu \to \nu_\mu$ 
(and their CP conjugates), the relevant parameters 
are $\varepsilon^s_{\mu\gamma}$, $\varepsilon^d_{\gamma e}$ 
and $\varepsilon^d_{\gamma\mu}$, where $\gamma \in \{ e,\mu,\tau \}$. 
Thus, the parameter space is reduced to 9 complex 
or 18 real parameters, in addition to the standard ones. 
In this work, 
we treat all of them as independent parameters, which is 
the most general case.

Deriving an analytical formula for the neutrino oscillation 
probabilities is difficult even in the standard three-flavour 
scenario. Typically, expressions for the probabilities 
are given as perturbative expansions in small 
parameters such as $\dm{21}/\dm{31}$ or 
$\sin\theta_{13}$~\cite{Cervera:2000kp,Freund:2001pn,Akhmedov:2004ny}. 
For the discussion in this 
section, we refer to the analytical formulae for vacuum 
oscillation probabilities derived in Ref.~\cite{Kopp:2007ne}, 
which include source and detector NSIs. These formulae 
are valid up to second order in $\dm{21}/\dm{31}$ and 
$\sin\theta_{13}$, and up to first order in the NSI parameters. It 
is easy to observe that linearizing the expressions in the 
NSI parameters and ignoring cubic and higher order 
terms overall, leaves 
only a few NSI parameters in the expressions.
For instance, in the case of the vacuum probability $P_{\mu e}$, only the 
NSI parameters $\epssme$, $\epsdme$ and $\epsdte$ are present up 
to linear order. 
While these approximate analytical formulae provide 
useful insights into the physics of NSIs in neutrino oscillations, we 
stress that all simulation results presented in this work make use of
numerically computed neutrino oscillation probabilities without approximations.

\begin{figure}
\begin{tabular}{ccc}
\qquad\qquad  $\epssme$ & \qquad\qquad  $\epssmm$ & \qquad\qquad  $\epssmt$ \\
\epsfig{file=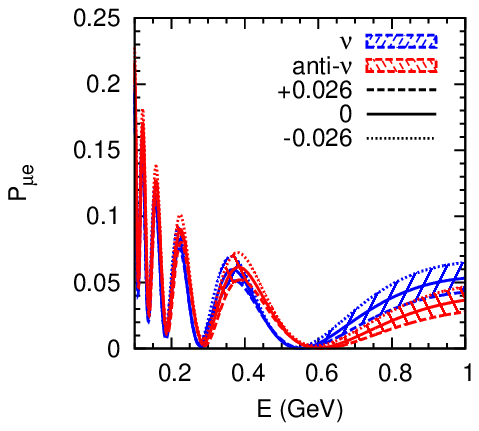, width=0.33\textwidth, bbllx=60, bblly=50, bburx=210, bbury=175,clip=} &
\epsfig{file=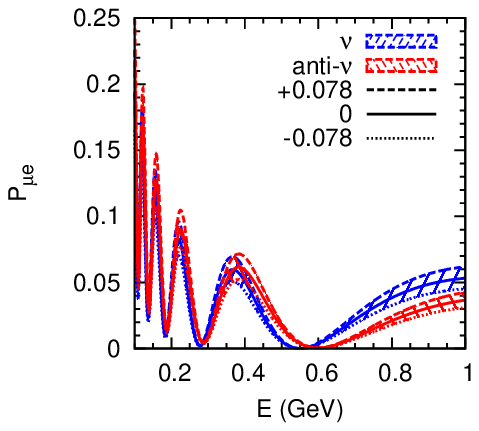, width=0.33\textwidth, bbllx=60, bblly=50, bburx=210, bbury=175,clip=} &
\epsfig{file=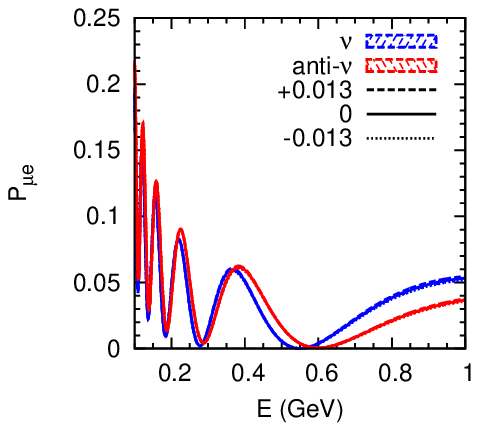, width=0.33\textwidth, bbllx=60, bblly=50, bburx=210, bbury=175,clip=} \\
\qquad\qquad  $\epsdee$ & \qquad\qquad  $\epsdme$ & \qquad\qquad  $\epsdte$ \\
\epsfig{file=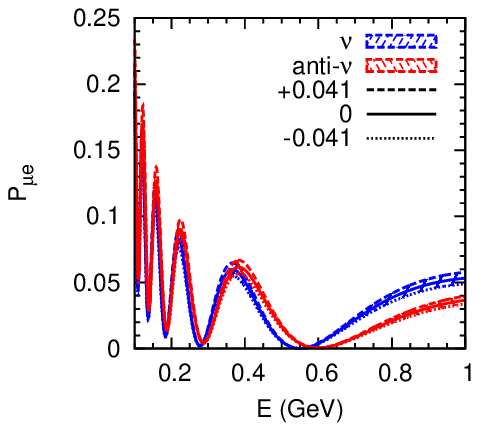, width=0.33\textwidth, bbllx=60, bblly=50, bburx=210, bbury=175,clip=} &
\epsfig{file=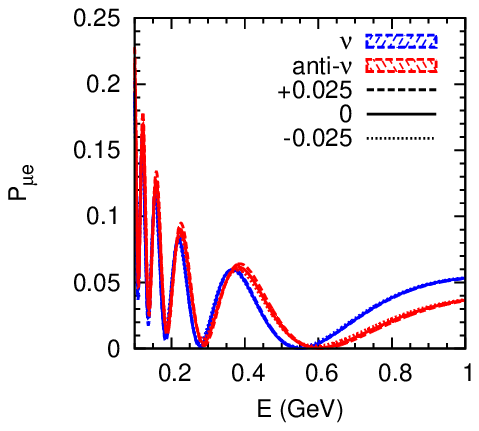, width=0.33\textwidth, bbllx=60, bblly=50, bburx=210, bbury=175,clip=} &
\epsfig{file=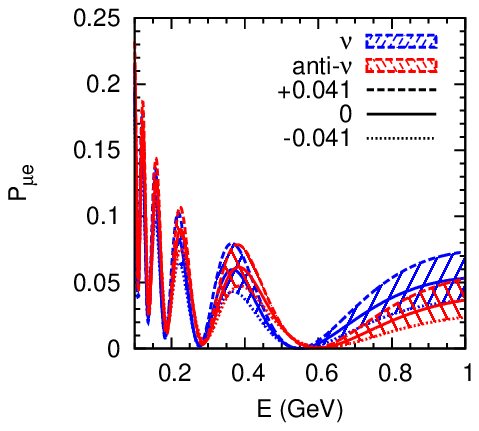, width=0.33\textwidth, bbllx=60, bblly=50, bburx=210, bbury=175,clip=} \\
\end{tabular}
\caption{\footnotesize Neutrino oscillation probability 
$P_{\mu e}$ as a function of the neutrino energy $E$ and its variation 
with each of the relevant NSI parameters. The values of the NSI parameters are chosen within their 90\% C.L.~bounds,
i.e.~assuming their phases to be either 0 or $\pi$. The variation is shown for both neutrinos and antineutrinos.
The values of the fundamental neutrino parameters are set to $\dm{21}=7.6\times10^{-5} \, {\rm eV}^2$, $\dm{31}=2.45\times10^{-3} \, {\rm eV}^2$, 
$\sin^2\theta_{12}=0.304$, $\theta_{23}=45^\circ$, $\sin^2 2\theta_{13}=0.09$ and $\dcp=0$.}
\label{fig:prob}
\end{figure}

In Fig.~\ref{fig:prob}, we have plotted the variation of 
the neutrino oscillation probability $P_{\mu e}$ with the amplitude 
of each of the relevant NSI parameters. The range of 
values chosen for the NSI parameters is the 90\% C.L.~bounds 
on them as listed in Eq.~(\ref{eq:bounds}). 
Each of the probabilities shown are calculated numerically, 
using $\dcp=0$, $\theta_{23}=45^\circ$ and normal 
neutrino mass ordering; and all other NSI parameters, 
including phases, set to zero. Out of the three NSI
parameters present up to linear order, the variation due 
to $\epsdte$ is the strongest, while that due to 
$\epsdme$ is the weakest. This pattern follows from the 
allowed range given by the current bounds.
Out of the remaining three, $\epssmm$ has the greatest 
effect, which is again because it is not very tightly 
constrained by current data. 

\begin{figure}
\begin{tabular}{ccc}
\qquad\qquad $\epssme$ & \qquad\qquad $\epssmm$ & \qquad\qquad $\epssmt$ \\
\epsfig{file=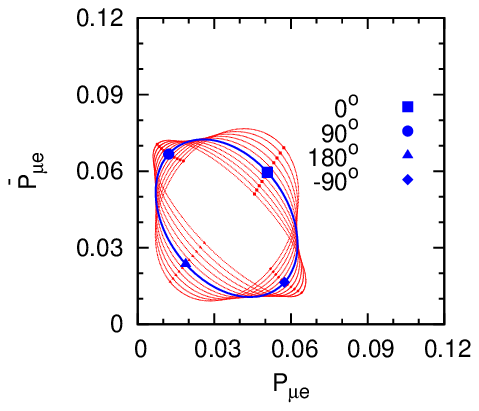, width=0.33\textwidth, bbllx=70, bblly=55, bburx=200, bbury=175,clip=} &
\epsfig{file=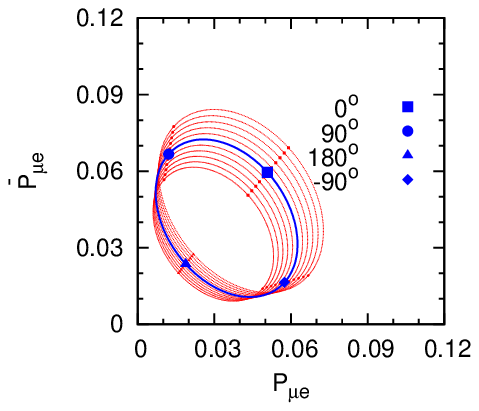, width=0.33\textwidth, bbllx=70, bblly=55, bburx=200, bbury=175,clip=} &
\epsfig{file=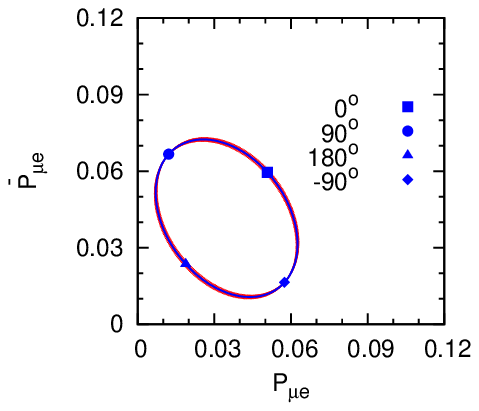, width=0.33\textwidth, bbllx=70, bblly=55, bburx=200, bbury=175,clip=} \\
\qquad\qquad $\epsdee$ & \qquad\qquad $\epsdme$ & \qquad\qquad $\epsdte$ \\
\epsfig{file=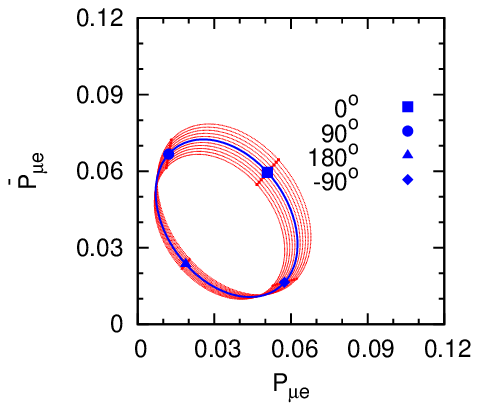, width=0.33\textwidth, bbllx=70, bblly=55, bburx=200, bbury=175,clip=} &
\epsfig{file=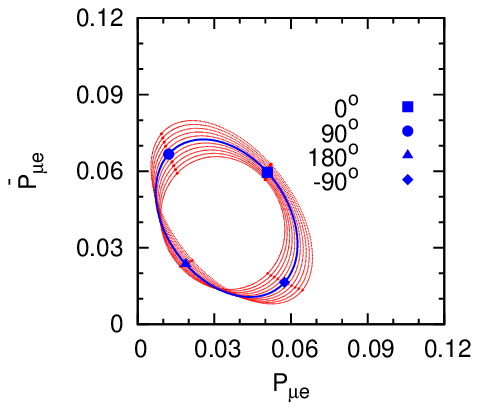, width=0.33\textwidth, bbllx=70, bblly=55, bburx=200, bbury=175,clip=} &
\epsfig{file=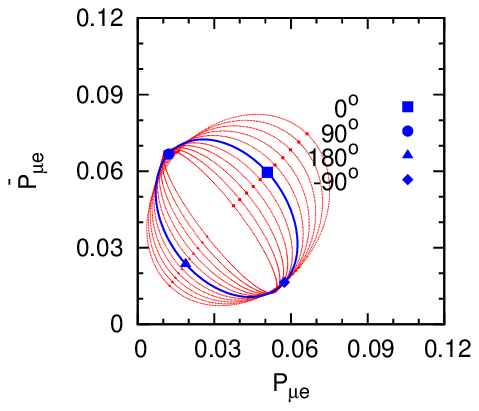, width=0.33\textwidth, bbllx=70, bblly=55, bburx=200, bbury=175,clip=} \\
\end{tabular}
\caption{\footnotesize Bi-probability ellipse for \essnusb\ and its 
variation with the 
relevant NSI parameters. The values of the fundamental neutrino parameters and the NSI parameters are the same as in Fig.~\ref{fig:prob}.}
\label{fig:bi}
\end{figure}

Figure~\ref{fig:prob} is plotted for a fixed value of $\dcp= 0$. For \essnusb, it is interesting to explore the 
interplay between $\dcp$ and the NSI parameters. To this end, 
we show in Fig.~\ref{fig:bi} bi-probability plots for \essnusb. 
This figure is presented for a fixed energy of 400 MeV, which corresponds 
to the second oscillation maximum for the \essnusb\ baseline. 
This is also the energy around which the unoscillated 
event rate is maximal. 
As $\dcp$ varies over its full range, the neutrino and 
antineutrino probabilities trace out an ellipse as shown. In 
the standard case, we obtain the central (blue) ellipse. In each 
of the panels of this figure, one NSI parameter is varied 
within its 90\% C.L.~bound, which gives the spread in the 
ellipse. 

In order to explain the features observed in Fig.~\ref{fig:bi}, 
we define the variation of the neutrino oscillation probability as
\begin{equation}
 \Delta P_{\mu e}^{\rm vac} (\varepsilon^x_{\alpha\beta}) = P_{\mu e}^{\rm vac} (\varepsilon^x_{\alpha\beta}) - P_{\mu e}^{\rm vac} (\varepsilon^x_{\alpha\beta}=0) ~,
\end{equation}
where $\alpha,\beta \in \{e,\mu,\tau \}$ 
and $x \in \{s,d \}$.
Using the perturbative analytical expression for $P_{\mu e}^{\rm vac}$ from 
Ref.~\cite{Kopp:2007ne}, we obtain for the cases of $\epssme$, $\epsdme$ and $\epsdmt$ 
\begin{eqnarray}
 \Delta P_{\mu e}^{\rm vac} (\epssme) & \simeq & - 4 |\epssme| \sin\theta_{13} \sin\theta_{23} \sin\left( \Delta+\dcp \right) \sin\Delta ~, \label{eq:delP1} \\
 \Delta P_{\mu e}^{\rm vac} (\epsdme) & \simeq & - 4 |\epsdme| \sin\theta_{13} \cos 2\theta_{23} \sin\theta_{23} \cos\dcp \sin^2\Delta \nonumber \\
 && - 2 |\epsdme| \sin\theta_{13} \sin\theta_{23} \sin\dcp \sin 2\Delta \nonumber \\
 && + |\epsdme| \frac{\dm{21}}{\dm{31}} \Delta \sin 2\theta_{12} \sin 2\theta_{23} \sin\theta_{23} \sin 2\Delta ~, \\
 \Delta P_{\mu e}^{\rm vac} (\epsdte) & \simeq & 4 |\epsdte| \sin\theta_{13} \sin 2\theta_{23} \sin\theta_{23} \cos\dcp \sin^2\Delta \nonumber \\
 && + |\epsdte| \frac{\dm{21}}{\dm{31}} \Delta \sin 2\theta_{12} \sin 2\theta_{23} \cos\theta_{23} \sin 2\Delta  ~, \label{eq:delP3}
 \end{eqnarray}
where $\Delta \equiv \dm{31}L/(4E)$. In deriving each of  
Eqs.~(\ref{eq:delP1})-(\ref{eq:delP3}), we have set all other 
NSI parameters to zero.
Note that for the cases of $\epssmm$, $\epssmt$ and $\epsdee$, 
there are no linear-order terms in the corresponding formulae,
and the dependence on the NSI parameters only appear at second
order and above.
First, we observe (as in Fig.~\ref{fig:prob})
that the variation of $P_{\mu e}$ is the 
largest for $\epsdte$ (due to linear variation and weakest upper bound) and
the smallest for $\epssmt$ (due to higher-order variation and strongest 
upper bound). For $\epssme$, $\epssmm$, $\epsdee$ and $\epsdme$, 
the variations are intermediate,
depending on a non-trivial combination between the value of 
the upper bound on the considered NSI parameter and if this NSI 
parameter appears at linear order or not in the variation.
Second, we can explain the structure of the band for each panel. 
We illustrate this for the case of $\epssme$. 
For the baseline and energy considered, $\Delta$ evaluates 
to around $-120^\circ$, close to the second oscillation maximum. 
It is then easy to see that the maximum `width' of the 
band occurs when $\Delta+\dcp=\pm 90^\circ$, i.e.~when 
$\dcp$ is around $30^\circ$ or $-150^\circ$. Likewise, 
for $\Delta+\dcp=0,180^\circ$, the probability becomes 
independent of $\epssme$, and the band `pinches off'. This 
occurs when $\dcp$ is around $120^\circ$ or $-60^\circ$. 
For antineutrinos, the sign of $\dcp$ is changed, and one 
can use similar arguments to find the broadest and 
narrowest points along the $\overline{P_{\mu e}}$ axis 
as well. 

\begin{figure}
\begin{tabular}{ccc}
\qquad\qquad $\epssme$ & \qquad\qquad $\epssmm$ & \qquad\qquad $\epssmt$ \\
\epsfig{file=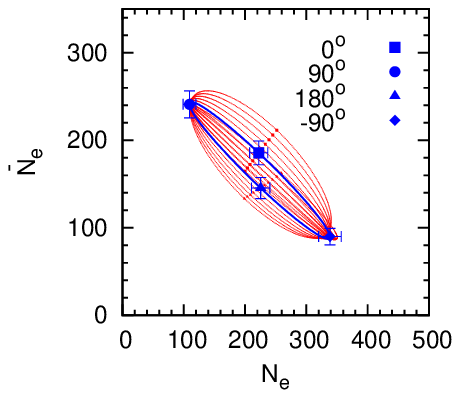, width=0.33\textwidth, bbllx=70, bblly=55, bburx=205, bbury=175,clip=} &
\epsfig{file=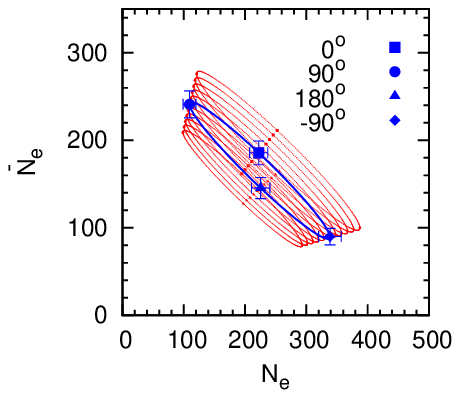, width=0.33\textwidth, bbllx=70, bblly=55, bburx=205, bbury=175,clip=} &
\epsfig{file=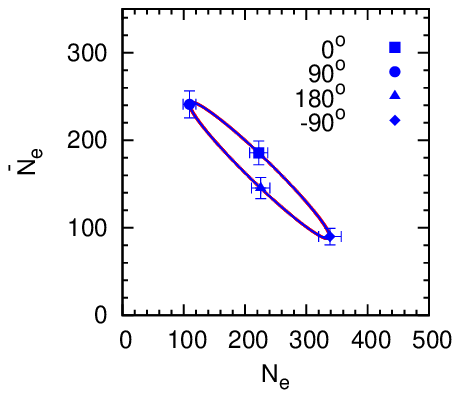, width=0.33\textwidth, bbllx=70, bblly=55, bburx=205, bbury=175,clip=} \\
\qquad\qquad $\epsdee$ & \qquad\qquad $\epsdme$ & \qquad\qquad $\epsdte$ \\
\epsfig{file=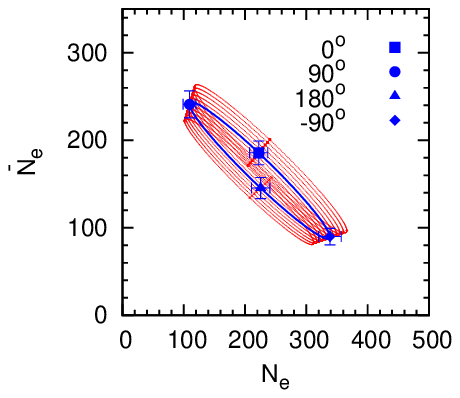, width=0.33\textwidth, bbllx=70, bblly=55, bburx=205, bbury=175,clip=} &
\epsfig{file=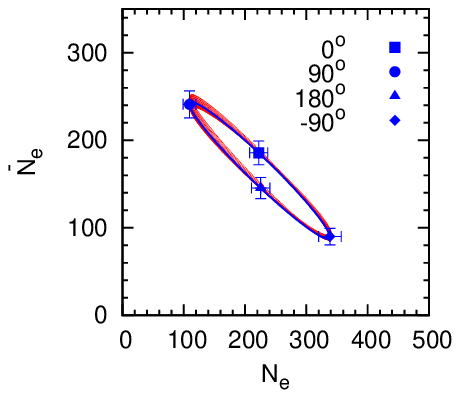, width=0.33\textwidth, bbllx=70, bblly=55, bburx=205, bbury=175,clip=} &
\epsfig{file=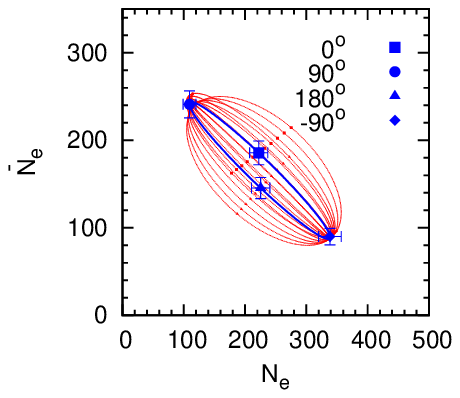, width=0.33\textwidth, bbllx=70, bblly=55, bburx=205, bbury=175,clip=} \\
\end{tabular}
\caption{\footnotesize Bi-rates ellipse for \essnusb\ and its variation with the 
relevant NSI parameters. The values of the fundamental neutrino parameters and the NSI parameters are the same as in Fig.~\ref{fig:prob}.}
\label{fig:birate}
\end{figure}

Finally, Fig.~\ref{fig:birate} shows the neutrino and antineutrino event rates for \essnusb\ in a bi-rate plot, using the same parameter values as for Fig.~\ref{fig:bi}. The event rates plotted are the 
total rates across all energy bins. Statistical error bars have been included for four representative values of $\dcp$. In addition to the variation of the probabilities in Fig.~\ref{fig:bi}, this figure gives a first indication of the impact of NSIs versus the possible experimental resolution of the \essnusb. Where the experimental error bars on the total event rates are smaller than the possible variation of the NSI parameters, the \essnusb\ will generally be sensitive to NSIs smaller than the current bounds. However, note that the converse is not necessarily true as the experimental results do not only include the total event rates, but also spectral information, which may also be used to constrain the NSIs. In particular, this will be apparent for our results on $\epsdme$, which does not change the event rates significantly. 

\section{Results on NSIs at \essnusb}
\label{sec:results}

The main goal of the proposed \essnusb\ experiment is to measure 
the CP-violating phase $\dcp$ with high precision. In this section, 
we examine both the impact of the NSI parameters on 
this $\dcp$ measurement and the ability of \essnusb\ to 
measure the NSI parameters.

The central values of the neutrino parameters $\dm{21}$, $|\dm{31}|$, 
$\theta_{12}$ and $\theta_{13}$ are taken close to their 
current best-fit values~\cite{Capozzi:2013csa,Forero:2014bxa,Gonzalez-Garcia:2014bfa}. 
We have also imposed 
Gaussian priors on these parameters with a width obtained 
from these global fits. The values of $\theta_{23}$ and $\dcp$ 
used are different in each case, and are specified in the 
text. In addition, we have assumed a 5\% prior on the true value of $\sin^2 2\theta_{23}$. The NSI parameters are of the form 
$\varepsilon^x_{\alpha\beta}$, where $\alpha,\beta \in \{e,\mu,\tau \}$ 
and $x \in \{s,d \}$, since the source and detector NSI 
parameters can be different in general. Thus, we have 18 
complex NSI parameters, or 36 real NSI parameters, in addition 
to the standard ones.
We have run our simulations for both normal (NO) and inverted (IO)
neutrino mass ordering. We find that there is very little qualitative 
difference between the results in these two cases. Therefore, 
in what follows, we show only the NO results.

\subsection{Effect on precision measurement at \essnusb\ }

In this subsection, we discuss the interplay 
between NSI parameters and the $\dcp$ precision of \essnusb. 
The results are shown in the form of precision contours in 
the $\theta_{23}-\dcp$ plane. This is performed for three 
representative values of $\theta_{23} \in \{ 42^\circ, 45^\circ, 48^\circ \}$; 
and four representative values of $\dcp \in \{-90^\circ, 0^\circ, 90^\circ, 180^\circ \}$. 

\begin{figure}[htb]
\begin{tabular}{cccc}
 & \qquad NO, $\theta_{23}=42^\circ$ & \qquad NO, $\theta_{23}=45^\circ$ & \qquad NO, $\theta_{23}=48^\circ$ \\
\rotatebox[origin=l]{90}{\qquad\qquad\qquad $\dcp=-90^\circ$} &
\epsfig{file=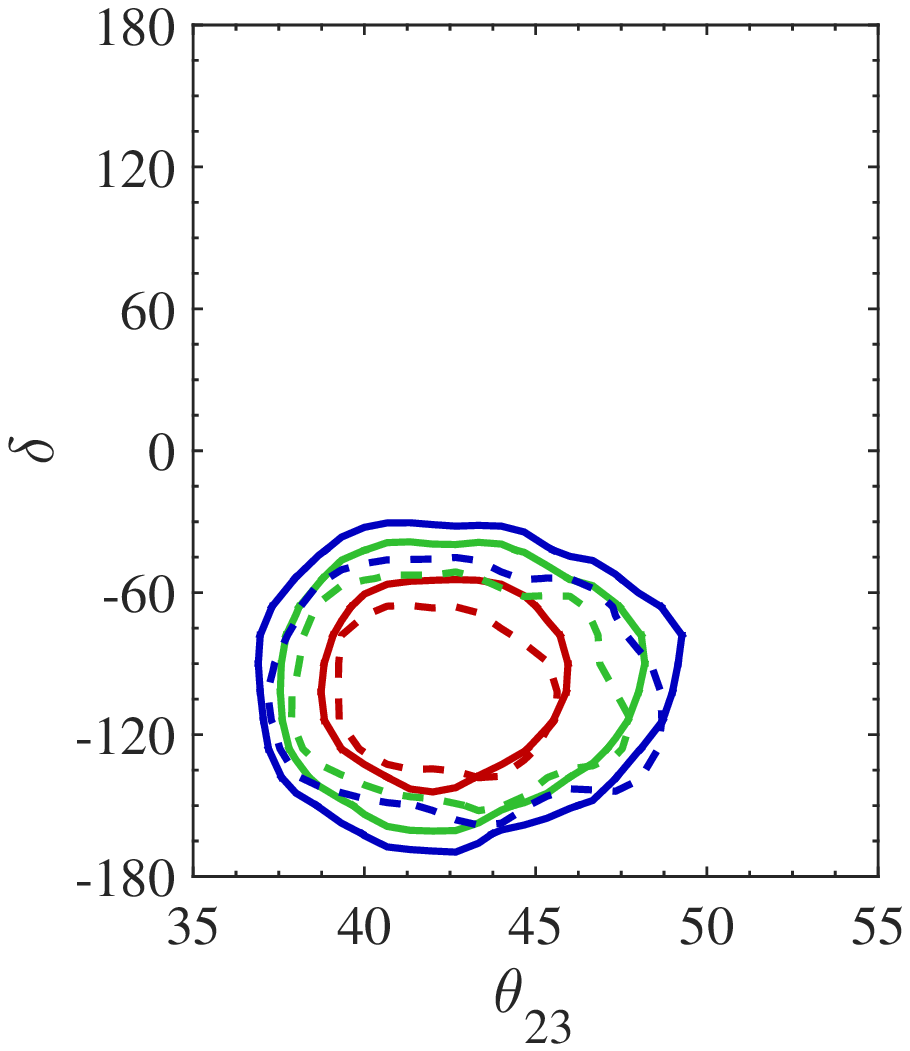, width=0.27\textwidth, bbllx=15, bblly=0, bburx=310, bbury=301,clip=} &
\epsfig{file=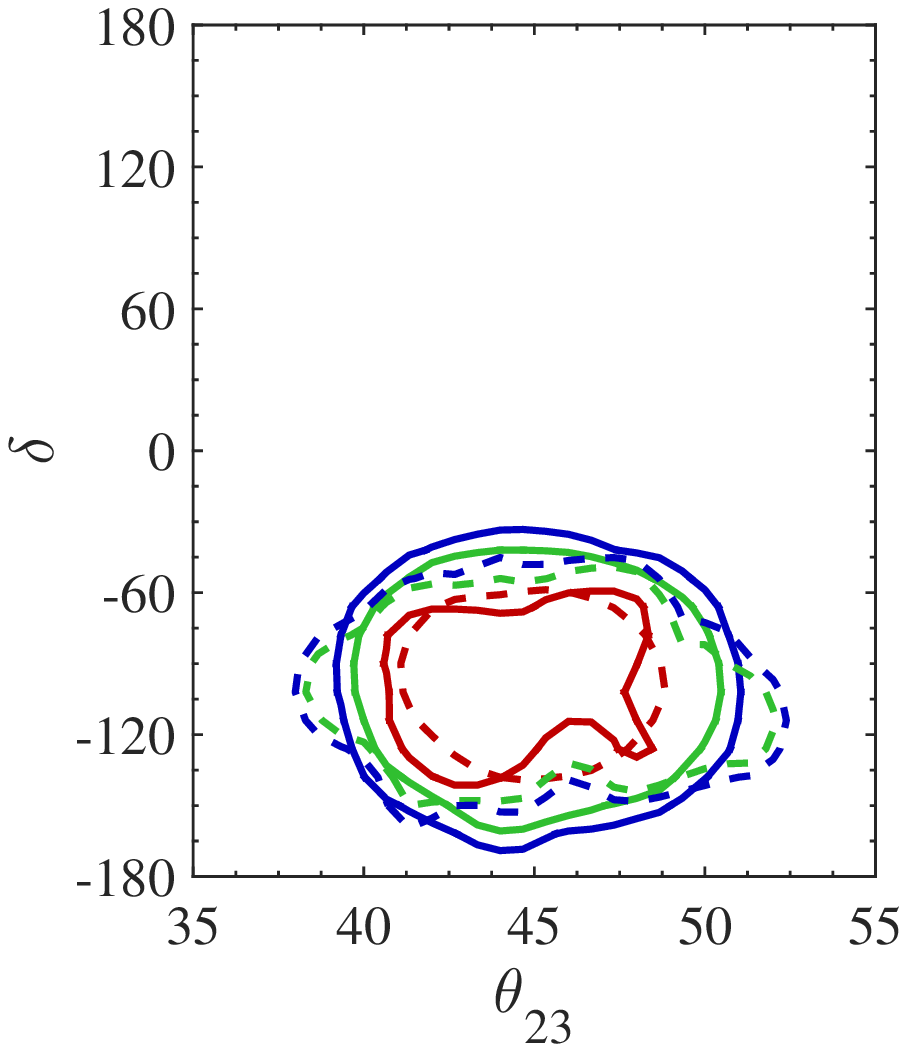, width=0.27\textwidth, bbllx=15, bblly=0, bburx=310, bbury=301,clip=} &
\epsfig{file=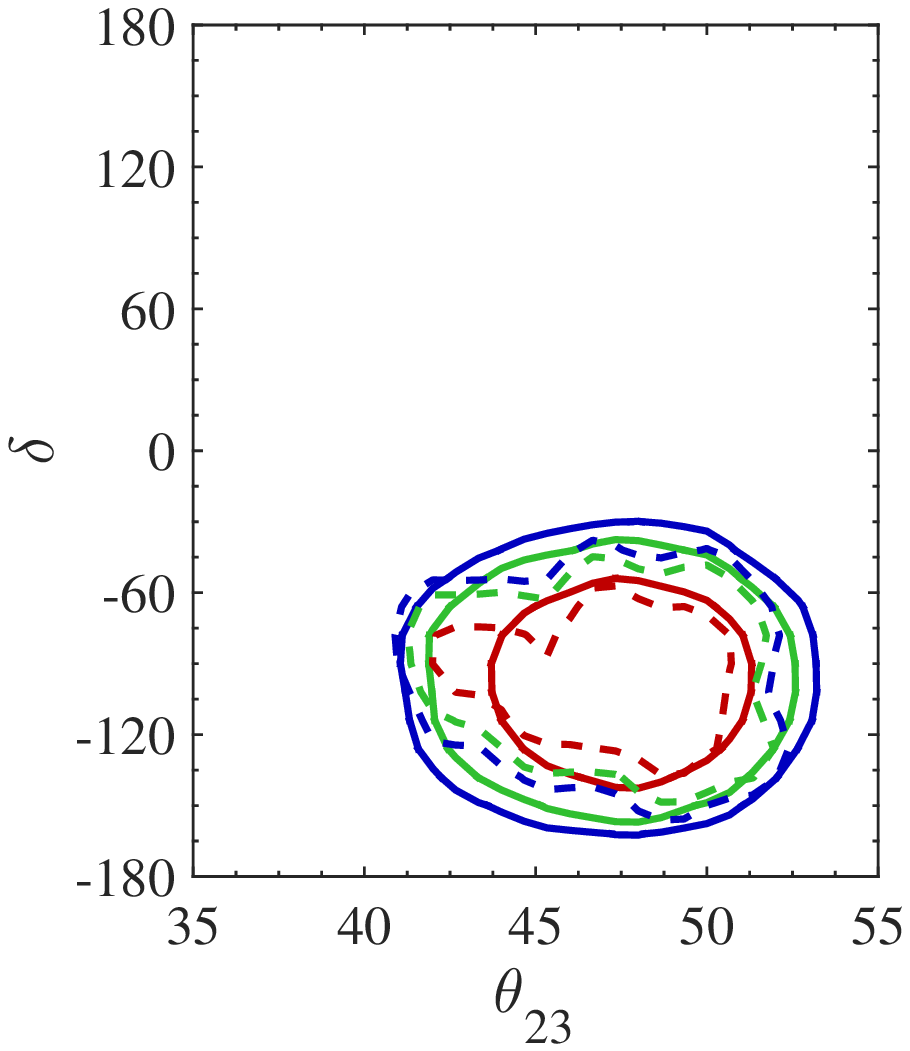, width=0.27\textwidth, bbllx=15, bblly=0, bburx=310, bbury=301,clip=} \\
\rotatebox[origin=l]{90}{\qquad\qquad\qquad $\dcp=0^\circ$} &
\epsfig{file=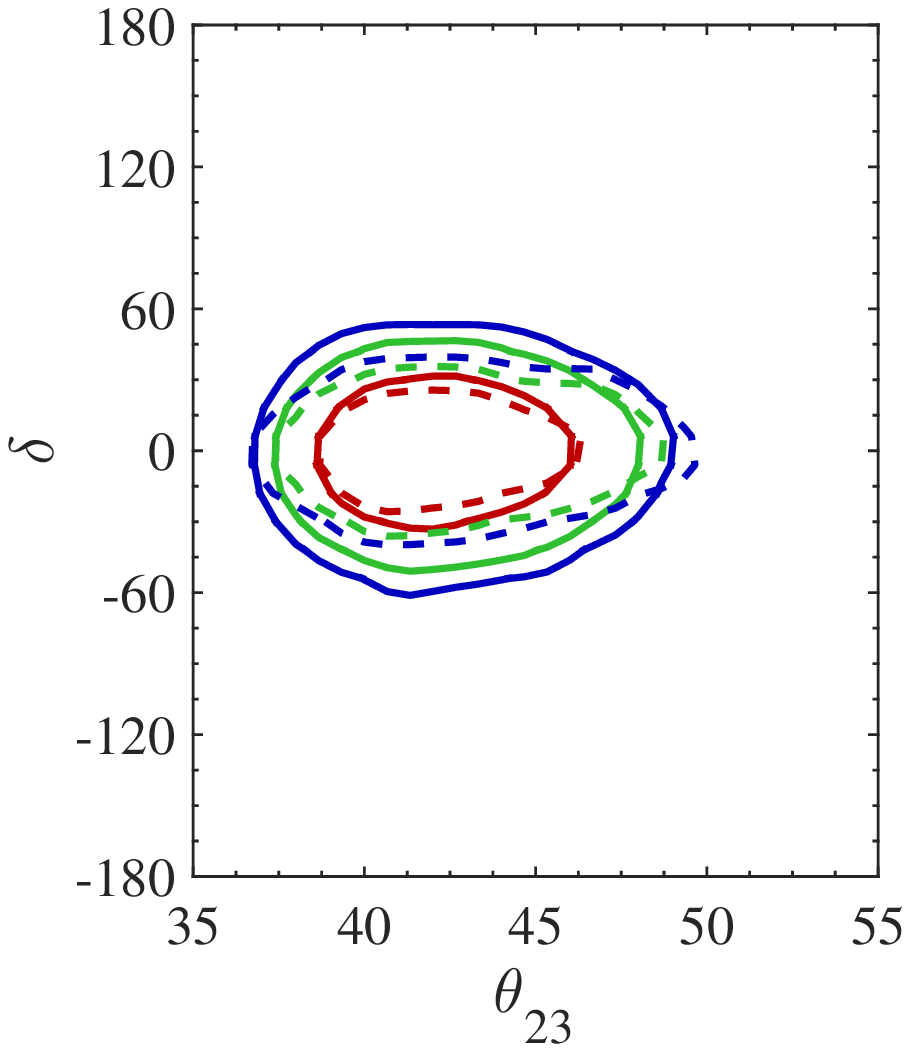, width=0.27\textwidth, bbllx=15, bblly=0, bburx=310, bbury=301,clip=} &
\epsfig{file=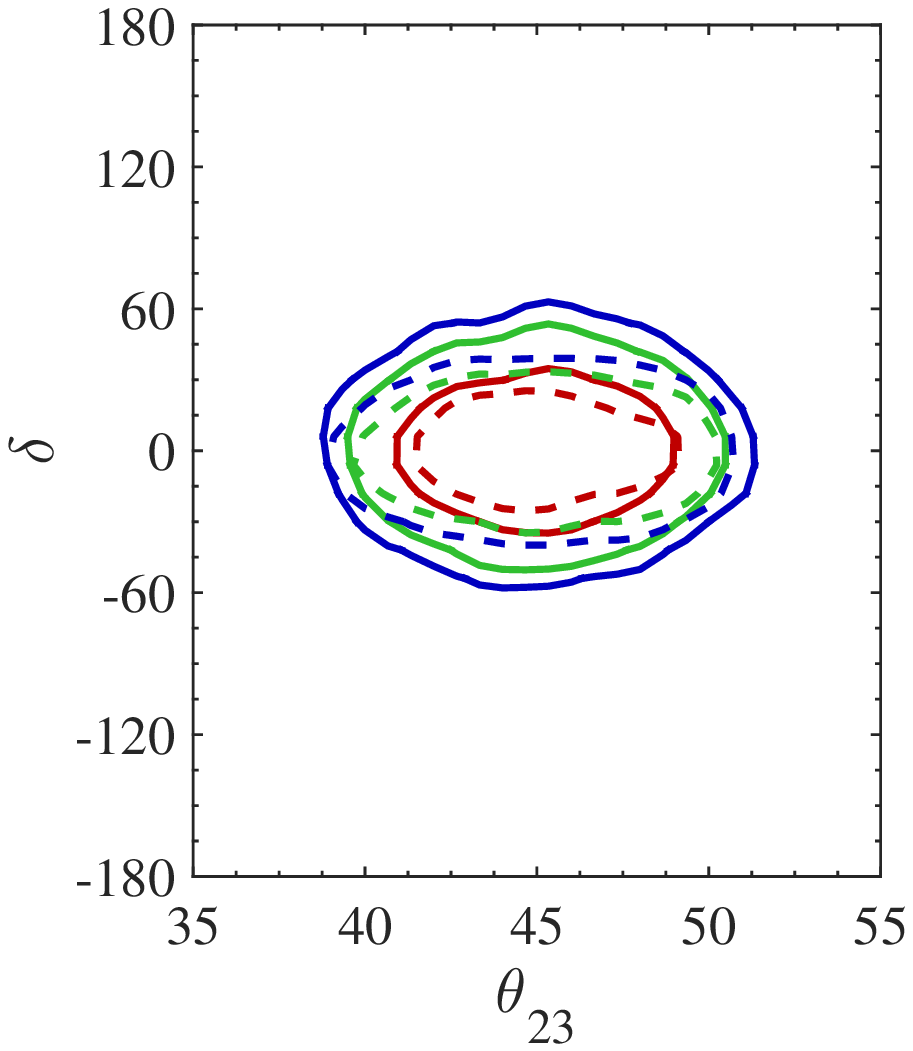, width=0.27\textwidth, bbllx=15, bblly=0, bburx=310, bbury=301,clip=} &
\epsfig{file=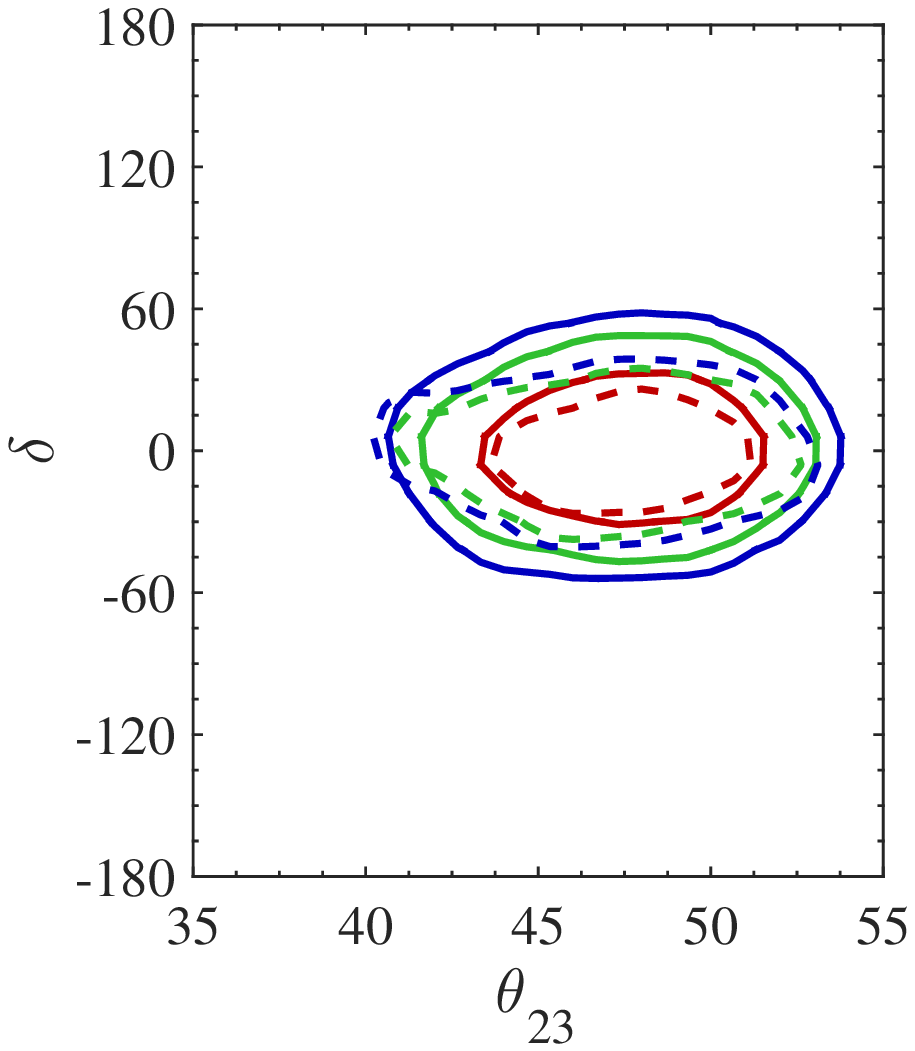, width=0.27\textwidth, bbllx=15, bblly=0, bburx=310, bbury=301,clip=} \\
\rotatebox[origin=l]{90}{\qquad\qquad\qquad $\dcp=90^\circ$} &
\epsfig{file=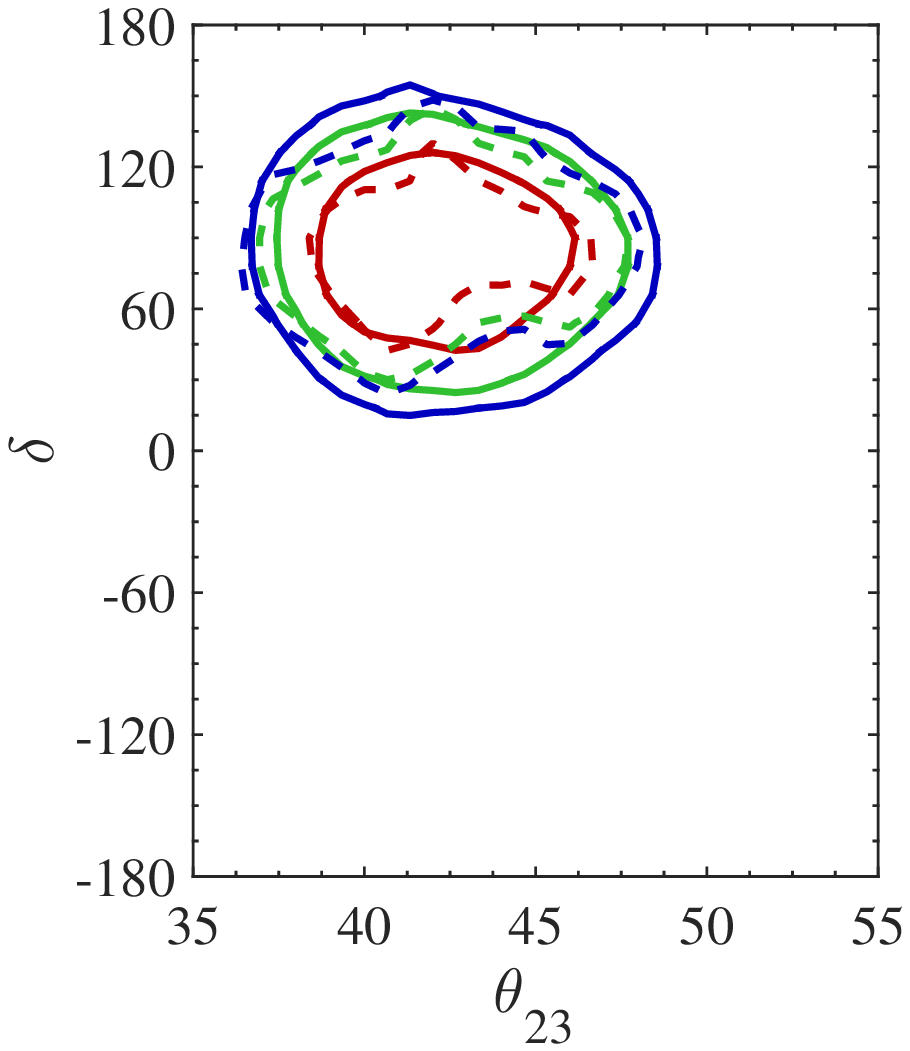, width=0.27\textwidth, bbllx=15, bblly=0, bburx=310, bbury=301,clip=} &
\epsfig{file=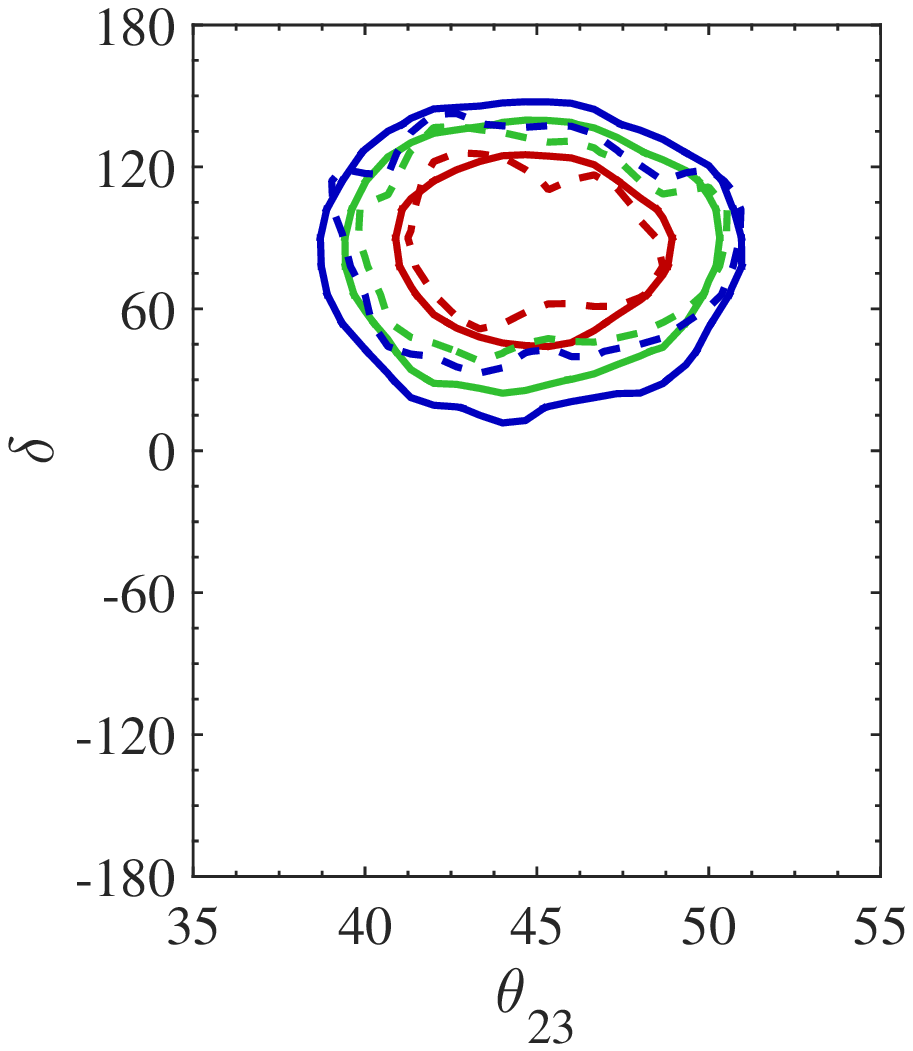, width=0.27\textwidth, bbllx=15, bblly=0, bburx=310, bbury=301,clip=} &
\epsfig{file=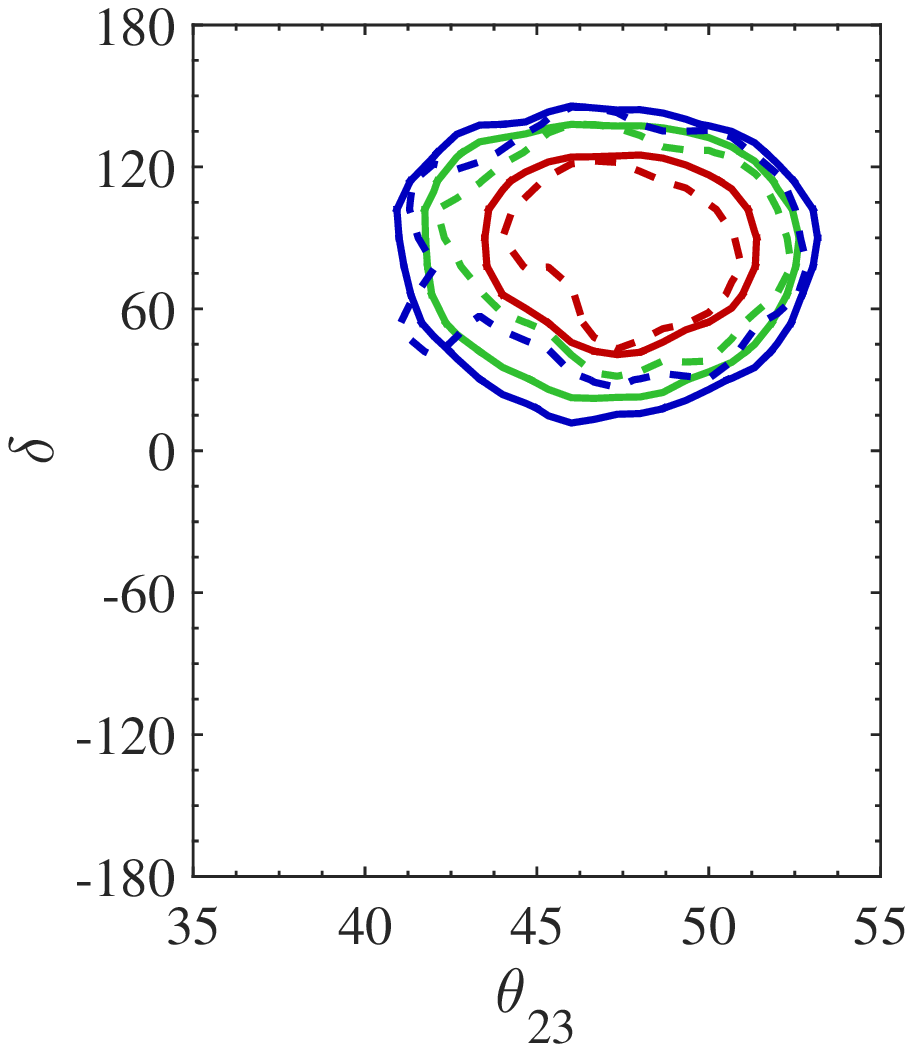, width=0.27\textwidth, bbllx=15, bblly=0, bburx=310, bbury=301,clip=} \\
\rotatebox[origin=l]{90}{\qquad\qquad\qquad $\dcp=180^\circ$} &
\epsfig{file=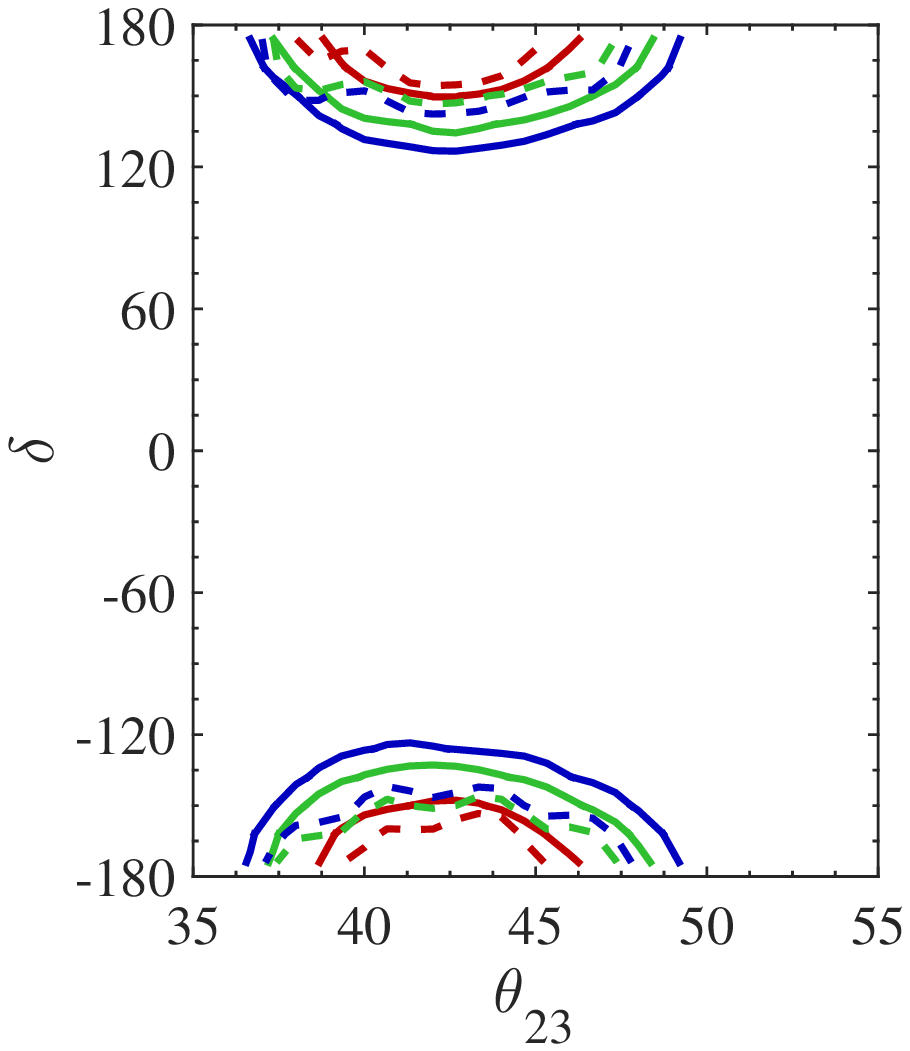, width=0.27\textwidth, bbllx=15, bblly=0, bburx=310, bbury=301,clip=} &
\epsfig{file=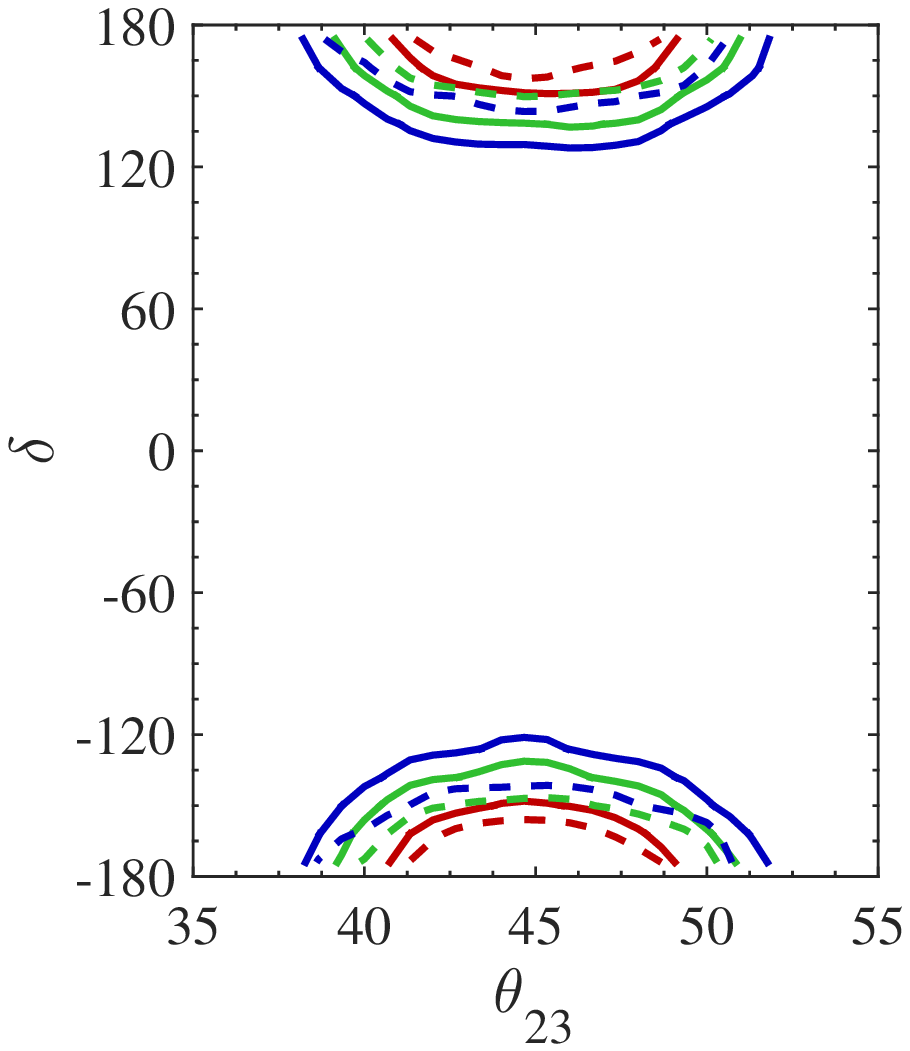, width=0.27\textwidth, bbllx=15, bblly=0, bburx=310, bbury=301,clip=} &
\epsfig{file=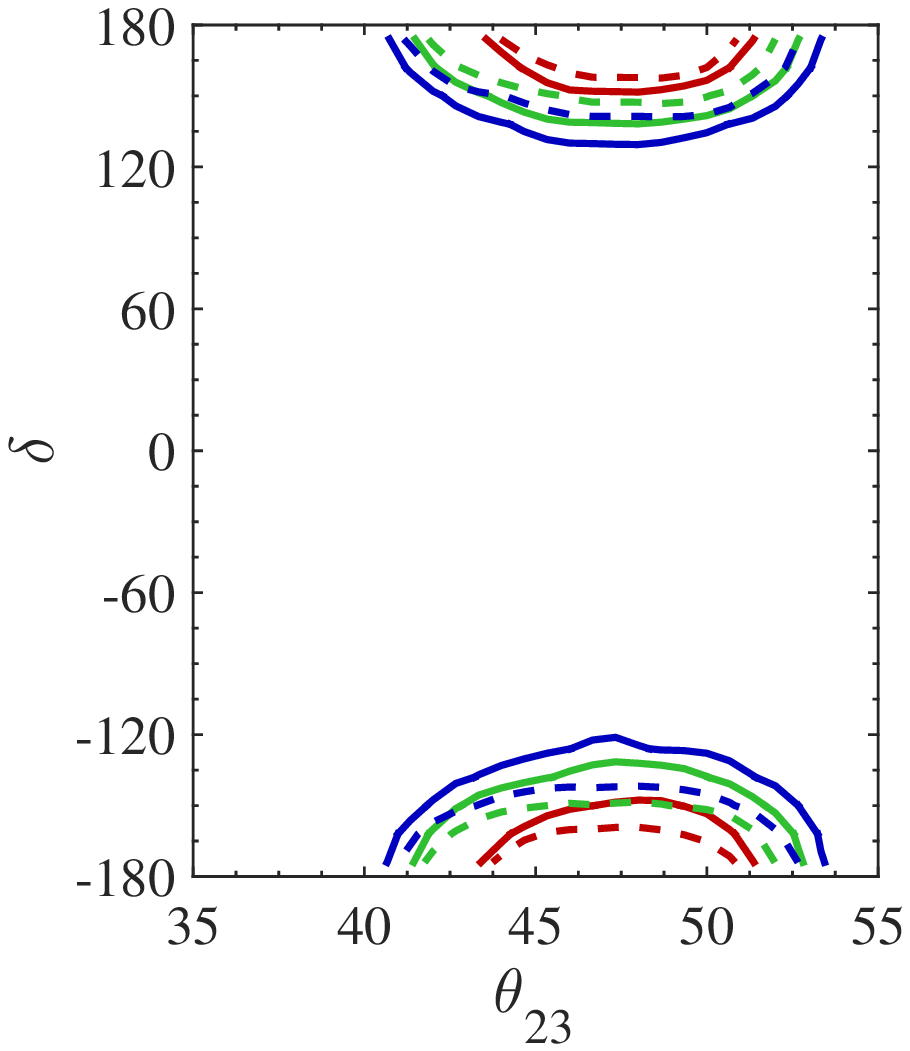, width=0.27\textwidth, bbllx=15, bblly=0, bburx=310, bbury=301,clip=} 
\end{tabular}
\caption{\footnotesize Effect of marginalizing over source and 
detector NSI parameters on precision measurements at \essnusb. 
Each panel shows the allowed region in the test $\theta_{23}-\dcp$ plane, when
the NSI parameters are taken to be zero in the data. The red, green and blue curves 
represent the 68\%, 90\% and 95\% C.L.~contours, respectively. The solid contours 
show the effect of marginalization over the NSI parameters, whereas the dashed contours 
are for the standard oscillation scenario in the absence of NSIs. }
\label{fig:effect_0}
\end{figure}

First, we explore the effect of marginalizing over the source and 
detector NSI parameters on precision measurements at \essnusb, 
in the special case the true NSI parameters are zero. 
In other words, we take all the NSI parameters to be zero 
when generating the mock data, but allow them all to vary in the fit. Thus, these 
plots show the robustness of the \essnusb\ measurements 
against a scan for NSIs. The results are 
shown in Fig.~\ref{fig:effect_0}. The solid 
curves show the 68\%, 90\% and 95\% C.L.~contours for the 
allowed region in the parameter space. The dashed contours 
are for the standard case where there are no NSI parameters 
in the data or the fit. 

We observe that the search for NSIs does not affect the 
$\theta_{23}$ precision of \essnusb\ much. The 
precision in $\dcp$ is worsened to at most twice its 
standard precision, in the worst case. For most cases, 
the precision is seen to be quite robust, even in spite of 
a severely enlarged parameter space. This is true, 
irrespective of the true value of $\theta_{23}$ or 
$\dcp$. 

\begin{figure}[htb]
\begin{tabular}{cccc}
 & \qquad NO, $\theta_{23}=42^\circ$ & \qquad NO, $\theta_{23}=45^\circ$ & \qquad NO, $\theta_{23}=48^\circ$ \\
\rotatebox[origin=l]{90}{\qquad\qquad\qquad $\dcp=-90^\circ$} &
\epsfig{file=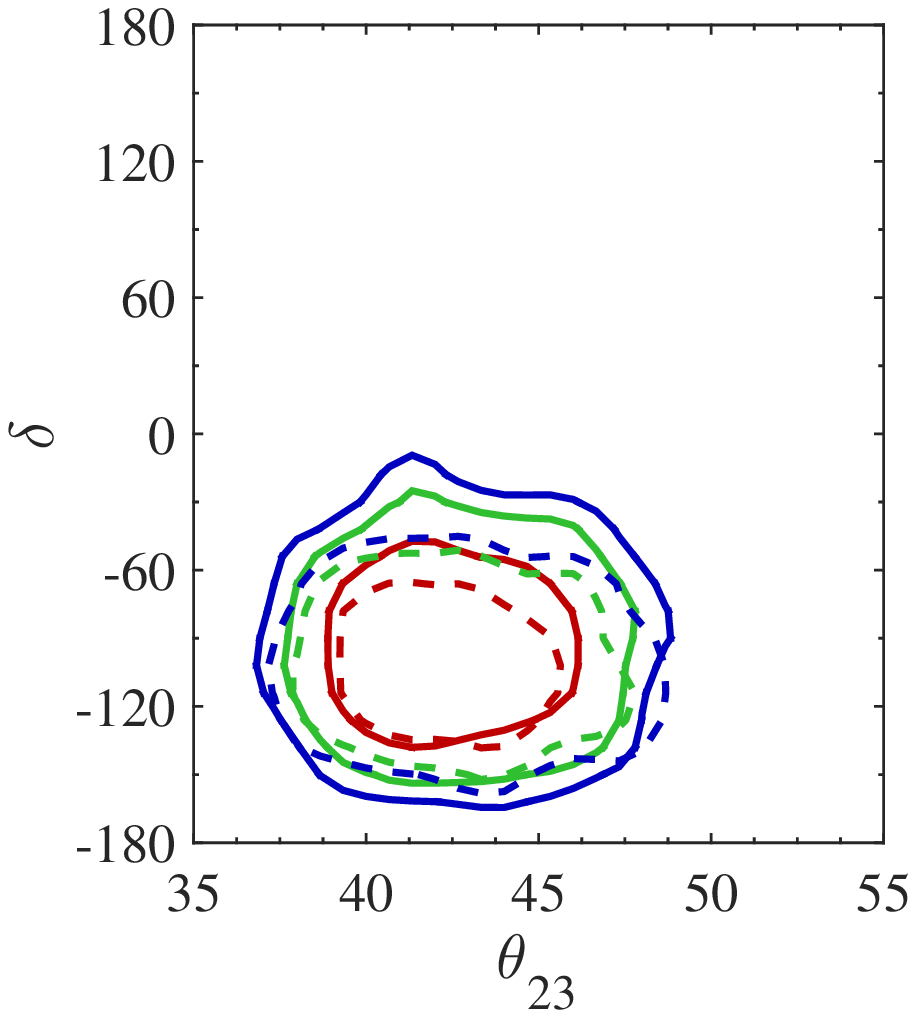, width=0.27\textwidth, bbllx=15, bblly=0, bburx=310, bbury=290,clip=} &
\epsfig{file=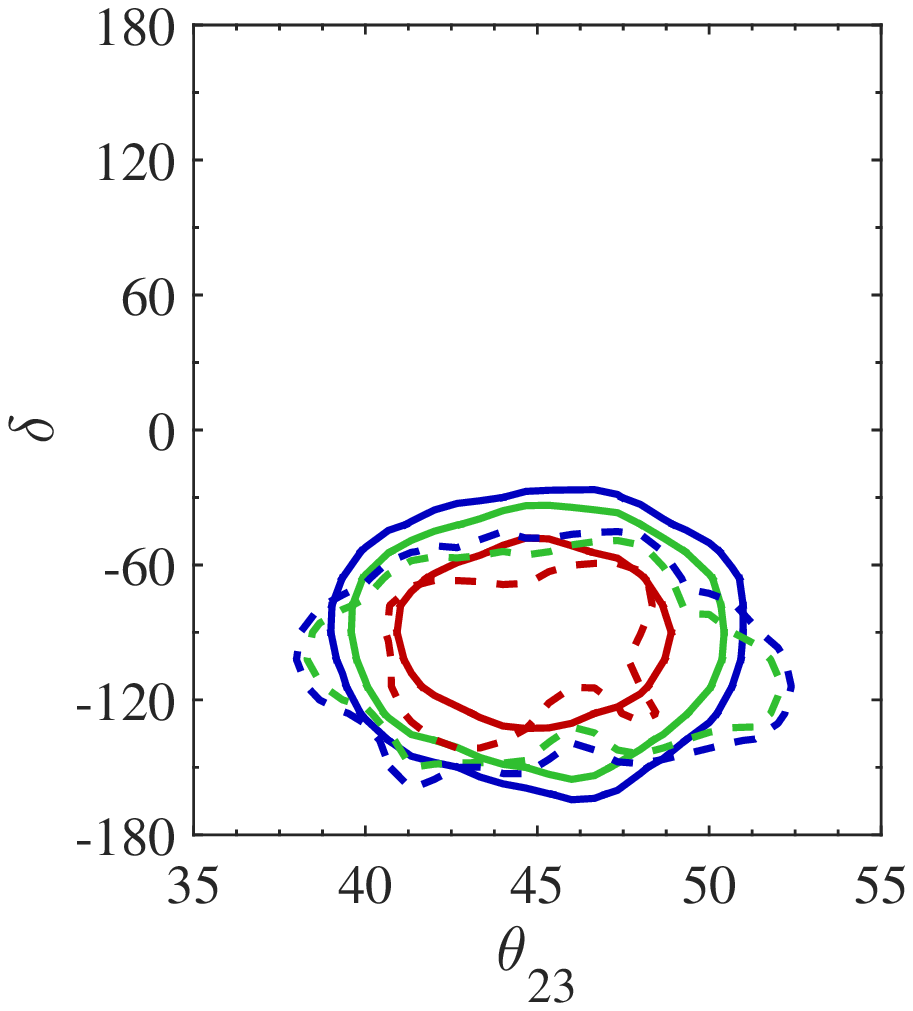, width=0.27\textwidth, bbllx=15, bblly=0, bburx=310, bbury=290,clip=} &
\epsfig{file=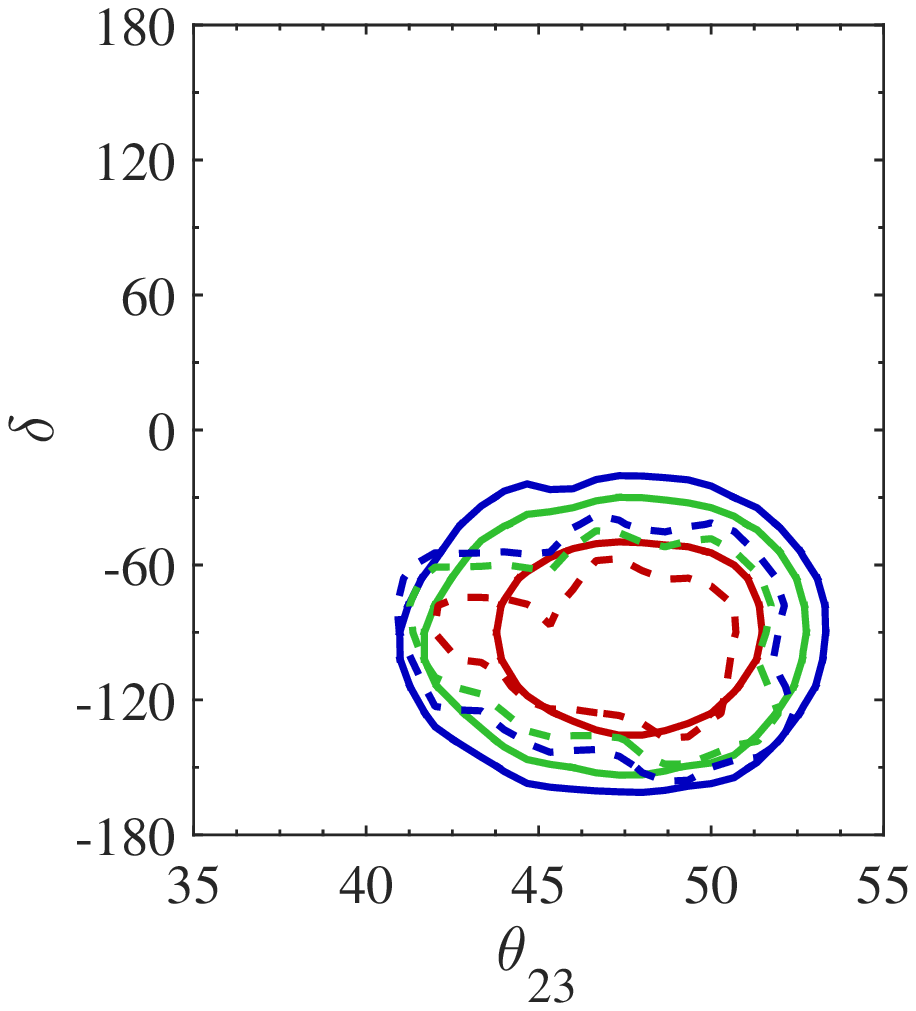, width=0.27\textwidth, bbllx=15, bblly=0, bburx=310, bbury=290,clip=} \\
\rotatebox[origin=l]{90}{\qquad\qquad\qquad $\dcp=0^\circ$} &
\epsfig{file=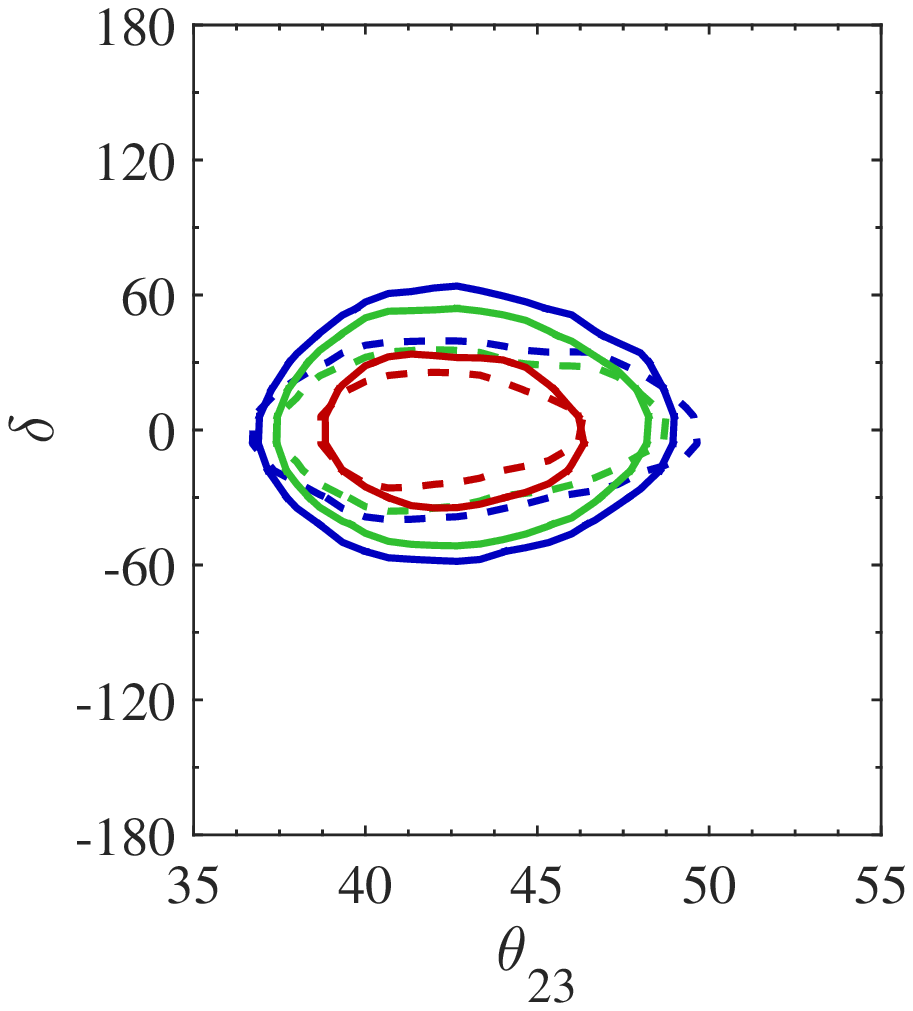, width=0.27\textwidth, bbllx=15, bblly=0, bburx=310, bbury=290,clip=} &
\epsfig{file=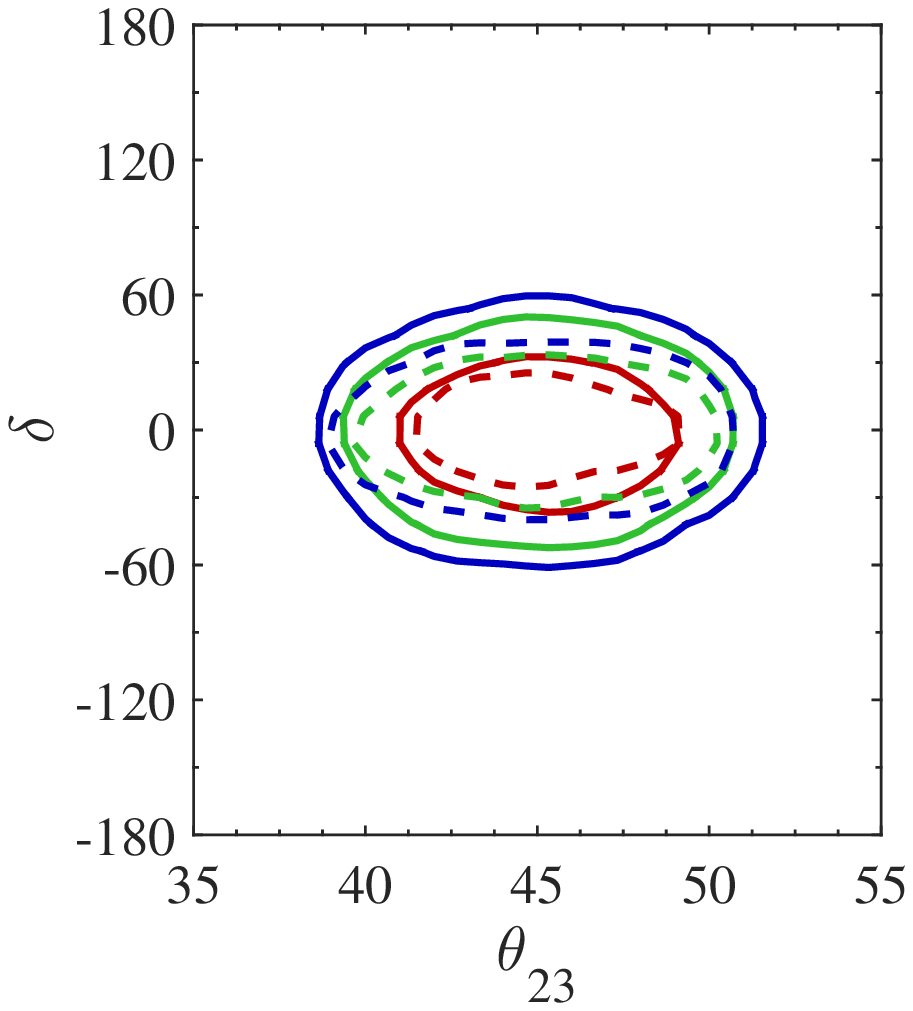, width=0.27\textwidth, bbllx=15, bblly=0, bburx=310, bbury=290,clip=} &
\epsfig{file=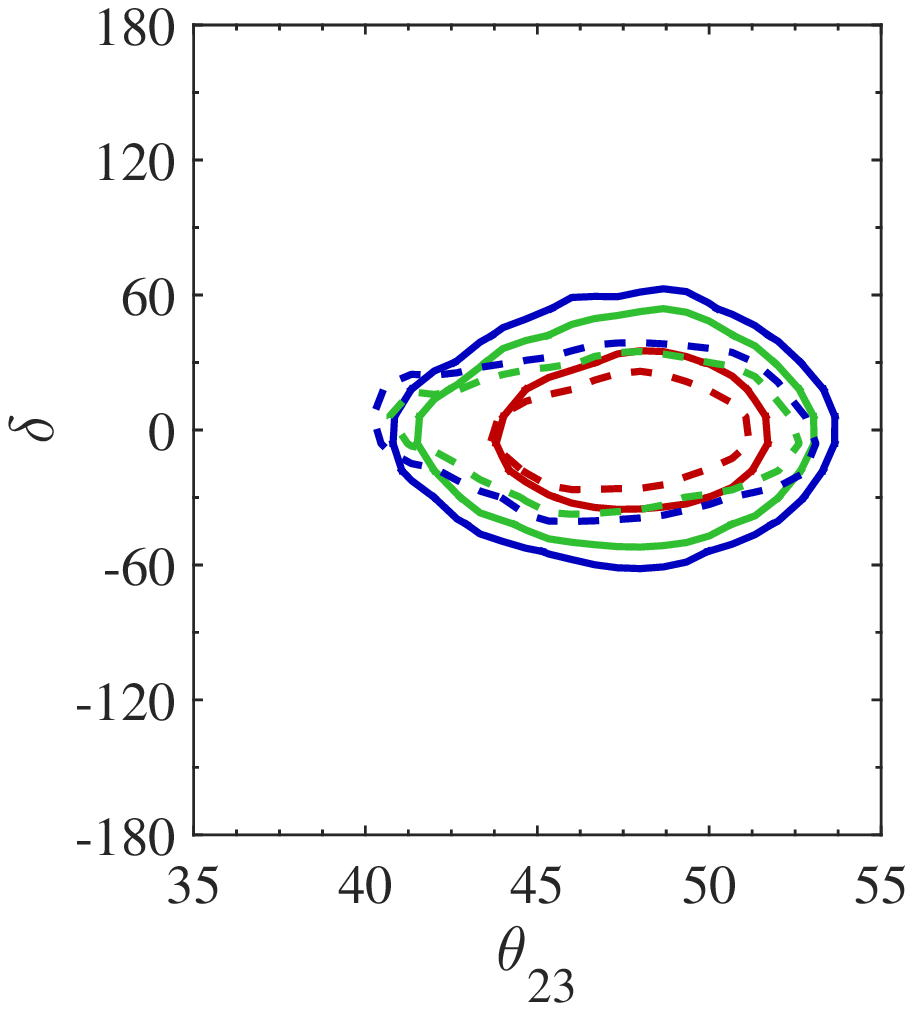, width=0.27\textwidth, bbllx=15, bblly=0, bburx=310, bbury=290,clip=} \\
\rotatebox[origin=l]{90}{\qquad\qquad\qquad $\dcp=90^\circ$} &
\epsfig{file=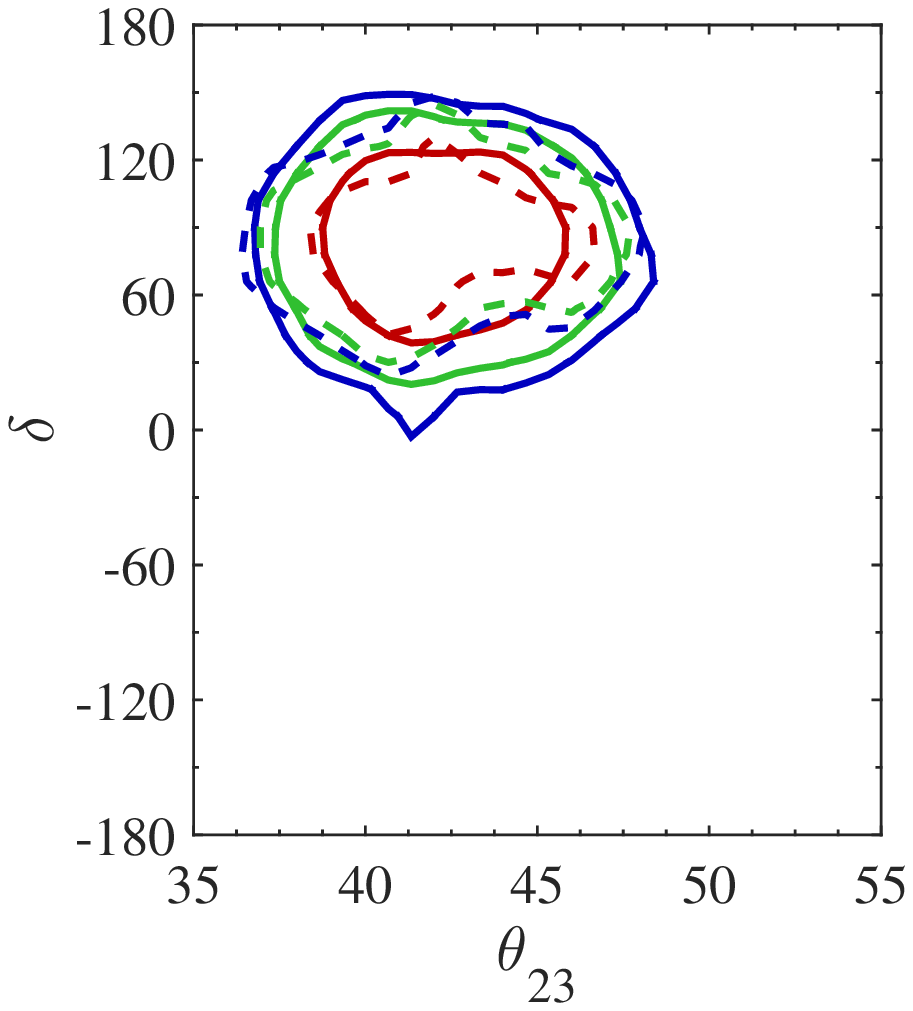, width=0.27\textwidth, bbllx=15, bblly=0, bburx=310, bbury=290,clip=} &
\epsfig{file=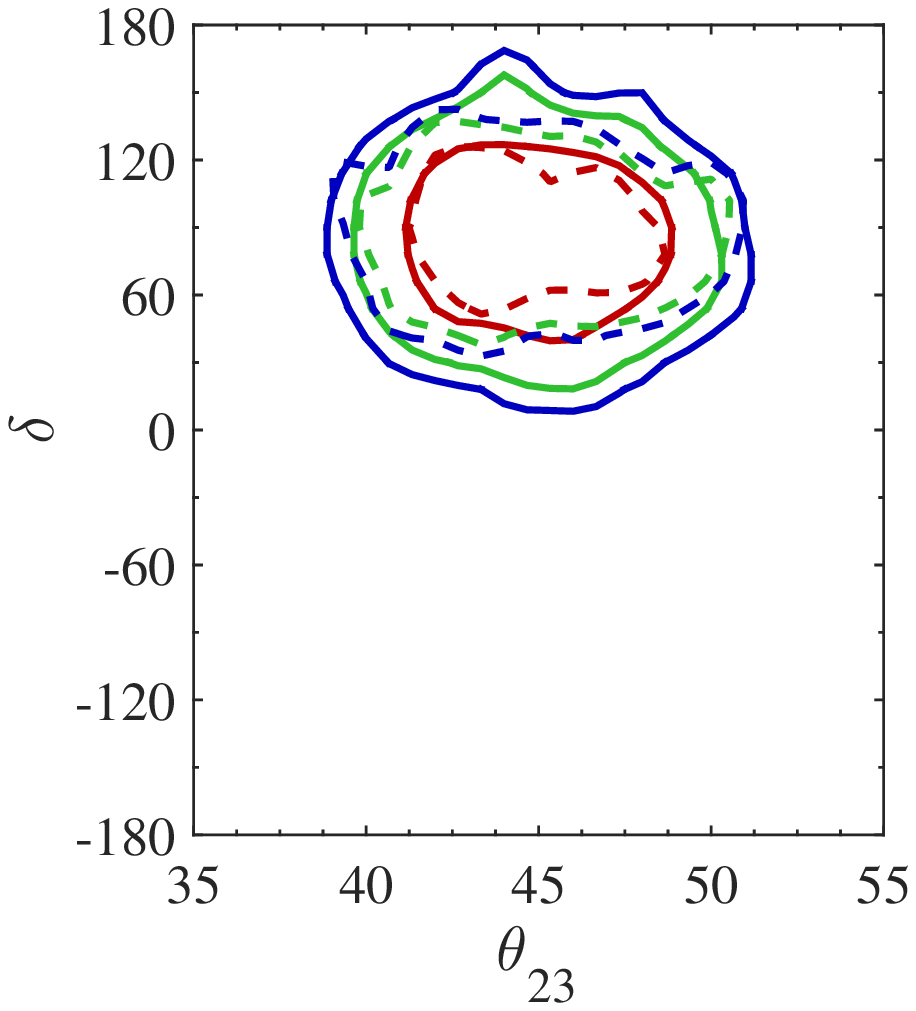, width=0.27\textwidth, bbllx=15, bblly=0, bburx=310, bbury=290,clip=} &
\epsfig{file=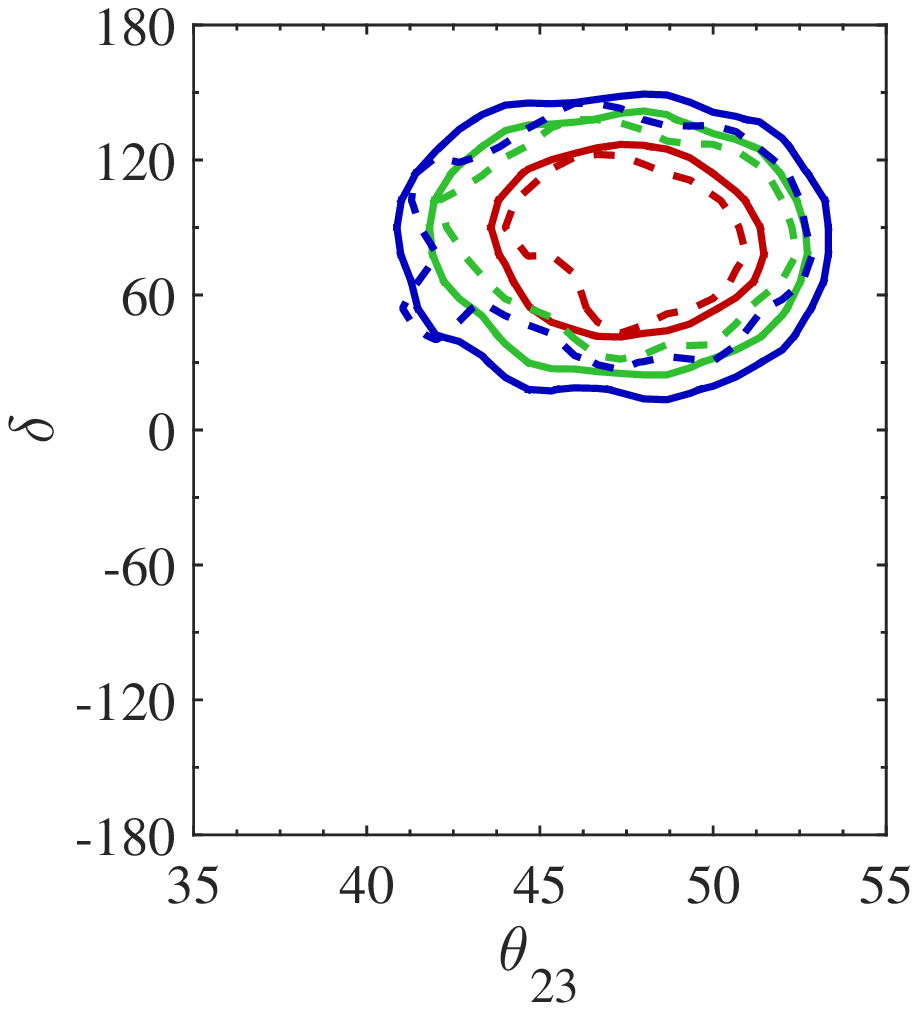, width=0.27\textwidth, bbllx=15, bblly=0, bburx=310, bbury=290,clip=} \\
\rotatebox[origin=l]{90}{\qquad\qquad\qquad $\dcp=180^\circ$} &
\epsfig{file=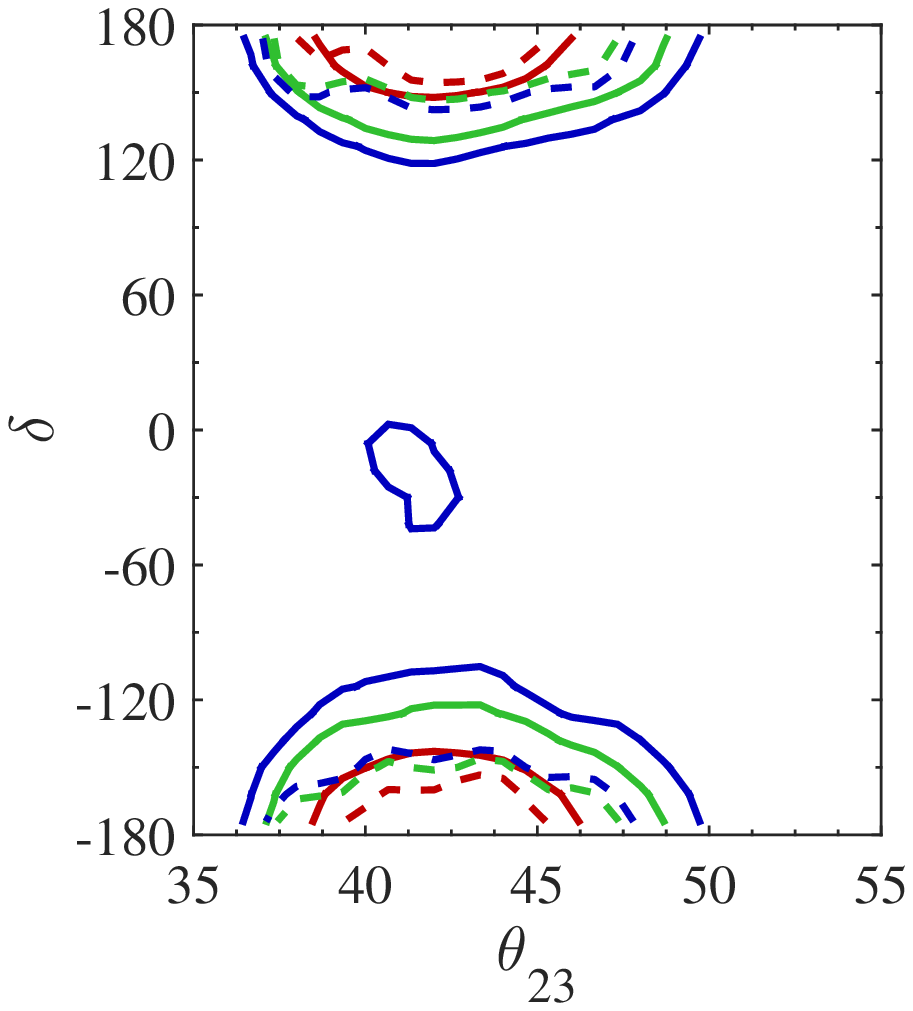, width=0.27\textwidth, bbllx=15, bblly=0, bburx=310, bbury=290,clip=} &
\epsfig{file=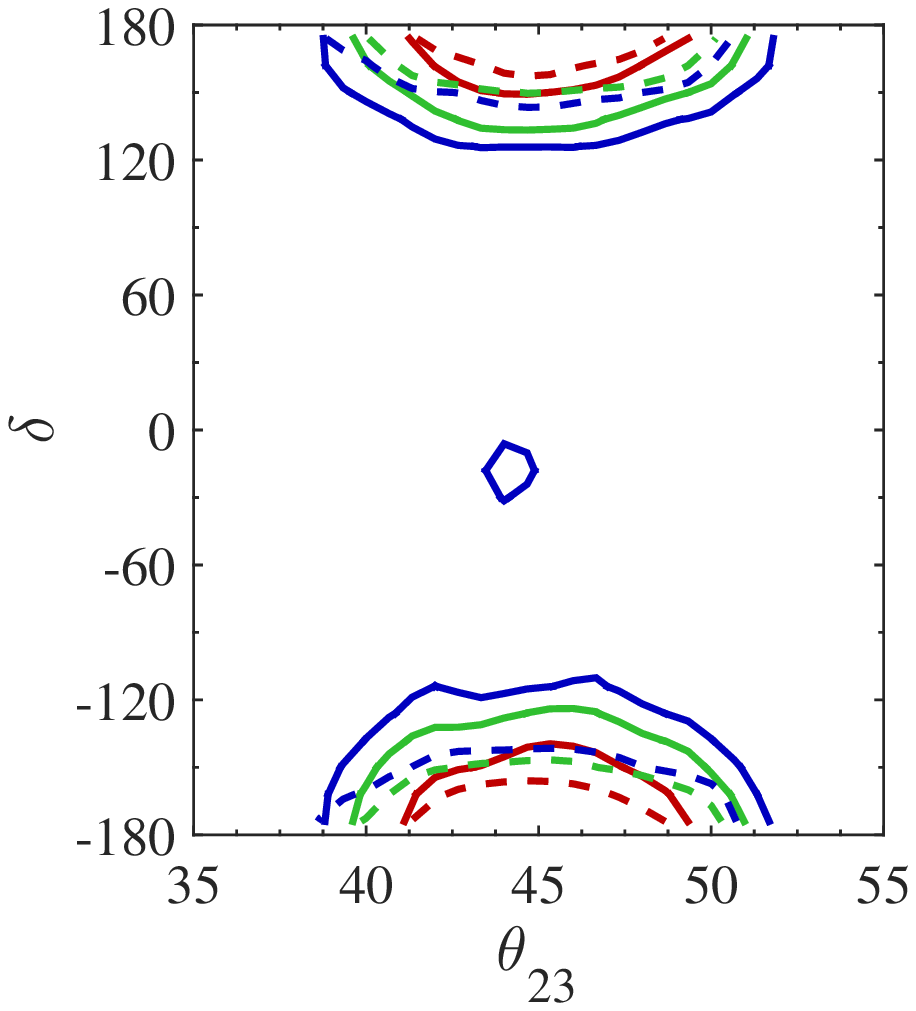, width=0.27\textwidth, bbllx=15, bblly=0, bburx=310, bbury=290,clip=} &
\epsfig{file=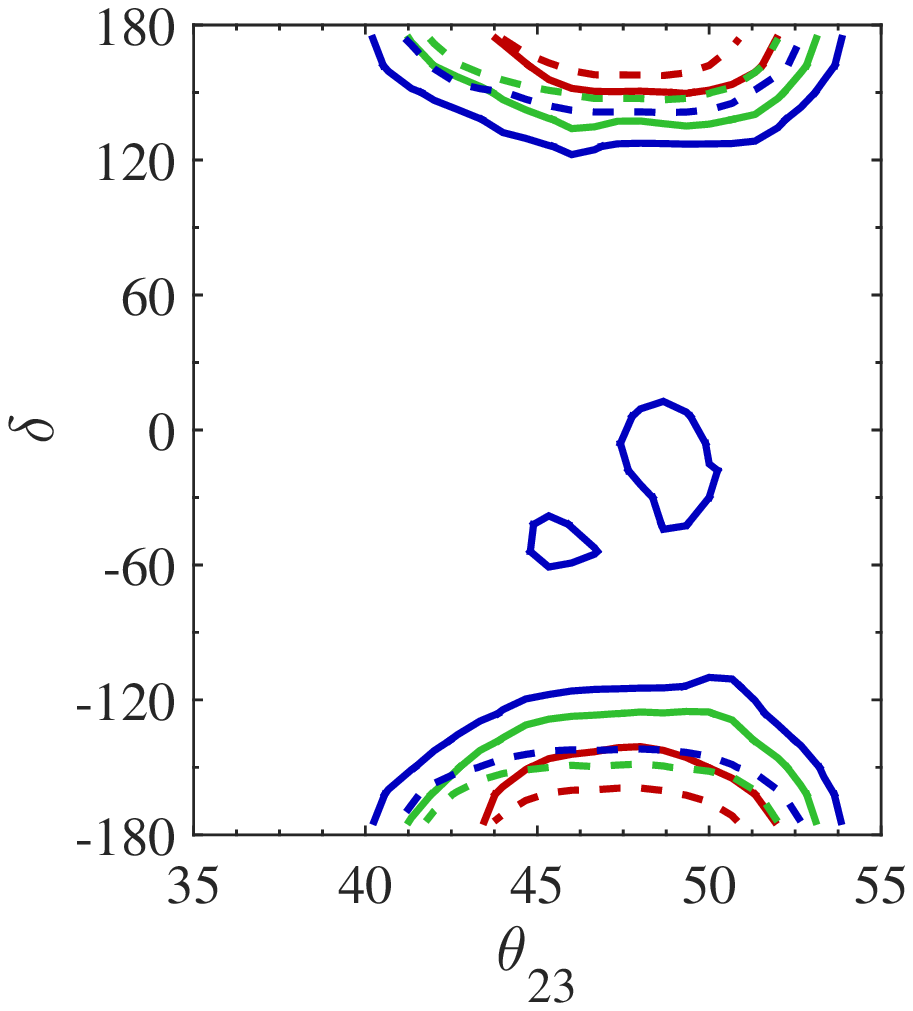, width=0.27\textwidth, bbllx=15, bblly=0, bburx=310, bbury=290,clip=} 
\end{tabular}
\caption{\footnotesize Effect of marginalizing over source and 
detector NSI parameters on precision measurements at \essnusb. 
Each panel shows the allowed region in the test $\theta_{23}-\dcp$ plane, when
the NSI parameters are taken to be non-zero in the data. 
The true values of the amplitudes of the NSI parameters are assumed to be half of their 
90\% C.L.~bounds from Ref.~\cite{Biggio:2009nt}.
The red, green and blue curves 
represent the 68\%, 90\% and 95\% C.L.~contours, respectively. The solid contours 
show the effect of marginalization over the NSI parameters, whereas the dashed contours 
are for the standard oscillation scenario in the absence of NSIs. }
\label{fig:effect_nonzero}
\end{figure}

Second, in Fig.~\ref{fig:effect_nonzero}, we 
investigate the same effect as in Fig.~\ref{fig:effect_0}, but
with a non-zero value for the NSI parameters in the mock data. 
These `true' values of the NSI parameters have been 
taken to be half of the bounds given in Eq.~(\ref{eq:bounds}) 
for the amplitudes. The true values of the non-standard 
phases are taken to be zero. In the fit, as before, all 
the standard as well as the NSI parameters are 
marginalized over. Thus, these plots show the robustness 
of measurements at \essnusb\ against a scan for NSI 
parameters, but in the presence of NSIs. 

As in the previous case, in the presence of NSIs, we observe that the $\theta_{23}$ measurement 
is not affected much, while the precision in $\dcp$ 
worsens. Here, the worsening depends significantly 
on the value of $\dcp$ in nature. When $\dcp=0$, the 
worsening of precision is least, whereas for $\dcp=180^\circ$, 
the precision is worst. 
This is seen uniformly across 
the range of $\theta_{23}$ values considered. 
The reason for this is as follows. A measure of the precision of 
$\dcp$ is ${\rm d}P^{\rm vac}_{\mu e}/{\rm d}\dcp$. In order to find the value of $\dcp$ 
at which this precision is minimal, we set the derivative 
of this quantity, i.e.~${\rm d}^2P^{\rm vac}_{\mu e}/{\rm d}\dcp^2$ to zero. 
Since the dependence of the probability on $\dcp$ is 
harmonic, the second derivative is proportional 
to the probability itself. As seen from the panels in 
Fig.~\ref{fig:bi}, the smallest probability 
for both neutrinos and antineutrinos is around $180^\circ$. 
This is why the precision of $\dcp$ is worst at $180^\circ$ in the presence of NSIs.

\begin{figure}[hbt]
\begin{tabular}{cccc}
 & \qquad NO, $\theta_{23}=42^\circ$ & \qquad NO, $\theta_{23}=45^\circ$ & \qquad NO, $\theta_{23}=48^\circ$ \\
\rotatebox[origin=l]{90}{\qquad\qquad\qquad $\dcp=-90^\circ$} &
\epsfig{file=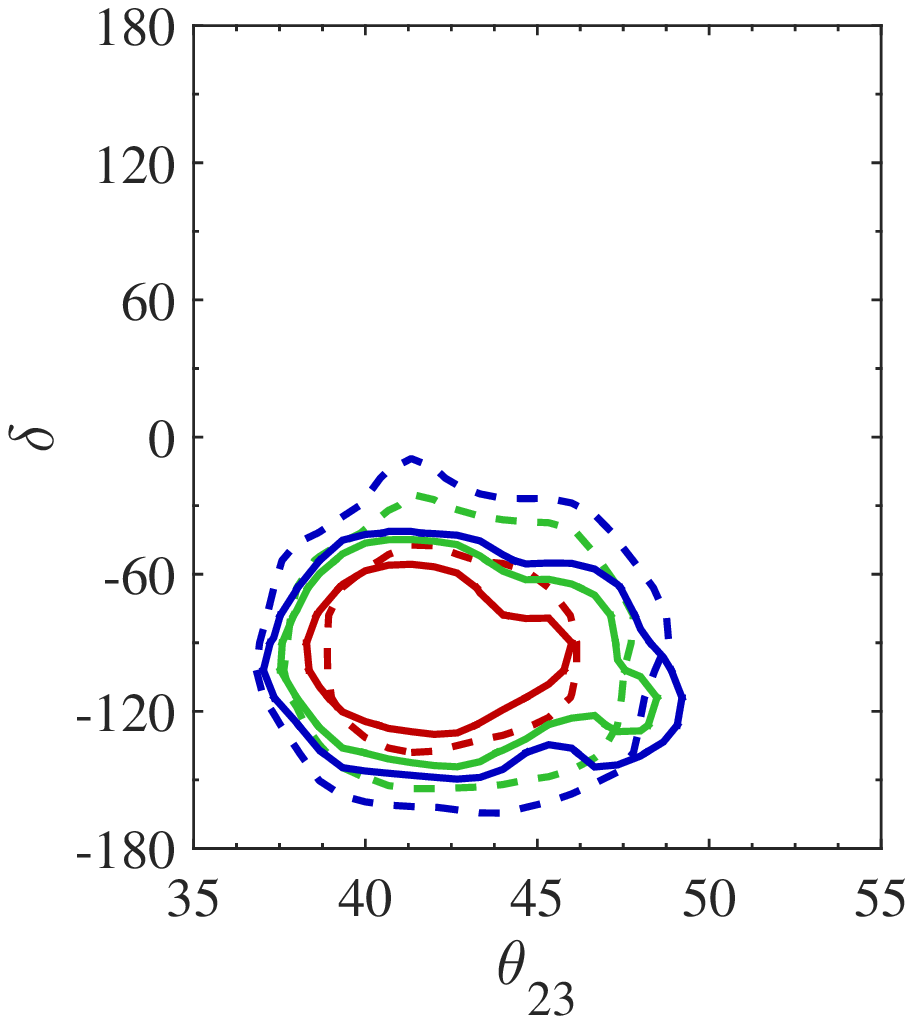, width=0.27\textwidth, bbllx=15, bblly=0, bburx=310, bbury=293,clip=} &
\epsfig{file=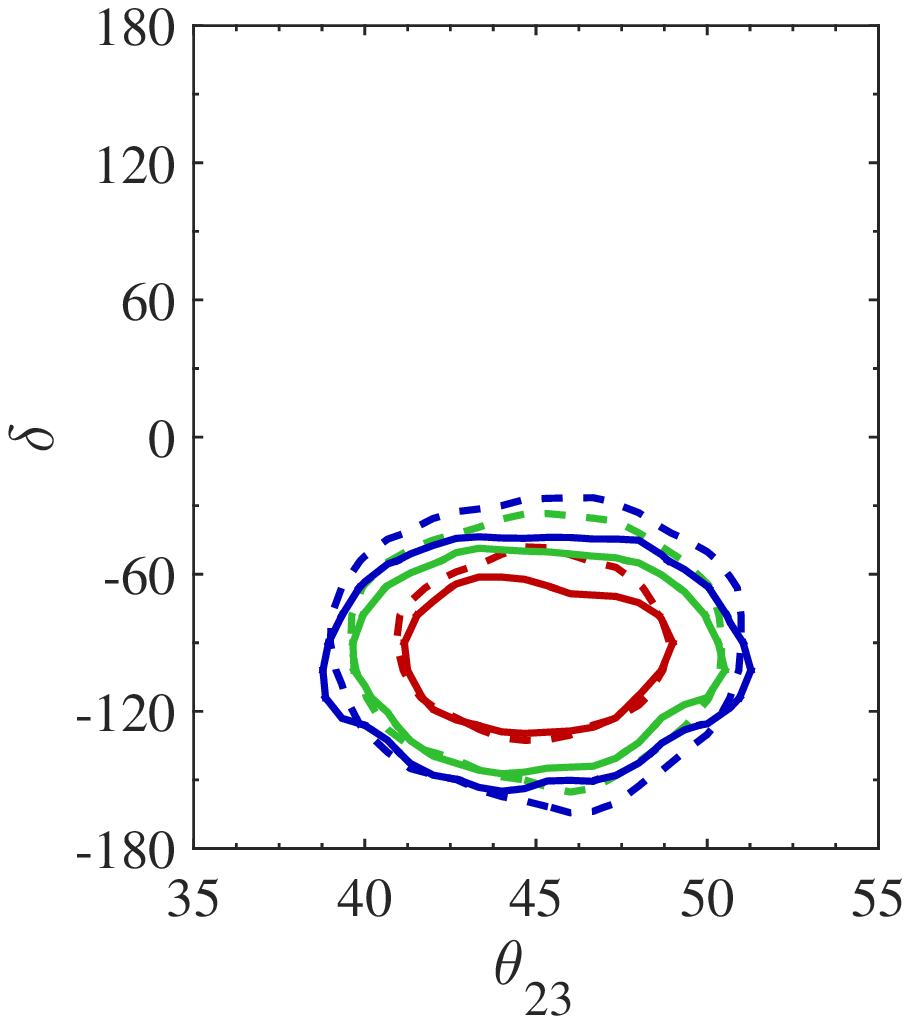, width=0.27\textwidth, bbllx=15, bblly=0, bburx=310, bbury=293,clip=} &
\epsfig{file=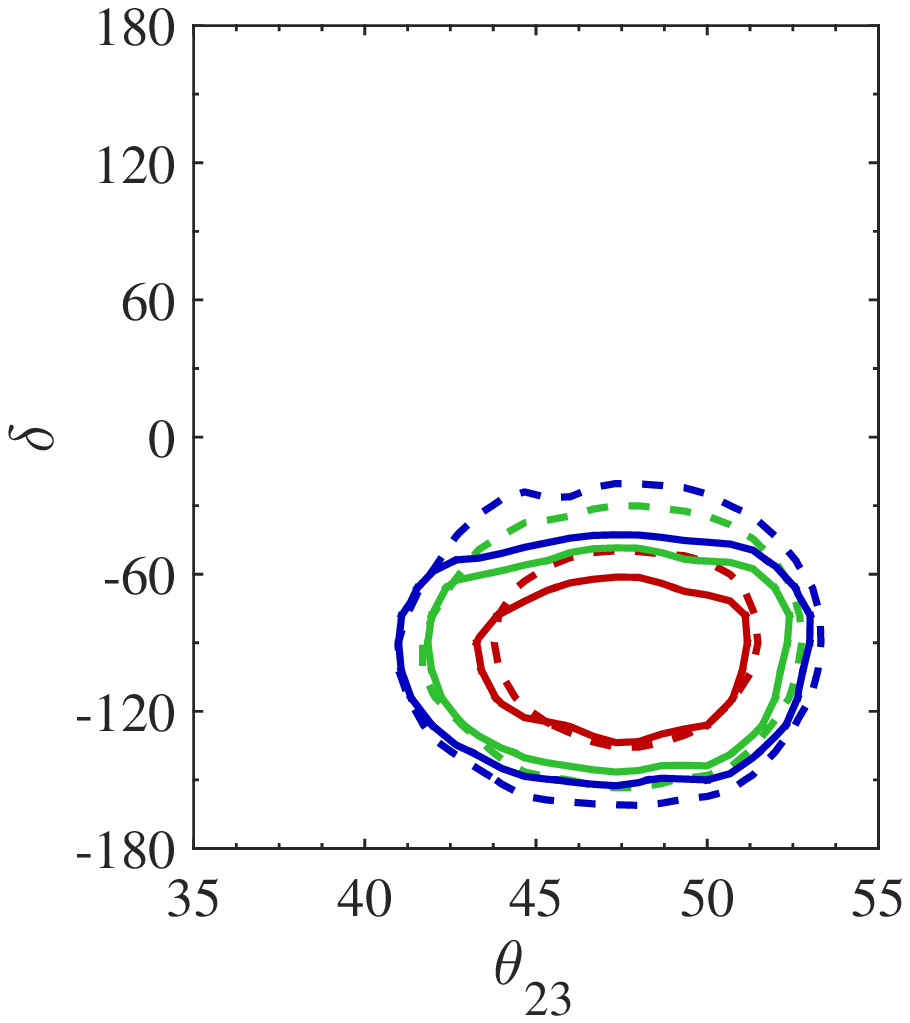, width=0.27\textwidth, bbllx=15, bblly=0, bburx=310, bbury=293,clip=} \\
\rotatebox[origin=l]{90}{\qquad\qquad\qquad $\dcp=0^\circ$} &
\epsfig{file=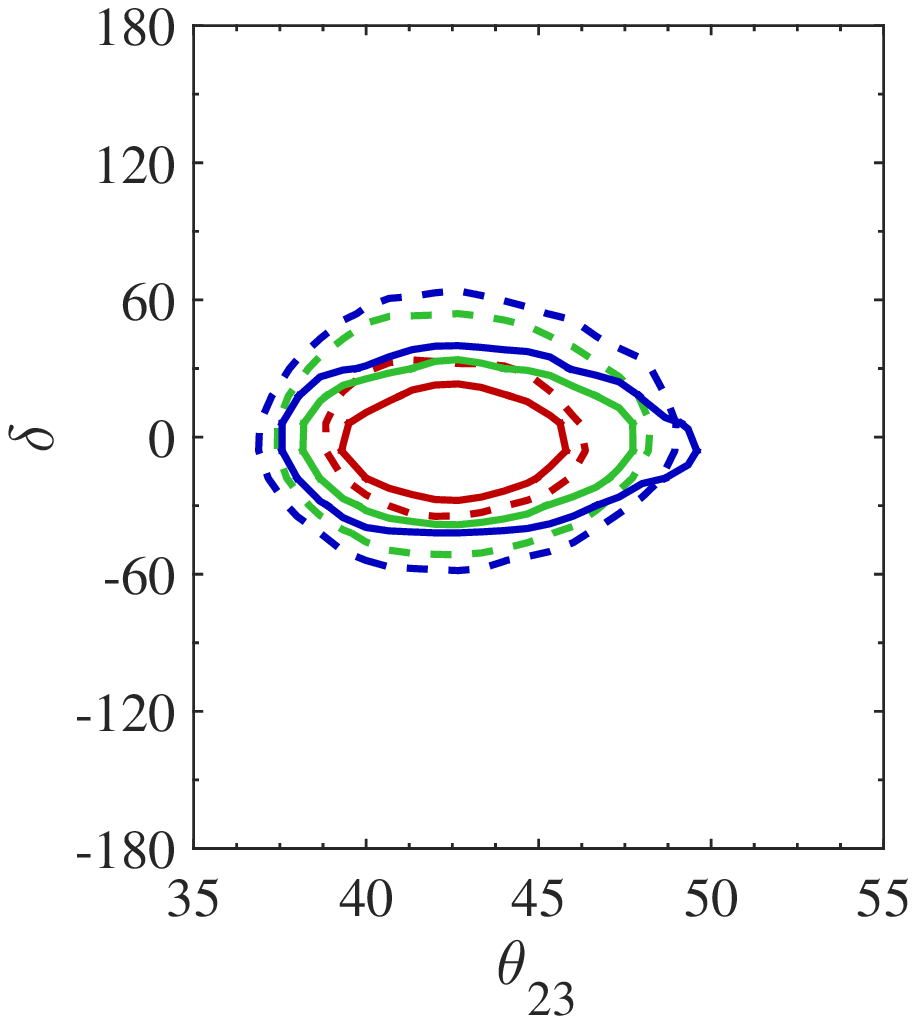, width=0.27\textwidth, bbllx=15, bblly=0, bburx=310, bbury=293,clip=} &
\epsfig{file=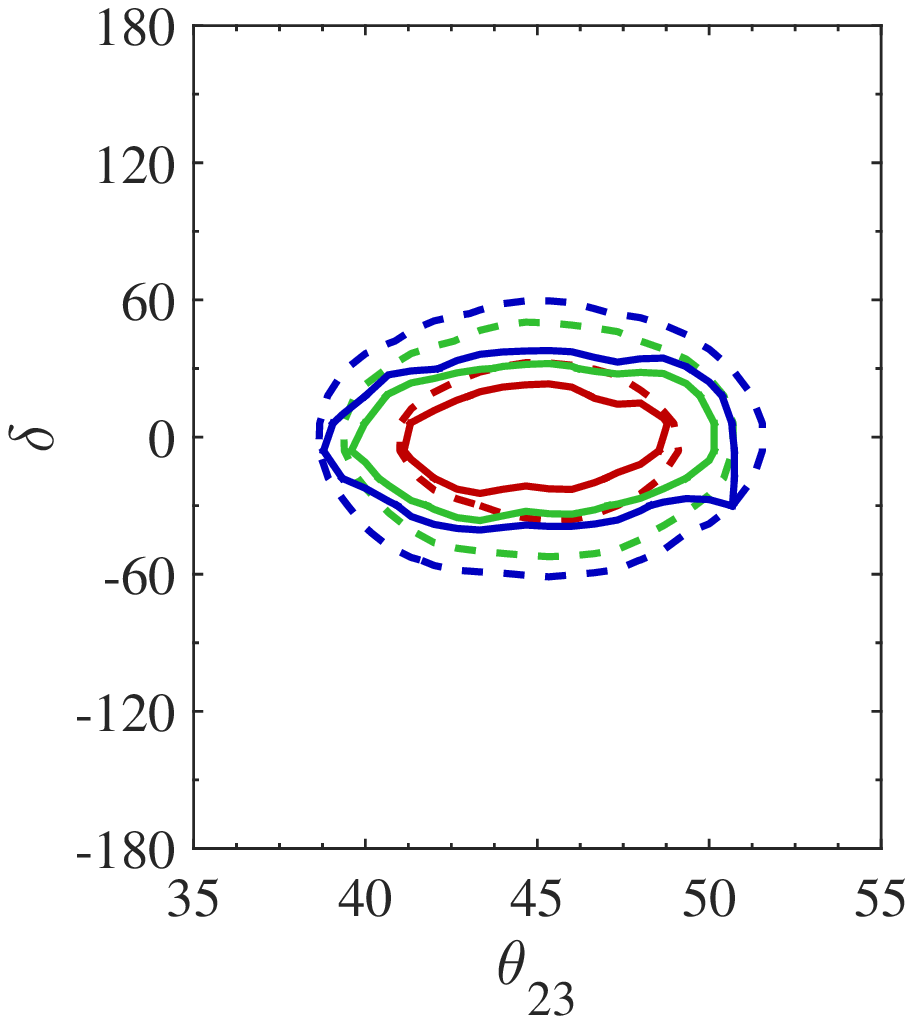, width=0.27\textwidth, bbllx=15, bblly=0, bburx=310, bbury=293,clip=} &
\epsfig{file=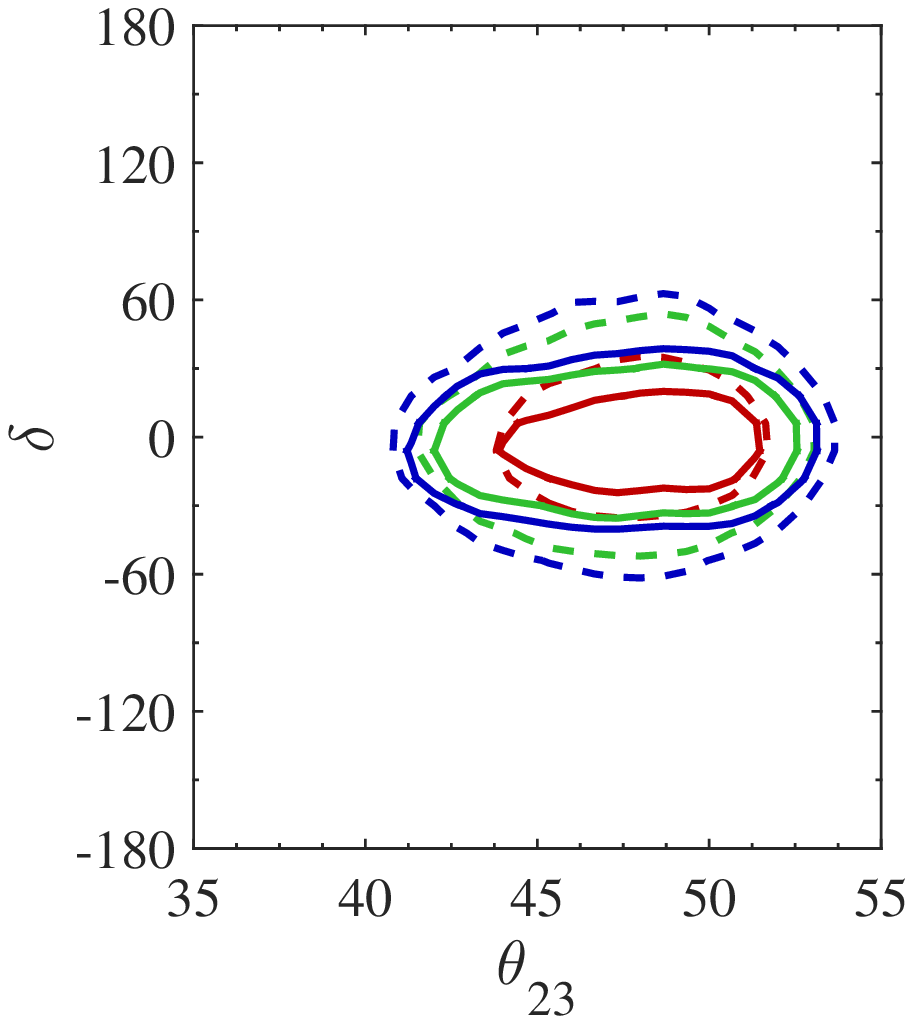, width=0.27\textwidth, bbllx=15, bblly=0, bburx=310, bbury=293,clip=} \\
\rotatebox[origin=l]{90}{\qquad\qquad\qquad $\dcp=90^\circ$} &
\epsfig{file=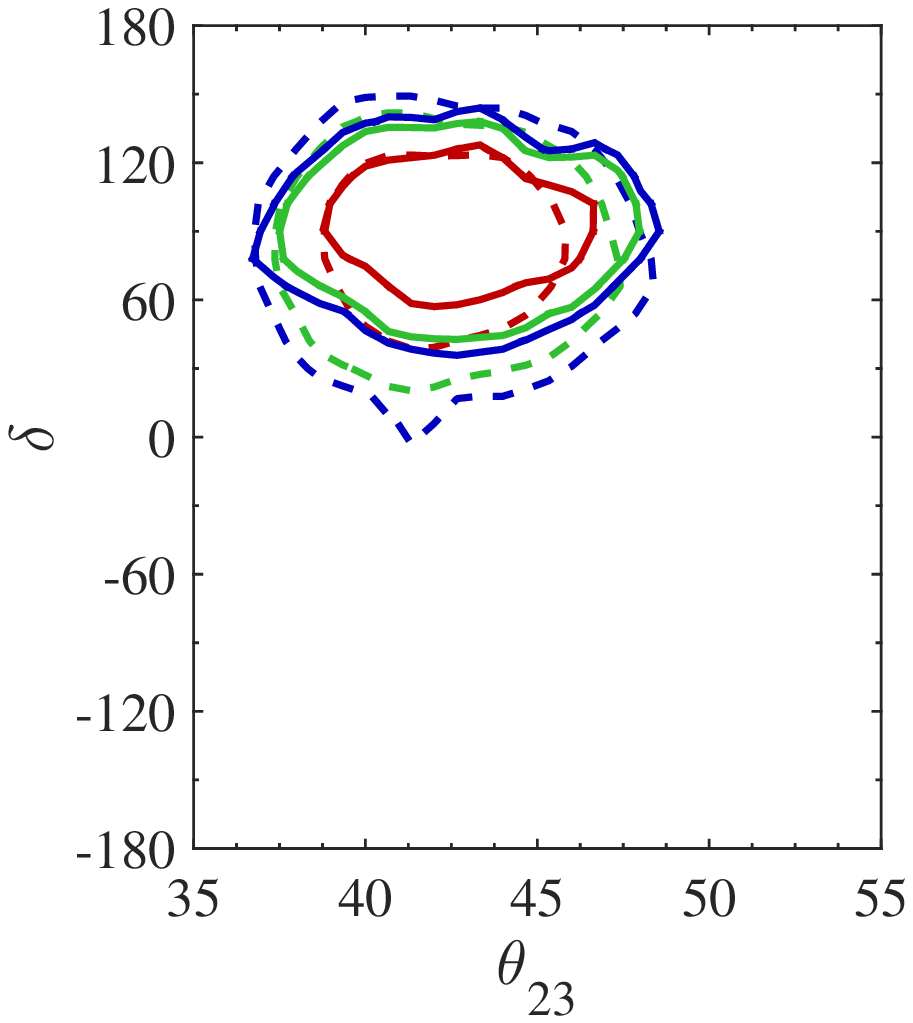, width=0.27\textwidth, bbllx=15, bblly=0, bburx=310, bbury=293,clip=} &
\epsfig{file=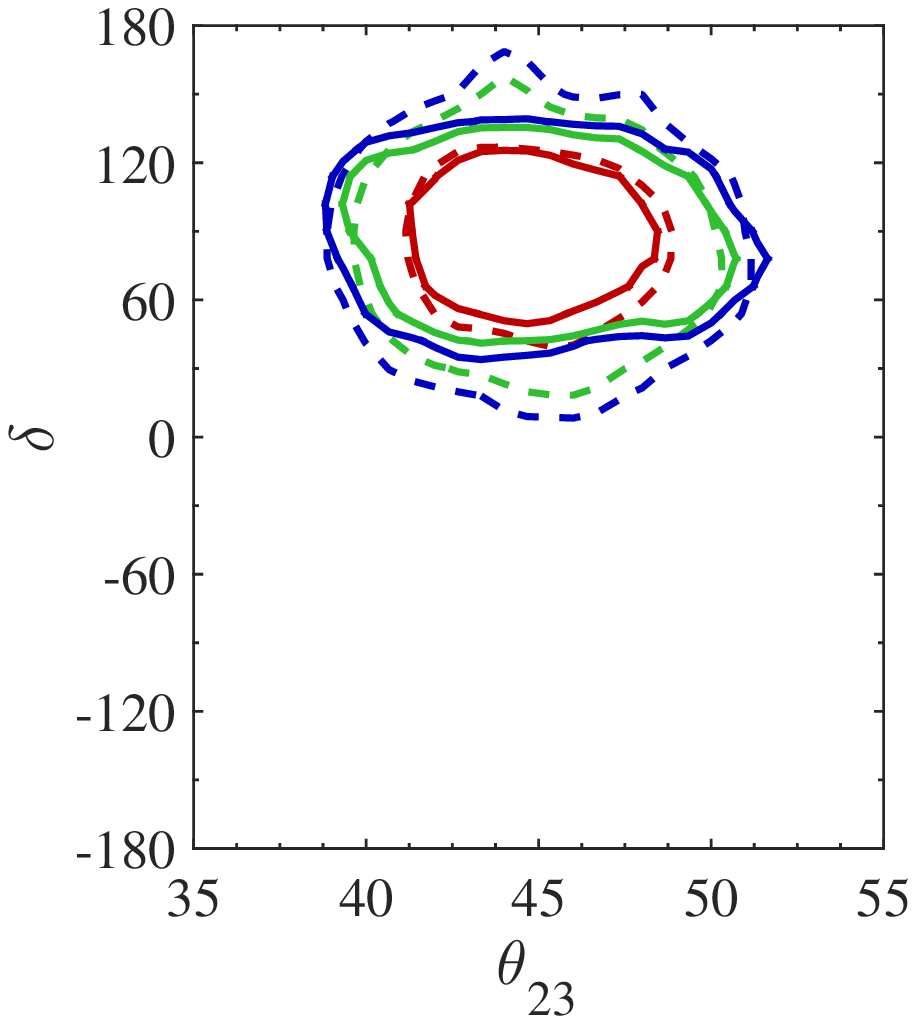, width=0.27\textwidth, bbllx=15, bblly=0, bburx=310, bbury=293,clip=} &
\epsfig{file=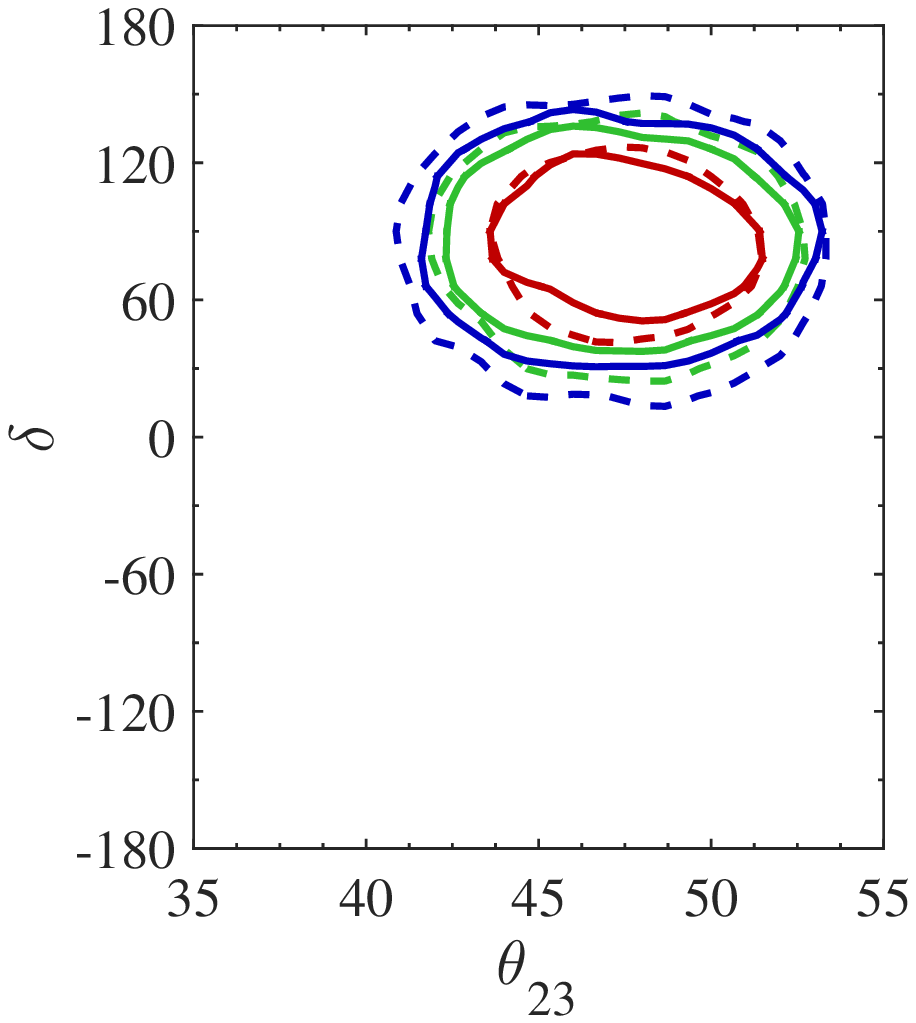, width=0.27\textwidth, bbllx=15, bblly=0, bburx=310, bbury=293,clip=} \\
\rotatebox[origin=l]{90}{\qquad\qquad\qquad $\dcp=180^\circ$} &
\epsfig{file=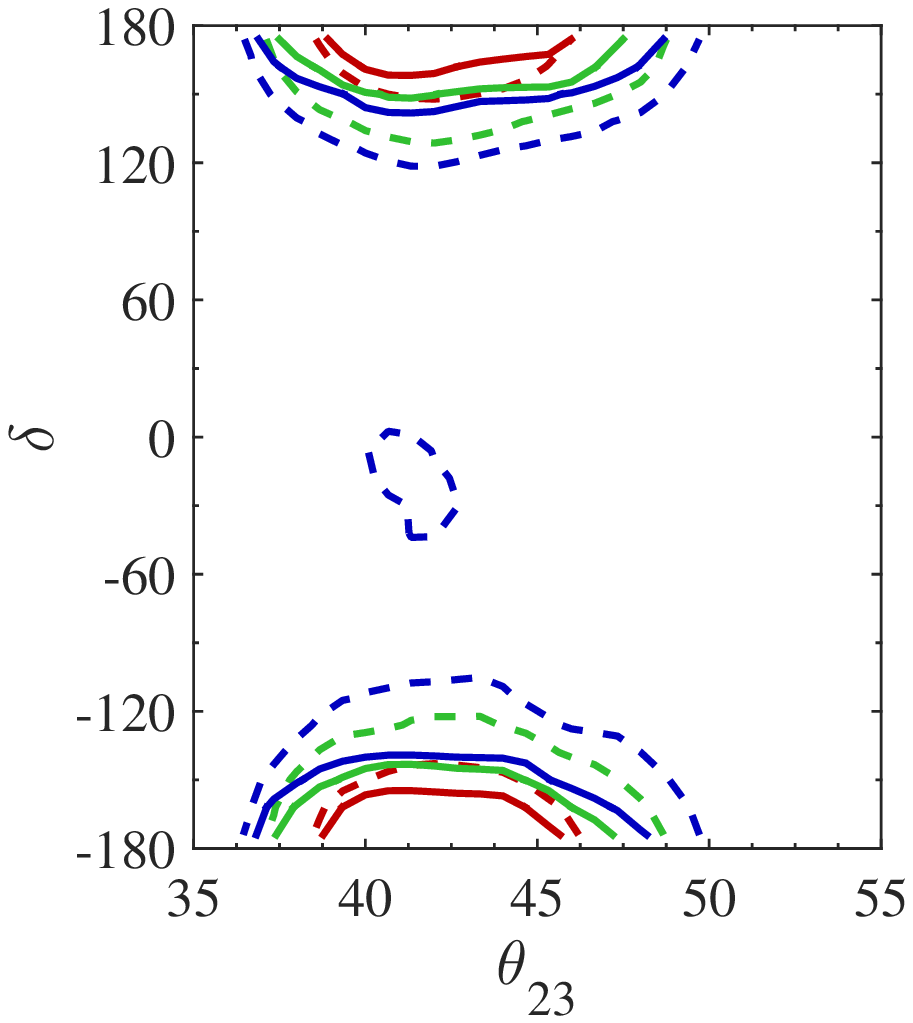, width=0.27\textwidth, bbllx=15, bblly=0, bburx=310, bbury=293,clip=} &
\epsfig{file=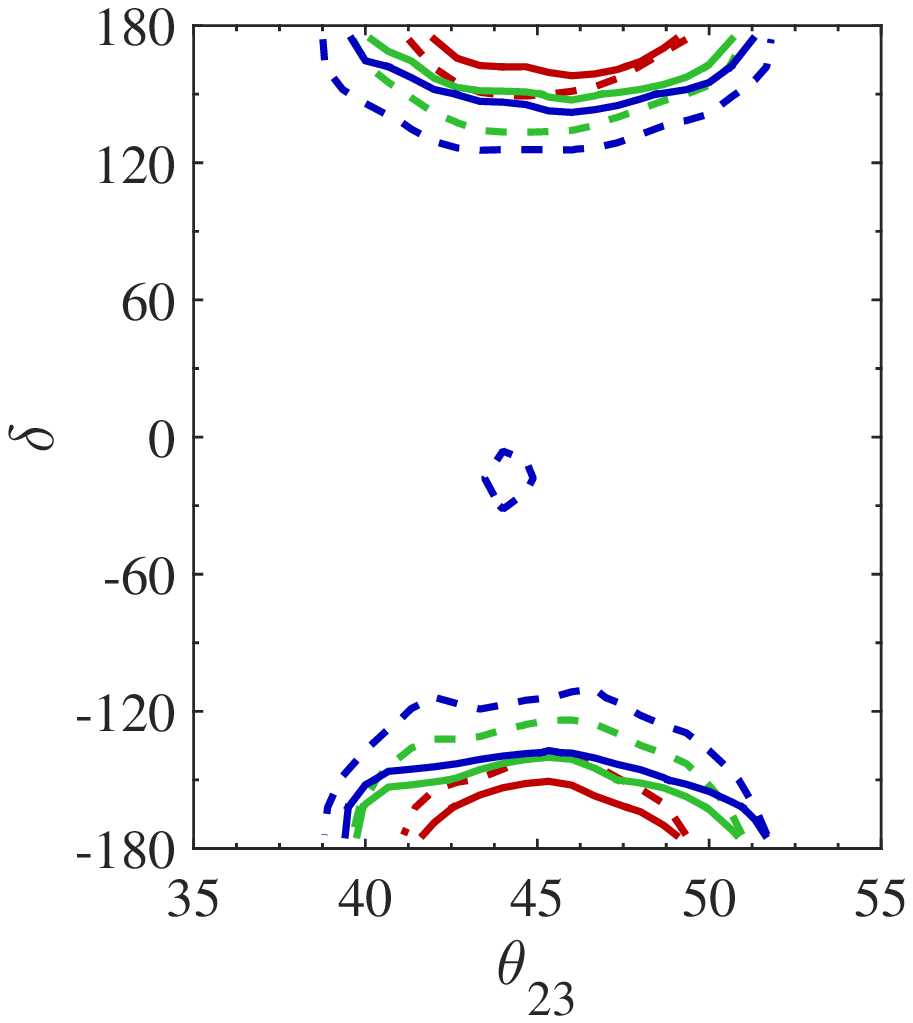, width=0.27\textwidth, bbllx=15, bblly=0, bburx=310, bbury=293,clip=} &
\epsfig{file=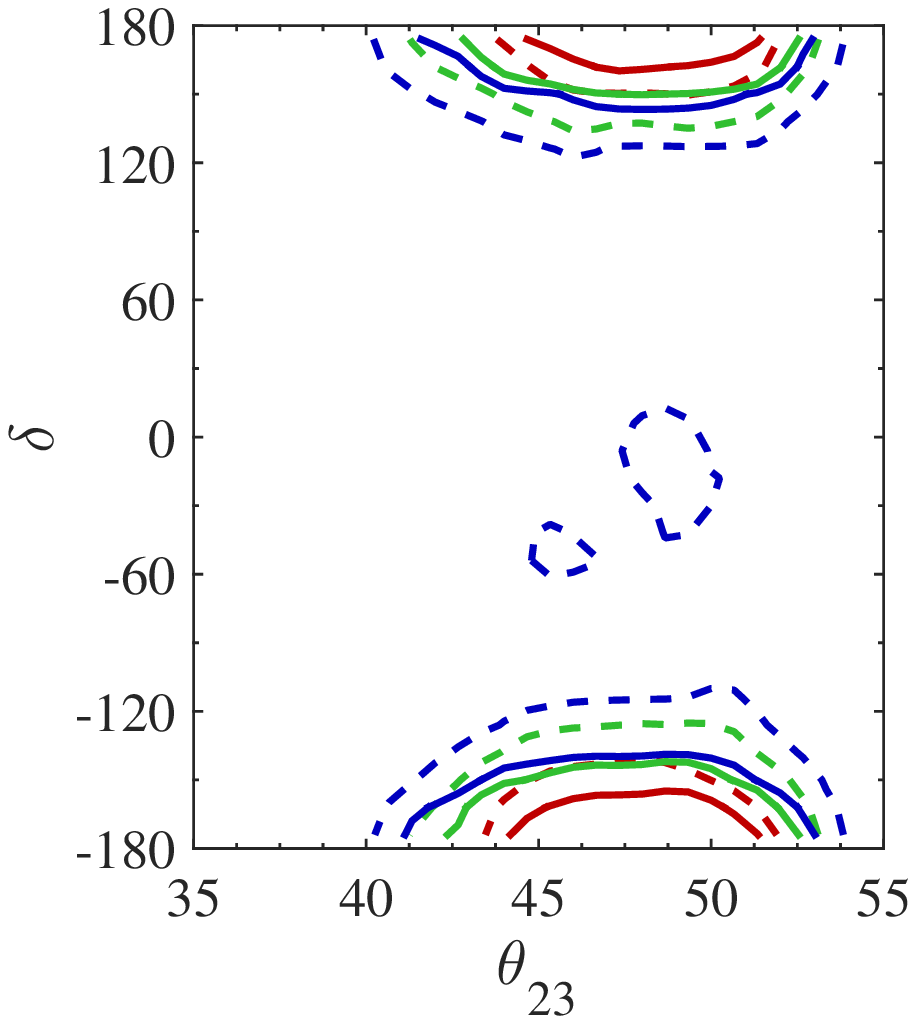, width=0.27\textwidth, bbllx=15, bblly=0, bburx=310, bbury=293,clip=} 
\end{tabular}
\caption{\footnotesize Precision measurements at \essnusb\
for the case where NSIs are present in nature, but are not 
scanned for. Each panel shows the allowed region in the test $\theta_{23}-\dcp$ plane, when
the NSI parameters are taken to be non-zero in the data. 
The true values of the amplitudes of the NSI parameters are assumed to be half of their 
90\% C.L.~bounds from Ref.~\cite{Biggio:2009nt}.
The red, green and blue curves 
represent the 68\%, 90\% and 95\% C.L.~contours, respectively. 
The solid (dashed) contours show the allowed region without (with)
marginalization over the NSI parameters. }
\label{fig:ignorance}
\end{figure}

Third, we study how 
the precision measurement at \essnusb\ would be affected if 
NSIs are present in nature, but are not accounted for in 
the scan of the parameter space. For this, we have taken 
non-zero values of the NSI parameters in the mock data (the 
same non-zero values as in the previous case of Fig.~\ref{fig:effect_nonzero}), but their 
values have been kept fixed at zero in the fit. The results 
are displayed in Fig.~\ref{fig:ignorance}. Here, the solid 
curves represent the 68\%, 90\% and 95\% C.L.~contours, 
when NSIs are present in the data, but not in the fit. 
The dashed curves are the corresponding contours for the 
case where the NSIs are marginalized 
over in the fit. Thus, the difference between the 
solid and dashed contours indicates the effect of our 
ignorance of the existence of NSIs. Our ignorance 
leads us to an over-optimistic precision in $\dcp$, 
as expected. The effect is more pronounced for the 
true value of $\dcp=180^\circ$. As before, the 
$\theta_{23}$ precision is not affected. 

\subsection{Constraining NSI parameters at ESS}

Having investigated the effect of NSIs on precision measurements 
at \essnusb, we explore the ability of this experiment 
to measure the NSI parameters themselves. As we have seen 
before, the effect of the NSI parameters on the probability 
is quite mild. Therefore, we do not expect to obtain very strong 
constraints on these parameters. 

\begin{figure}[hbt]
\begin{tabular}{lll}
\epsfig{file=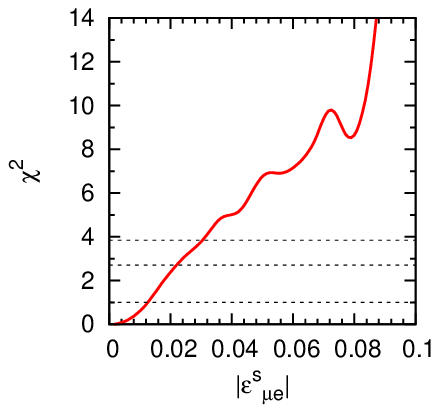, width=0.33\textwidth, bbllx=70, bblly=55, bburx=200, bbury=190,clip=} &
\epsfig{file=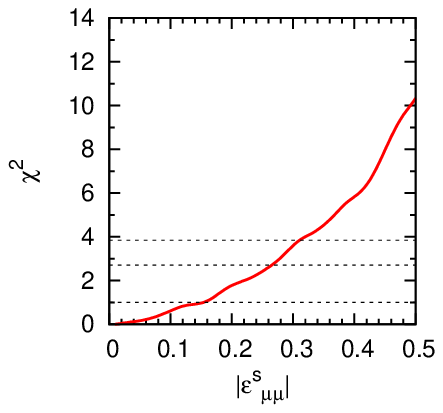, width=0.33\textwidth, bbllx=70, bblly=55, bburx=200, bbury=190,clip=} &
\epsfig{file=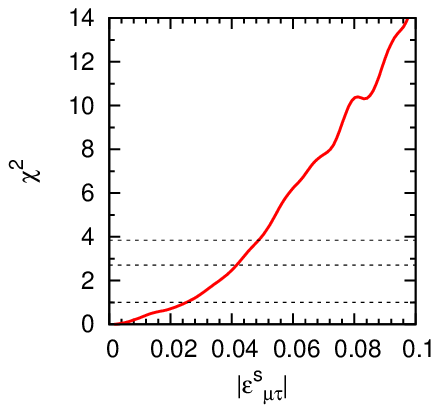, width=0.33\textwidth, bbllx=70, bblly=55, bburx=200, bbury=190,clip=} \\
\epsfig{file=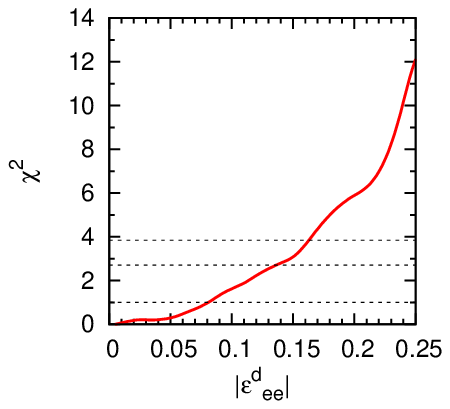, width=0.33\textwidth, bbllx=70, bblly=55, bburx=200, bbury=190,clip=} &
\epsfig{file=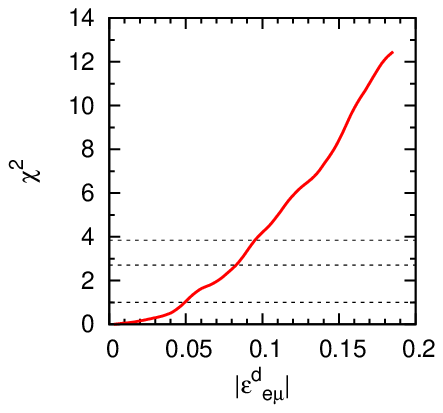, width=0.33\textwidth, bbllx=70, bblly=55, bburx=200, bbury=190,clip=} &
\epsfig{file=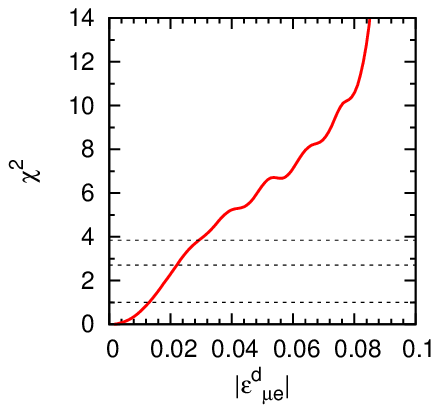, width=0.33\textwidth, bbllx=70, bblly=55, bburx=200, bbury=190,clip=} \\
\epsfig{file=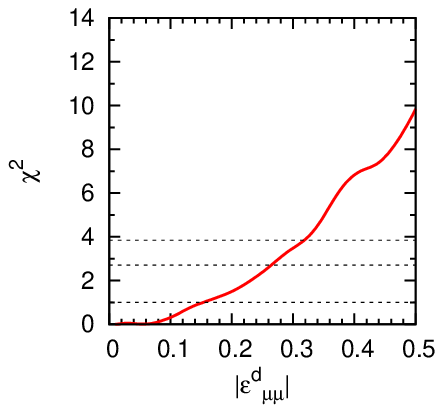, width=0.33\textwidth, bbllx=70, bblly=55, bburx=200, bbury=190,clip=} &
\epsfig{file=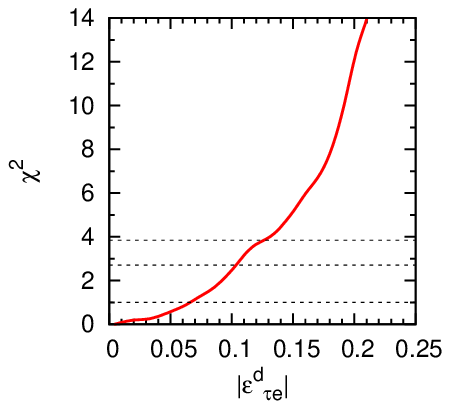, width=0.33\textwidth, bbllx=70, bblly=55, bburx=200, bbury=190,clip=} &
\epsfig{file=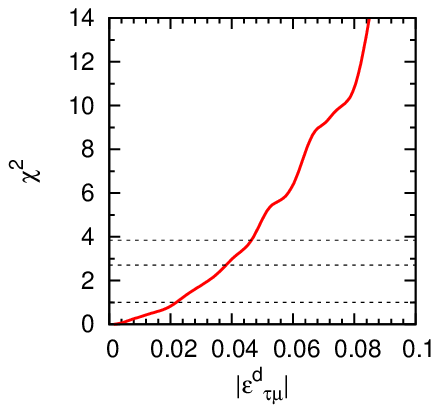, width=0.33\textwidth, bbllx=70, bblly=55, bburx=200, bbury=190,clip=} 
\end{tabular}
\caption{\footnotesize Limits on the amplitudes 
of the NSI parameters imposed 
by \essnusb\ data: $\chi^2$ as a function of the test 
value of the amplitudes of the NSI parameters, when the true value is zero. 
In each panel, all neutrino parameters (apart from the one 
indicated) have been marginalized over. The dotted lines 
from bottom to top show the 68\%, 90\% and 95\% 
C.L., respectively. }
\label{fig:limits}
\end{figure}

Figure~\ref{fig:limits} shows the limits which \essnusb\ can 
set on the amplitudes of the
NSI parameters for NO, $\theta_{23} = 45^\circ$ and $\dcp = 0$.
Consider the top-left panel, 
corresponding to the parameter $\epssme$. In generating 
this plot, we have set the true values of the NSI parameters 
to be zero. We show the $\chi^2$ as a function of the test 
value of $|\epssme|$, when all the other neutrino parameters, 
including the NSI ones are marginalized over. 
Horizontal lines have been drawn in the plots, corresponding 
to 68\%, 90\% and 95\% C.L., assuming a $\chi^2$ 
distribution. One can read off the limits that \essnusb\ can 
impose on these parameters from this plot.
Similarly, the other panels show the limits for the 
other relevant parameters. 

The 90\% C.L.~limits on the NSI parameters using data from 
\essnusb\ are summarized in Table~\ref{tab:limits}. The first column 
gives the limits 
when all the other NSI parameters are kept free in the fit, 
which can be simply read off from Fig.~\ref{fig:limits}. 
These limits should be interpreted as being realistic, since they 
are derived without making any assumptions on the values of 
the other NSI parameters. 
We have also computed the limits when the NSI parameters are only 
considered one at a time, i.e.~all other NSI parameters are fixed 
to zero. These limits, which are given in the second column, are 
more optimistic. The realistic limits on $|\epssme|$ and $|\epsdme|$ 
are comparable to the ones in Ref.~\cite{Biggio:2009nt}, 
which are listed in the third column for ease of 
comparison.\footnote{Note that the limit on a given NSI parameter 
in Ref.~\cite{Biggio:2009nt} has been computed considering only that 
parameter and assuming all other NSI parameters to be zero, 
which corresponds to the optimistic case.} 
This is because these parameters have the maximum effect on 
$P_{\mu e}$, as seen from the analytical expressions. In the 
optimistic case, the limits on $|\epssme|$ and $|\epsdme|$ improve 
by a factor of two compared to the existing bounds. For all the other NSI 
parameters (except $|\epssme|$ and $|\epsdme|$), the 
realistic and optimistic limits basically coincide and are 
less stringent than the limits in Ref.~\cite{Biggio:2009nt}. 

\begin{center}
\begin{table}[hbt]
\begin{tabular}{cccc}
 \hline
 Parameter & \parbox{0.25\textwidth}{\centering Limits with all other \\ NSI parameters free} & \parbox{0.25\textwidth}{\centering Limits with all other \\ NSI parameters zero} & \parbox{0.25\textwidth}{\centering Limits from Ref.~\cite{Biggio:2009nt} } \\
 \hline
 $|\epssme|$ & 0.025 & 0.014 & 0.026 \\
 $|\epssmm|$ & 0.27 & 0.27 & 0.078 \\
 $|\epssmt|$ & 0.040 & 0.040 & 0.013 \\
 \\[-5mm]
 $|\epsdee|$ & 0.15 & 0.15 & 0.041 \\
 $|\epsdem|$ & 0.087 & 0.082 & 0.026 \\
 $|\epsdme|$ & 0.025 & 0.014 & 0.025 \\
 $|\epsdmm|$ & 0.28 & 0.27 & 0.078 \\
 $|\epsdte|$ & 0.11 & 0.12 & 0.041 \\
 $|\epsdtm|$ & 0.040 & 0.033 & 0.013 \\
 \hline
\end{tabular}
\caption{\footnotesize 90\% C.L.~sensitivities of \essnusb\ to the NSI parameters.}
\label{tab:limits}
\end{table}
\end{center}

\begin{center}
\begin{figure}[hbt]
\begin{tabular}{lll}
\epsfig{file=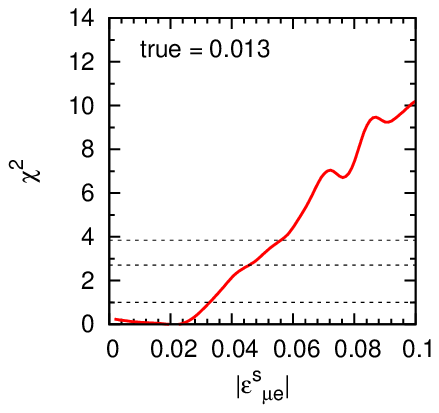, width=0.33\textwidth, bbllx=70, bblly=55, bburx=200, bbury=190,clip=} &
\epsfig{file=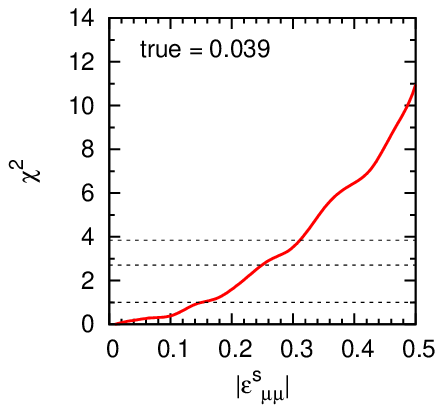, width=0.33\textwidth, bbllx=70, bblly=55, bburx=200, bbury=190,clip=} &
\epsfig{file=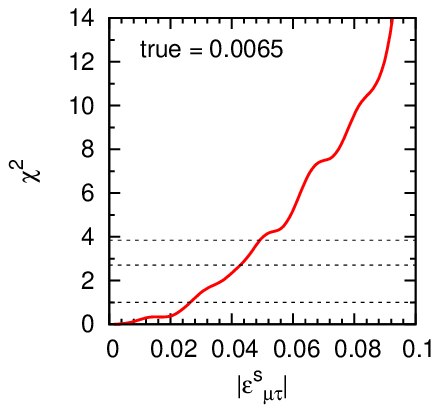, width=0.33\textwidth, bbllx=70, bblly=55, bburx=200, bbury=190,clip=} \\
\epsfig{file=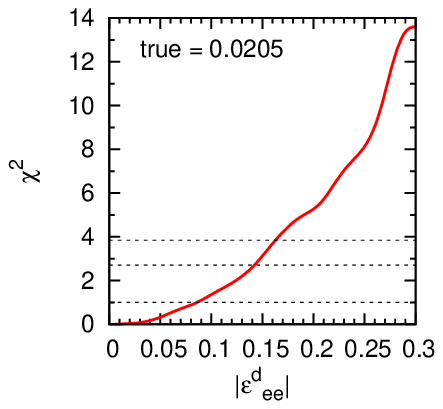, width=0.33\textwidth, bbllx=70, bblly=55, bburx=200, bbury=190,clip=} &
\epsfig{file=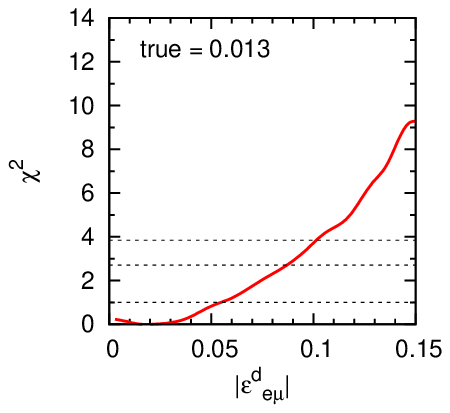, width=0.33\textwidth, bbllx=70, bblly=55, bburx=200, bbury=190,clip=} &
\epsfig{file=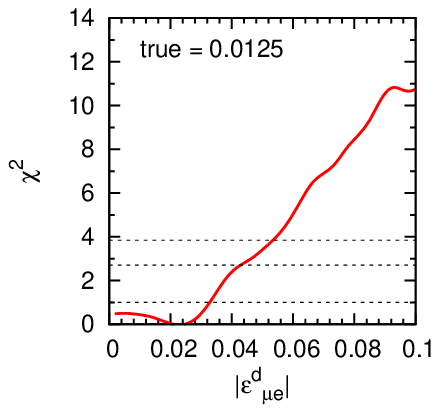, width=0.33\textwidth, bbllx=70, bblly=55, bburx=200, bbury=190,clip=} \\
\epsfig{file=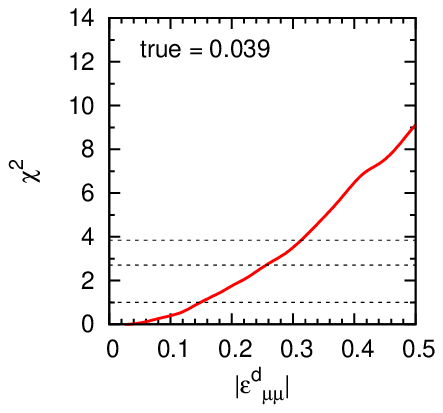, width=0.33\textwidth, bbllx=70, bblly=55, bburx=200, bbury=190,clip=} &
\epsfig{file=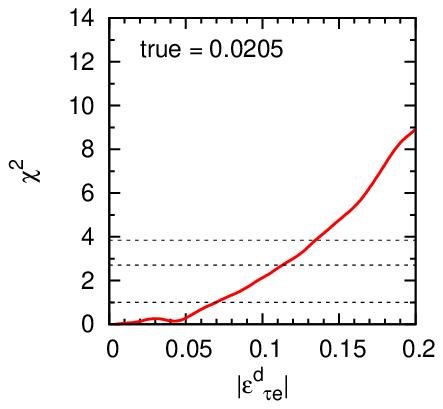, width=0.33\textwidth, bbllx=70, bblly=55, bburx=200, bbury=190,clip=} &
\epsfig{file=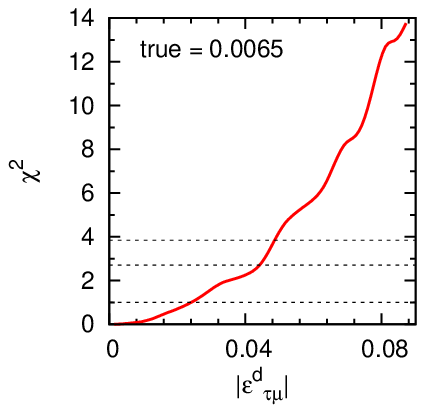, width=0.33\textwidth, bbllx=70, bblly=55, bburx=200, bbury=190,clip=} 
\end{tabular}
\caption{\footnotesize Precision on the amplitudes 
of the NSI parameters from 
\essnusb\ data: $\chi^2$ as a function of the test 
value of the amplitudes of the NSI parameters, when the true value is non-zero. 
The true values of the amplitudes of the NSI parameters are assumed to be half of their 
90\% C.L.~bounds from Ref.~\cite{Biggio:2009nt}.
In each panel, all neutrino parameters (apart from the one 
indicated) have been marginalized over. The dotted lines 
from bottom to top show the 68\%, 90\% and 95\% 
C.L., respectively.
}
\label{fig:precision}
\end{figure}
\end{center}

\begin{center}
\begin{figure}[hbt]
\begin{tabular}{lll}
\epsfig{file=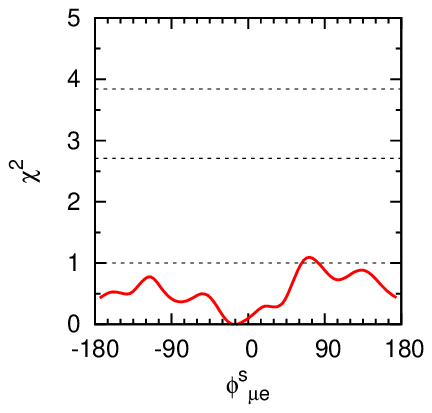, width=0.33\textwidth, bbllx=70, bblly=55, bburx=200, bbury=190,clip=} &
\epsfig{file=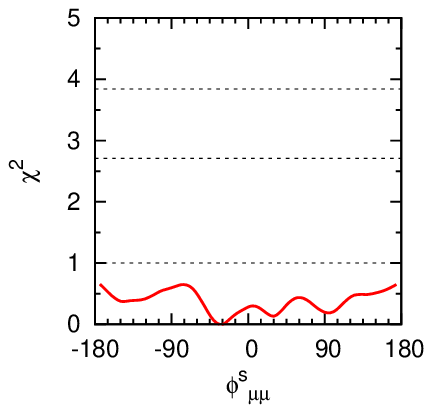, width=0.33\textwidth, bbllx=70, bblly=55, bburx=200, bbury=190,clip=} &
\epsfig{file=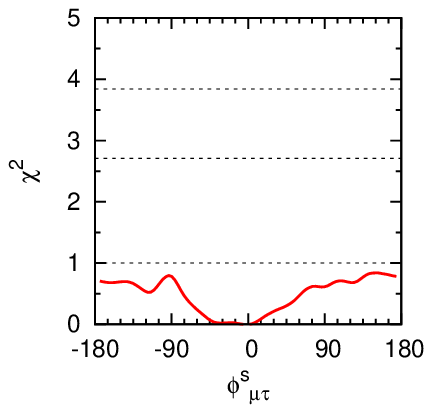, width=0.33\textwidth, bbllx=70, bblly=55, bburx=200, bbury=190,clip=} \\
\epsfig{file=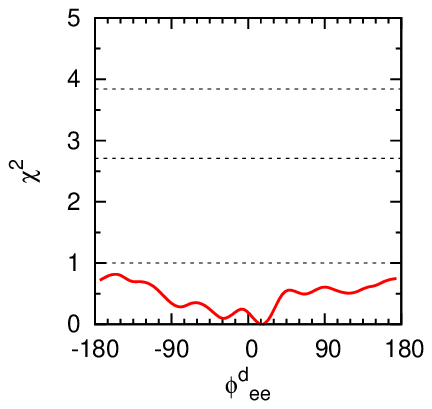, width=0.33\textwidth, bbllx=70, bblly=55, bburx=200, bbury=190,clip=} &
\epsfig{file=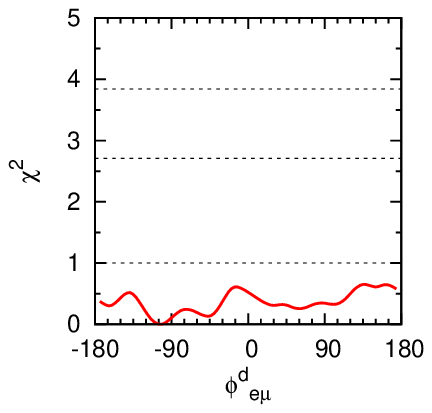, width=0.33\textwidth, bbllx=70, bblly=55, bburx=200, bbury=190,clip=} &
\epsfig{file=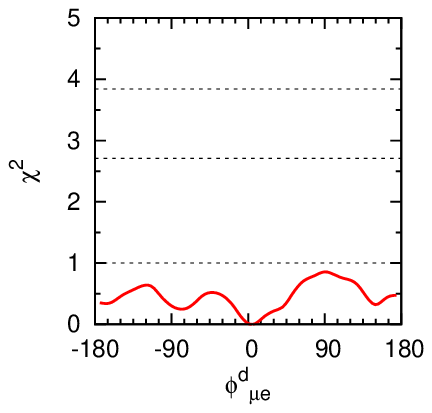, width=0.33\textwidth, bbllx=70, bblly=55, bburx=200, bbury=190,clip=} \\
\epsfig{file=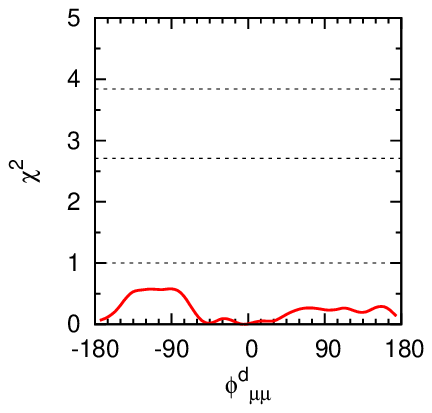, width=0.33\textwidth, bbllx=70, bblly=55, bburx=200, bbury=190,clip=} &
\epsfig{file=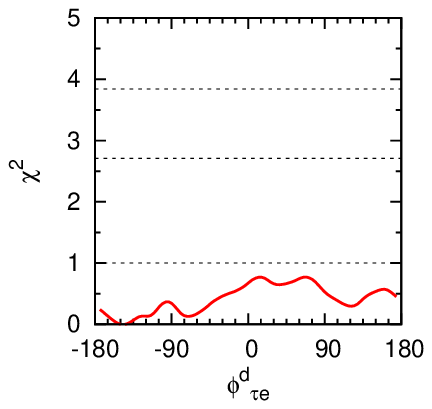, width=0.33\textwidth, bbllx=70, bblly=55, bburx=200, bbury=190,clip=} &
\epsfig{file=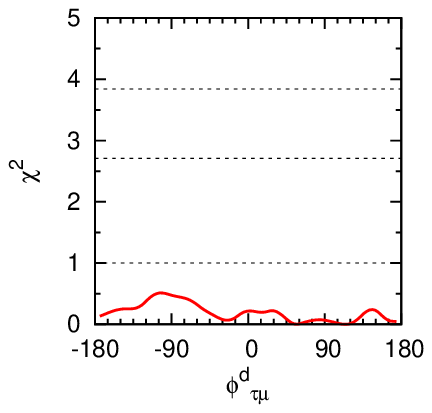, width=0.33\textwidth, bbllx=70, bblly=55, bburx=200, bbury=190,clip=} 
\end{tabular}
\caption{\footnotesize Precision on the phases of the 
NSI parameters from 
\essnusb\ data: $\chi^2$ as a function of the test 
value of the NSI phases. 
The true values of the amplitudes of the NSI parameters are assumed to be half of their 
90\% C.L.~bounds from Ref.~\cite{Biggio:2009nt}, 
while the true values of the phases are taken to be zero.
In each panel, all neutrino parameters (apart from the one 
indicated) have been marginalized over. The dotted lines 
from bottom to top show the 68\%, 90\% and 95\% 
C.L., respectively.
}
\label{fig:precisionphase}
\end{figure}
\end{center}

We have also checked whether \essnusb\ can measure the values of 
the NSI parameters with any reasonable precision. The procedure 
for this is the same as for Fig.~\ref{fig:limits}, except 
that the true values of the NSI parameters are non-zero. We 
have chosen these true values to be half of the 90\% C.L.~bounds 
given in Ref.~\cite{Biggio:2009nt}. The $\chi^2$ resulting 
from these computations is shown in 
Figs.~\ref{fig:precision} and \ref{fig:precisionphase} for 
the amplitudes and the phases of the NSI parameters, respectively. 
We find that \essnusb\ is not capable 
of distinguishing the chosen non-zero values of the parameters 
from zero, even at 68\% C.L., nor is it able to significantly 
constrain any of the NSI phases.

\subsection{Role of near detector and systematics}

Throughout this study we have used systematic errors of 
9\% in the signal and 18\% in the background events~\cite{Agostino:2012fd}. These
are typical values for a superbeam experiment with a 
megaton-scale water Cherenkov detector. In order to 
study the role of systematic errors on our results, we 
have also simulated our experiment with a smaller 
systematic error of 5\% in both signal and background. 
This is of course a very optimistic value. We have 
found that the limits on NSIs from \essnusb\ do 
not change appreciably with this drastic reduction of 
systematic errors. This is because in spite of having a 
large detector and intense source, \essnusb\ is still 
statistics-dominated due to the lower event rate at the second oscillation maximum.

\begin{figure}[htb]
\begin{tabular}{cccc}
 & \qquad NO, $\theta_{23}=42^\circ$ & \qquad NO, $\theta_{23}=45^\circ$ & \qquad NO, $\theta_{23}=48^\circ$ \\
\rotatebox[origin=l]{90}{\qquad\qquad\qquad $\dcp=-90^\circ$} &
\epsfig{file=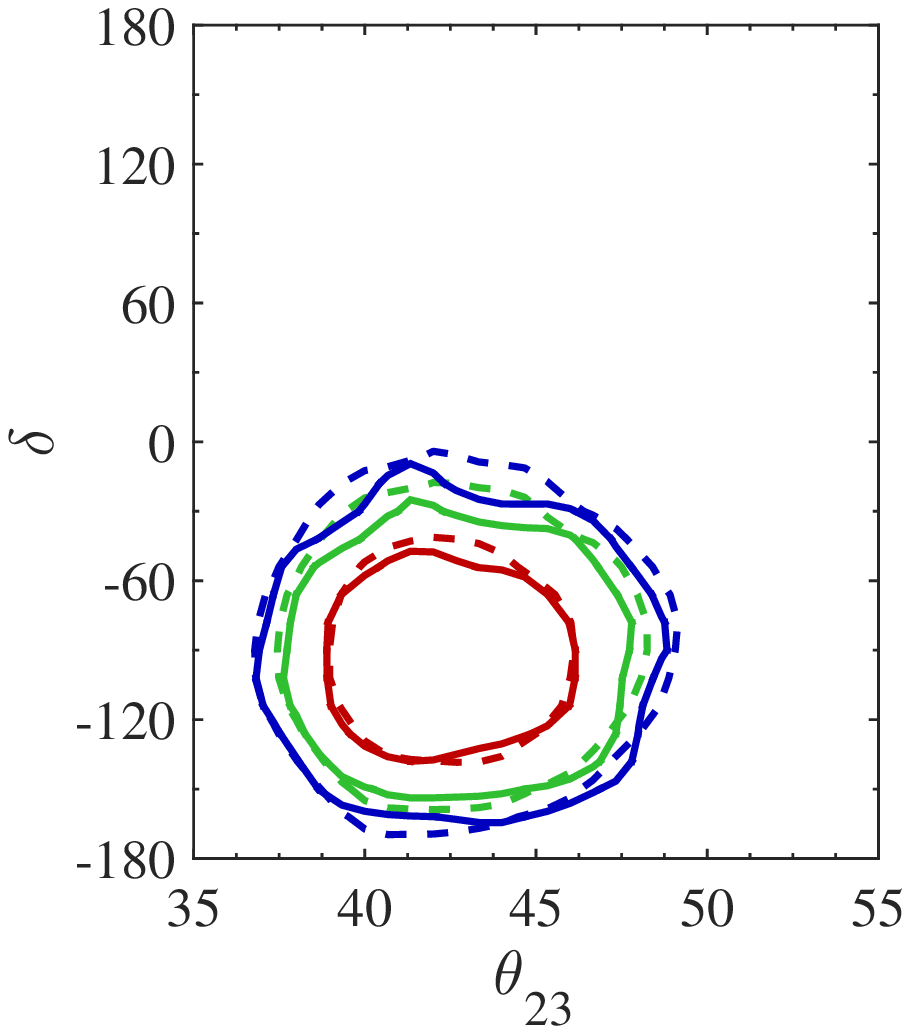, width=0.27\textwidth, bbllx=15, bblly=0, bburx=310, bbury=296,clip=} &
\epsfig{file=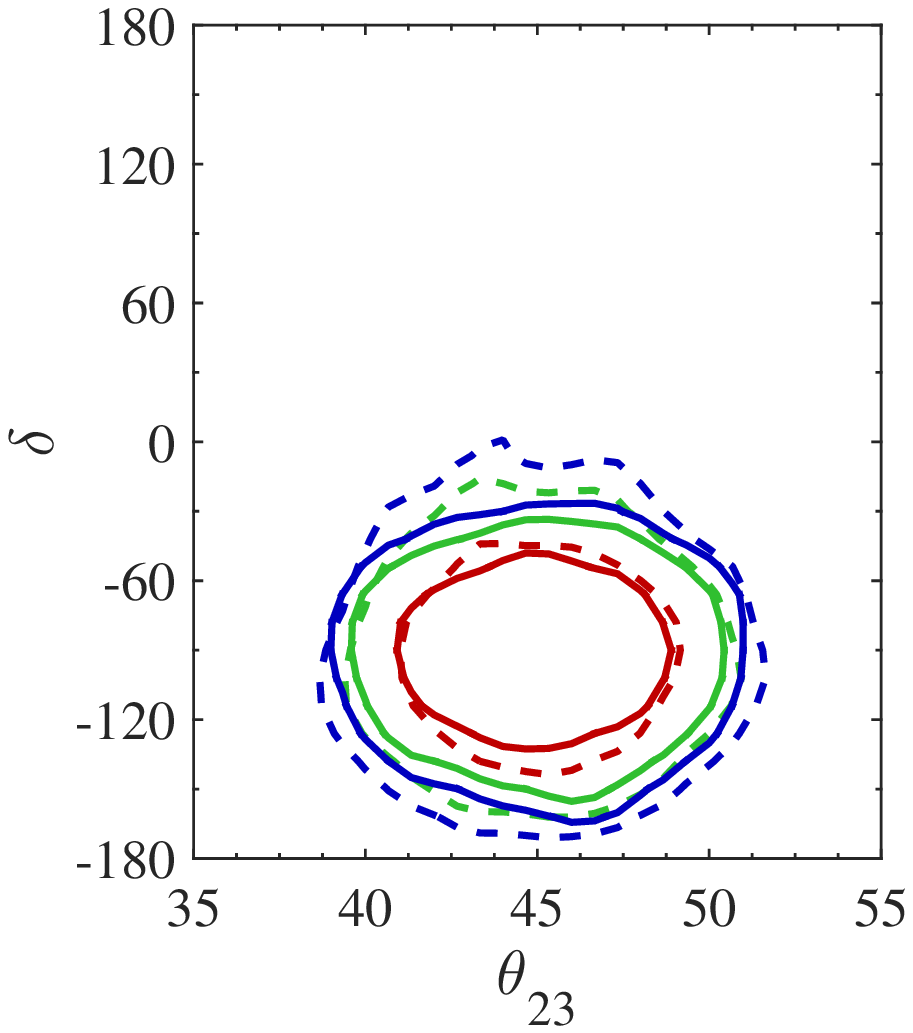, width=0.27\textwidth, bbllx=15, bblly=0, bburx=310, bbury=296,clip=} &
\epsfig{file=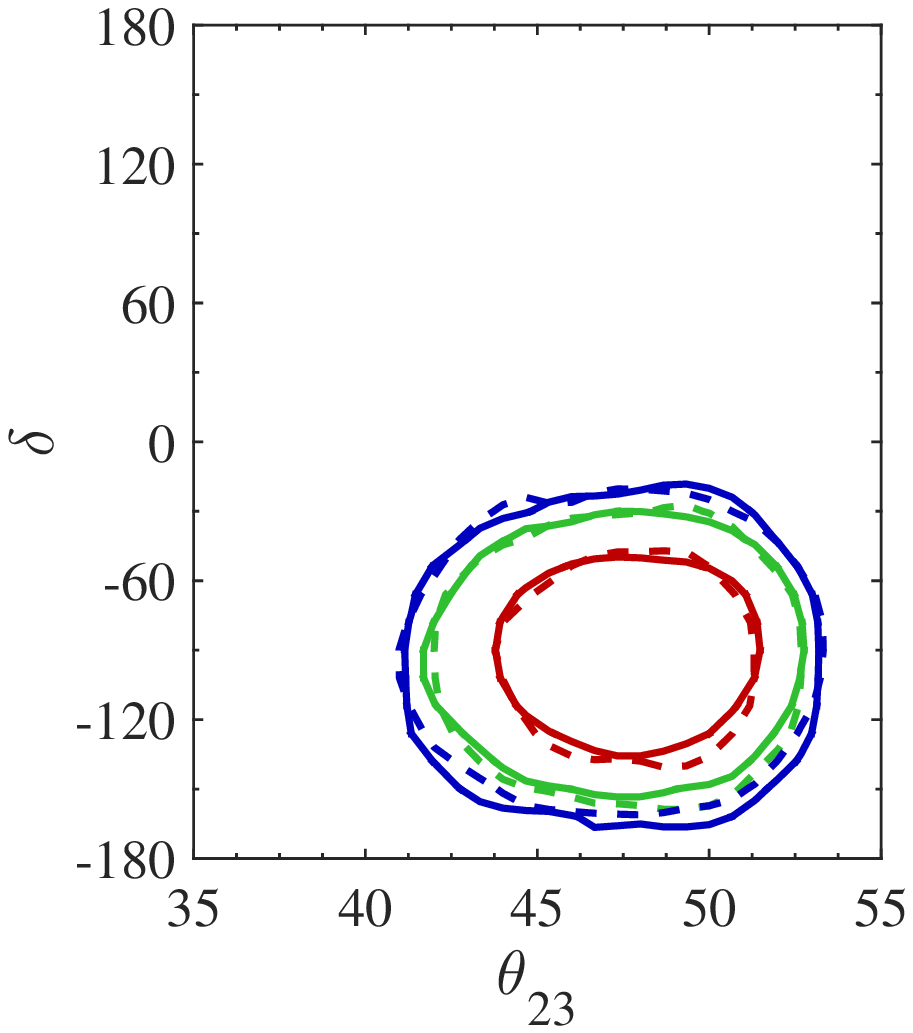, width=0.27\textwidth, bbllx=15, bblly=0, bburx=310, bbury=296,clip=} \\
\rotatebox[origin=l]{90}{\qquad\qquad\qquad $\dcp=0^\circ$} &
\epsfig{file=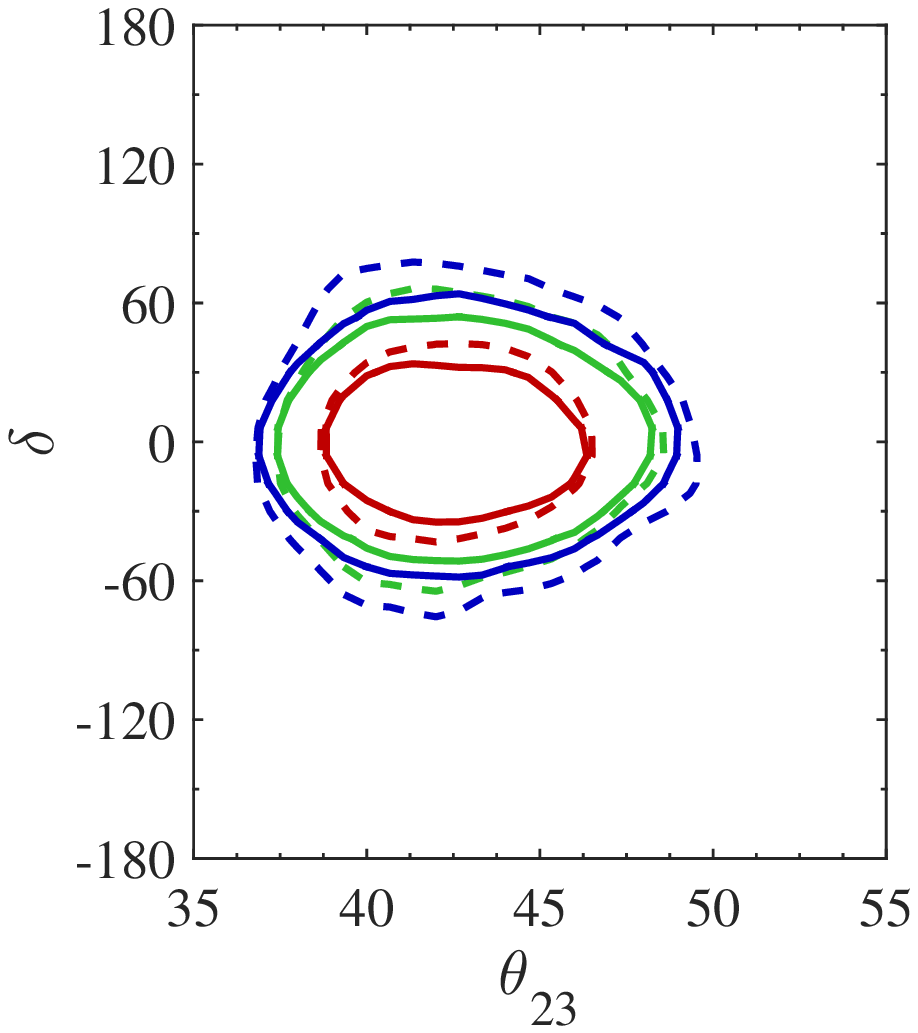, width=0.27\textwidth, bbllx=15, bblly=0, bburx=310, bbury=296,clip=} &
\epsfig{file=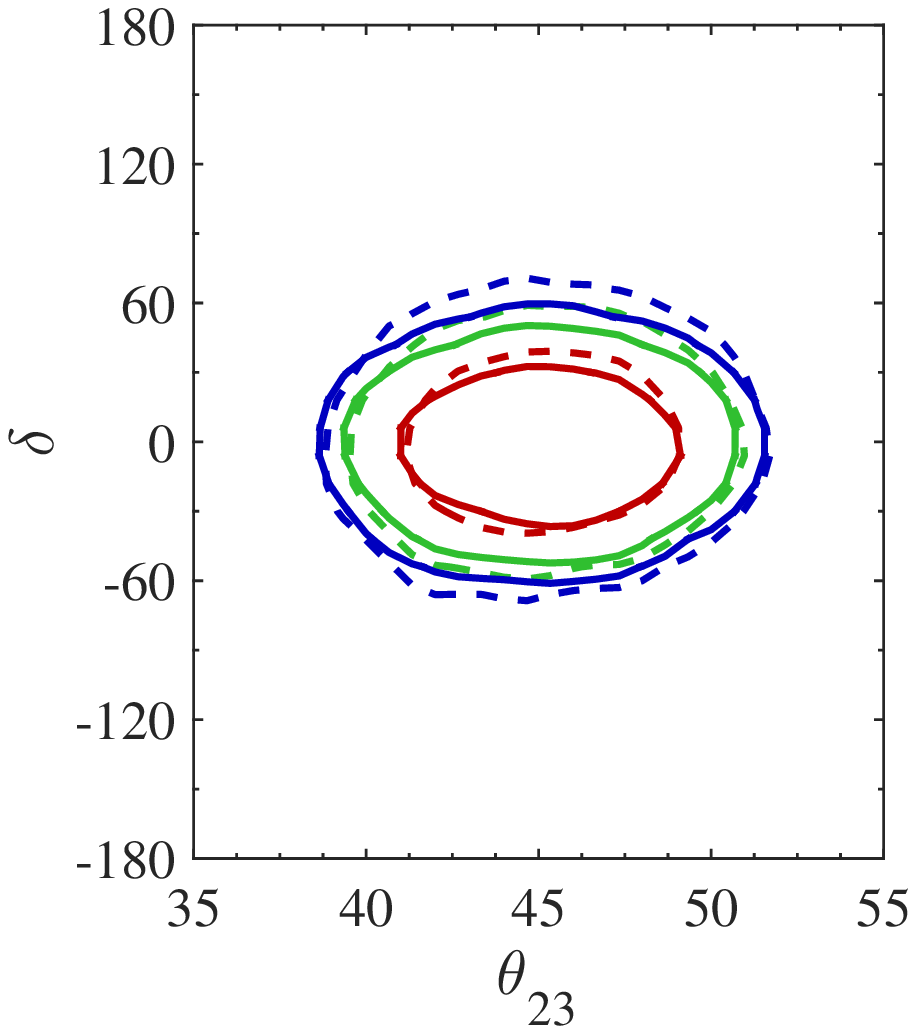, width=0.27\textwidth, bbllx=15, bblly=0, bburx=310, bbury=296,clip=} &
\epsfig{file=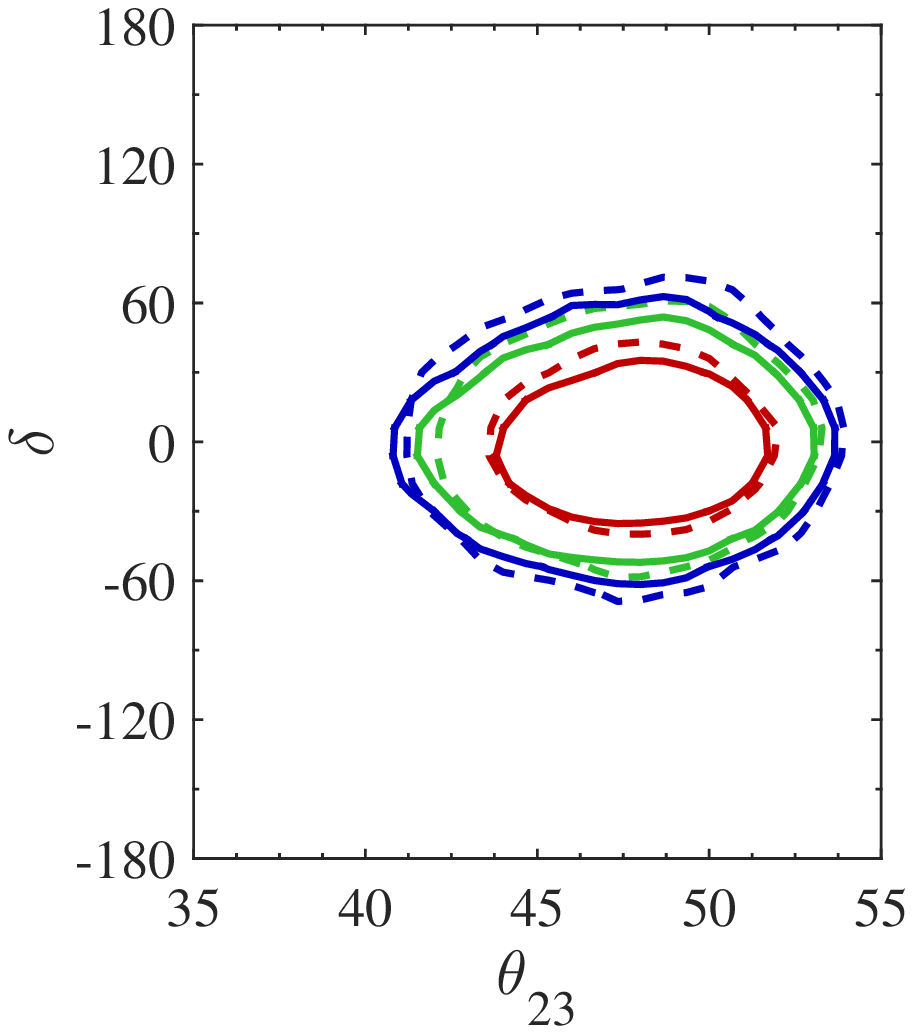, width=0.27\textwidth, bbllx=15, bblly=0, bburx=310, bbury=296,clip=} \\
\rotatebox[origin=l]{90}{\qquad\qquad\qquad $\dcp=90^\circ$} &
\epsfig{file=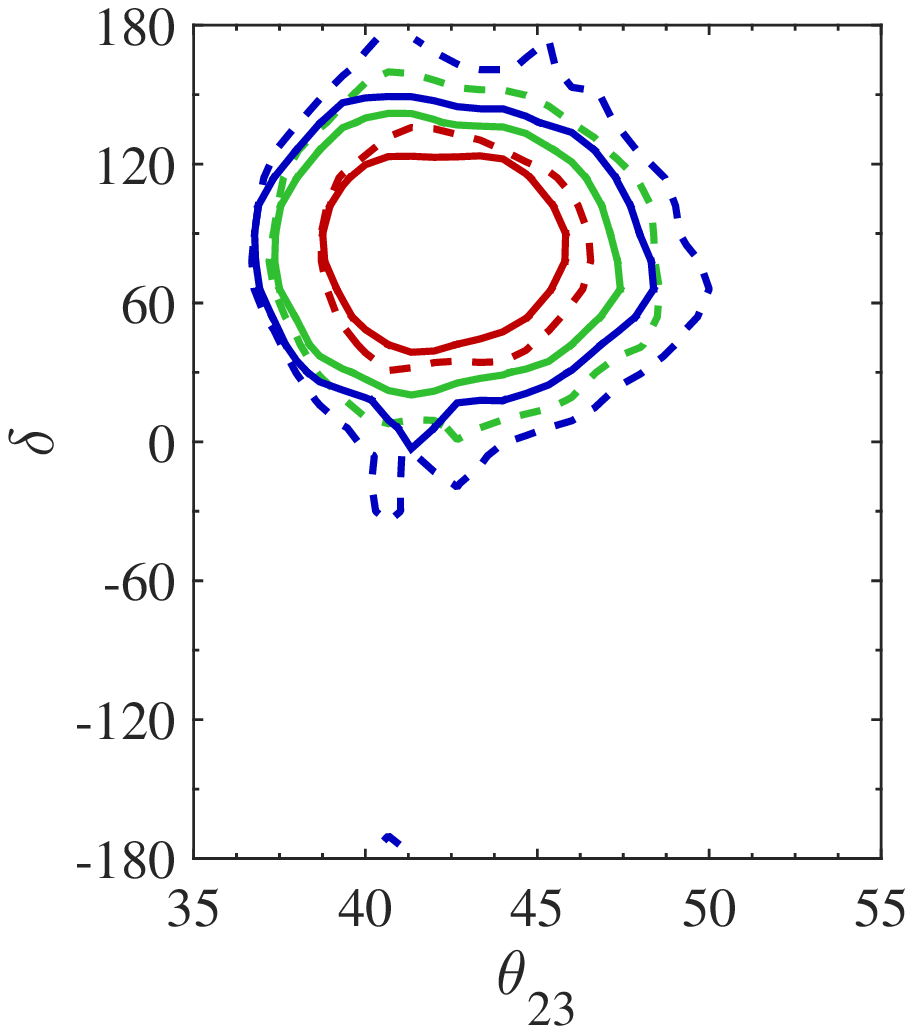, width=0.27\textwidth, bbllx=15, bblly=0, bburx=310, bbury=296,clip=} &
\epsfig{file=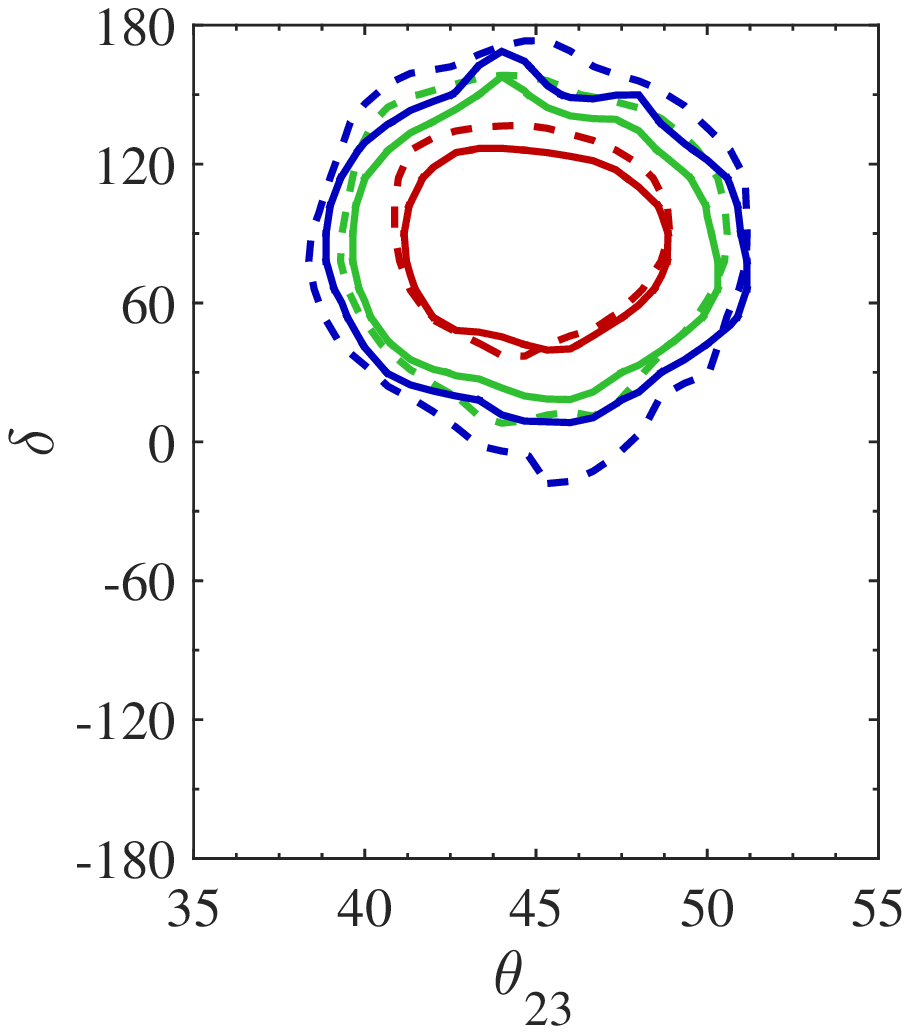, width=0.27\textwidth, bbllx=15, bblly=0, bburx=310, bbury=296,clip=} &
\epsfig{file=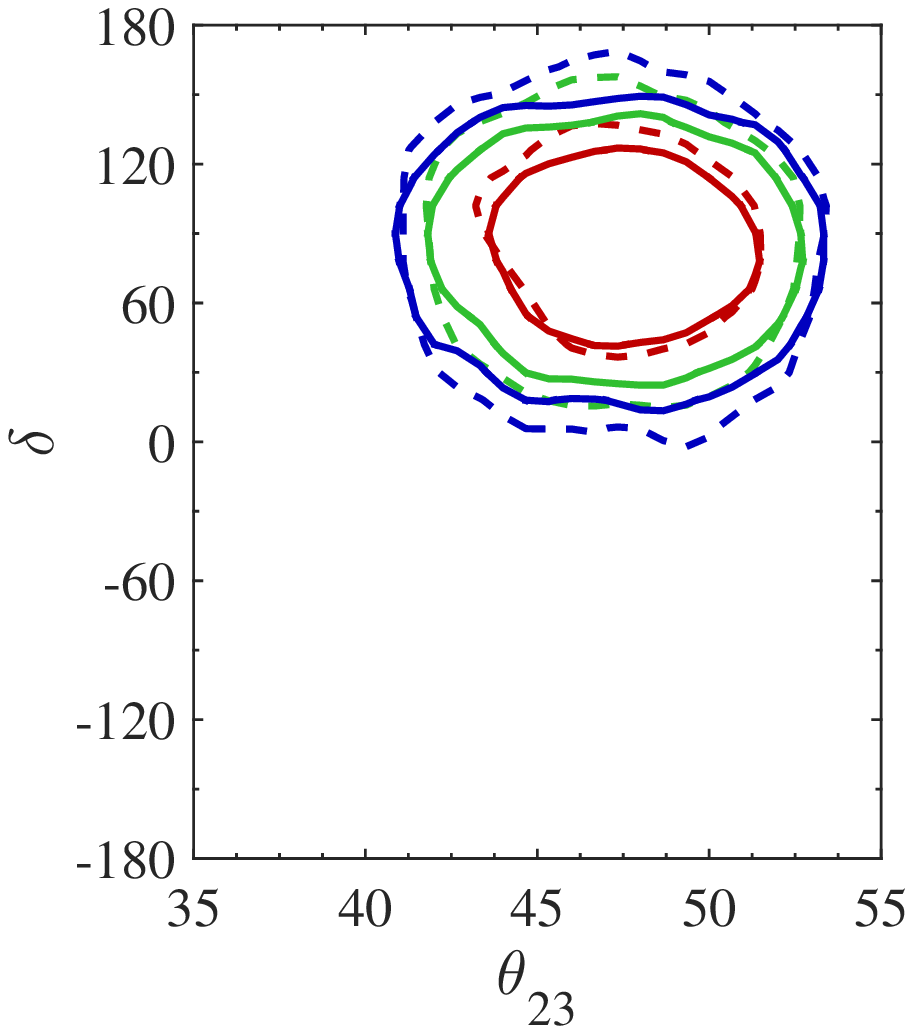, width=0.27\textwidth, bbllx=15, bblly=0, bburx=310, bbury=296,clip=} \\
\rotatebox[origin=l]{90}{\qquad\qquad\qquad $\dcp=180^\circ$} &
\epsfig{file=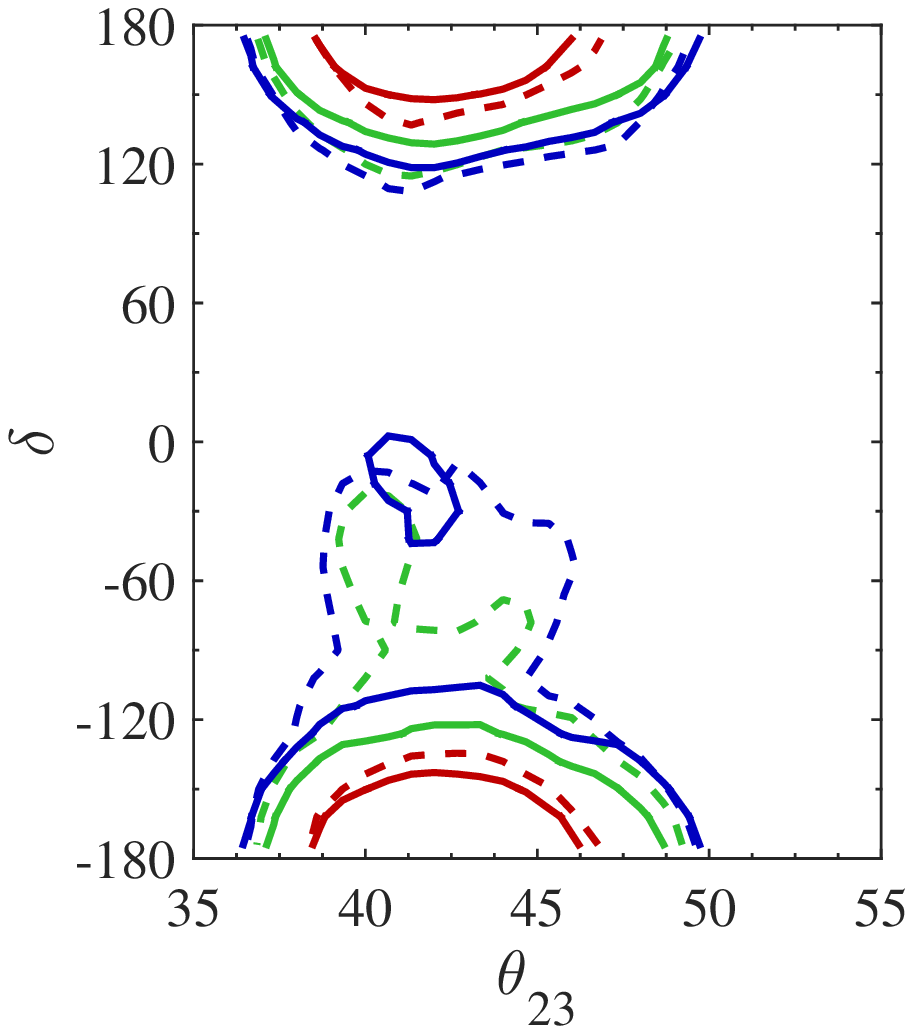, width=0.27\textwidth, bbllx=15, bblly=0, bburx=310, bbury=296,clip=} &
\epsfig{file=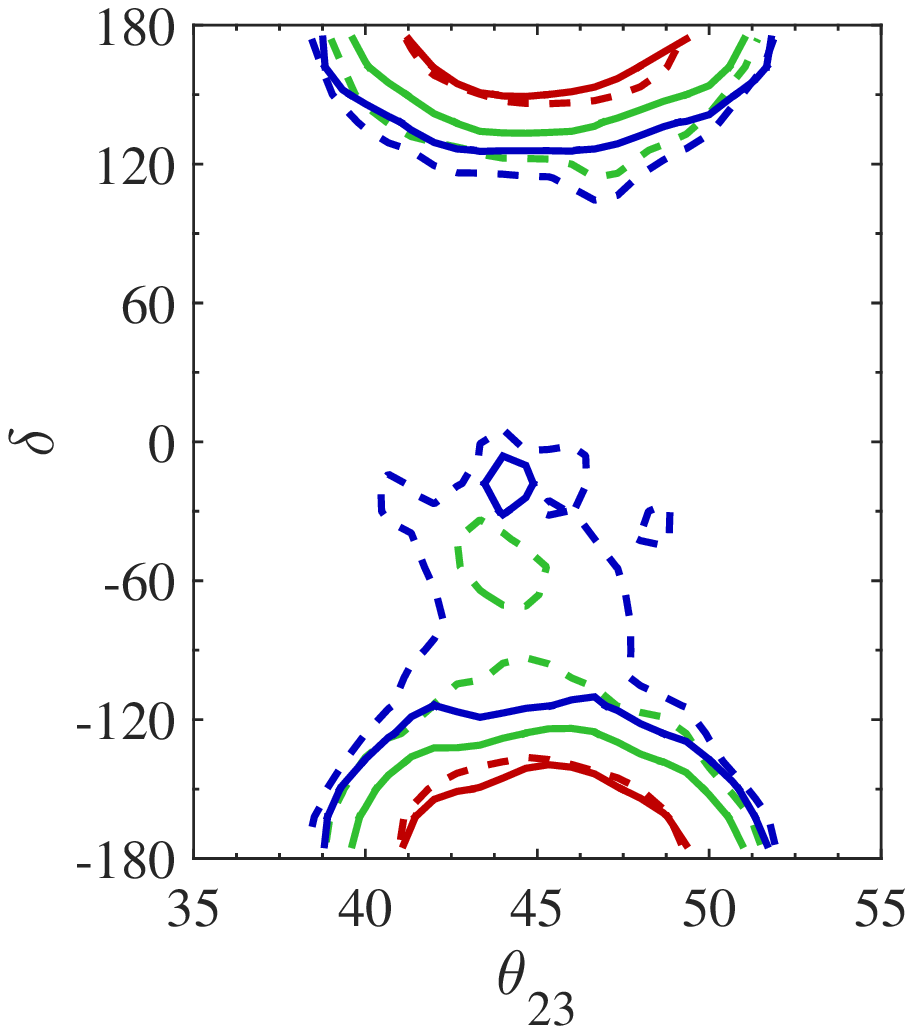, width=0.27\textwidth, bbllx=15, bblly=0, bburx=310, bbury=296,clip=} &
\epsfig{file=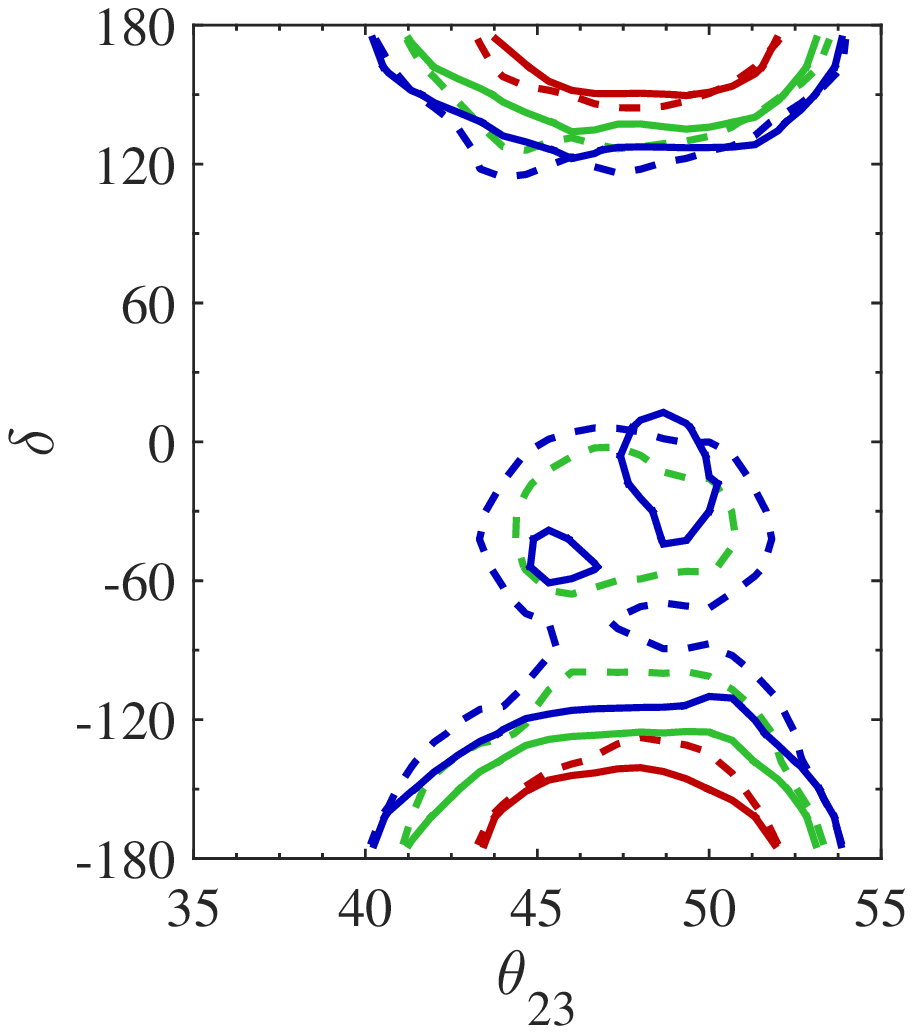, width=0.27\textwidth, bbllx=15, bblly=0, bburx=310, bbury=296,clip=} 
\end{tabular}
\caption{\footnotesize Role of the near detector in 
precision measurements at \essnusb. 
Each panel shows the allowed region in the test $\theta_{23}-\dcp$ plane, when
the NSI parameters are taken to be non-zero in the data. 
The true values of the amplitudes of the NSI parameters are assumed to be half of their 
90\% C.L.~bounds from Ref.~\cite{Biggio:2009nt}.
The red, green and blue curves 
represent the 68\%, 90\% and 95\% C.L.~contours, respectively. 
The solid (dashed) contours show the allowed region with (without)
the near detector.}
\label{fig:nd}
\end{figure}

Finally, we examine the role played by the near detector 
in the sensitivity of \essnusb. As described before, we 
have used a crude simulation of the near detector 
throughout this work. Here, we compare the results of 
our simulation with and without the near detector. We 
show in Fig.~\ref{fig:nd} a recomputed version of 
Fig.~\ref{fig:effect_nonzero}, both with and without 
the near detector. The solid contours are the same as 
before, but the dashed contours show the same allowed 
regions, if only the far detector is used. We observe that 
in the absence of a near detector, the $\dcp$-sensitivity 
of \essnusb\ is worsened. The limits on NSI parameters 
are also worse without the near detector. 

\section{Summary and conclusions}
\label{sec:sc}

In this work, we have investigated the effects of source and 
detector NSIs at the proposed neutrino oscillation experiment 
\essnusb, with a baseline of 540 km -- the source being the ESS in Lund, 
Sweden and a MEMPHYS-like detector in Garpenberg, Sweden. The \essnusb\ 
experiment is designed to determine the leptonic CP-violating phase 
$\dcp$ at the second oscillation maximum. However, it may also be 
able to probe source and detector NSIs. Due to the short baseline length 
and low neutrino energy of this experiment, matter NSIs 
will not be of importance, and are therefore not considered in this work. 

First, we have studied the three-flavour neutrino oscillation probabilities with source and detector NSIs,
which depend on six relevant NSI parameters -- $\epssme$, $\epssmm$, $\epssmt$, $\epsdee$, $\epsdme$ and $\epsdte$.
We used perturbative analytical expressions for the $\nu_\mu \to \nu_e$ channel that is the important channel for \essnusb\
in which $\epssme$, $\epsdme$ and $\epsdte$ are the dominating NSI parameters
in order to observe the impact of these parameters. We have found that, for the range of values allowed by the current data, 
the NSI parameter $\epsdte$ affect this probability the most,
whereas the NSI parameter $\epsdme$ the least. All other four NSI parameters have intermediate influence on the probability.

Second, we have explored the effect of marginalizing over the NSI parameters on precision measurements at \essnusb\ using two cases: (i) 
The true values of the NSI parameters are set to zero and (ii) the true values are set to half of the current 90\% C.L.\ bounds.
In both cases, the precision of measuring $\dcp$ is reduced by at most a factor of two. In addition, a measurement of the leptonic mixing angle $\theta_{23}$ is not affected by NSIs.
If we do not take the effect of NSIs into account when determining the value of $\dcp$, we obtain over-optimistic results.
The effect is most pronounced for a true value of $\dcp = 180^\circ$. Note that the impact of NSIs on the results are qualitatively same for both NO and IO.

Third, we have determined the possibility of \essnusb\ to measure the values of the NSI parameters.
In a realistic case with all NSI parameters free, we have found limits on $\epssme$ and $\epsdme$ at 90\% C.L.~that are
similar to the existing limits in the literature, whereas in a optimistic case with only one NSI parameter free and the rest set to zero,
the limits on $\epssme$ and $\epsdme$ are improved by a factor of two. Furthermore, we have set the true values of the NSI parameters
to half of their existing bound, and found that \essnusb\ is not able to differentiate the set values from zero at 68\% C.L. 
or impose any significant constraints on the phases of the NSI parameters.

Finally, we have examined the influence of the presence of a near detector at the \essnusb\ experimental setup.
Indeed, we show that without a near detector the results would be more pessimistic
concerning both the sensitivity of $\dcp$ and the limits on the NSI parameters.
Note that the results are not changed significantly by reducing the systematic errors.

In conclusion, using \essnusb\ with a near detector, the presence of NSIs will at most reduce the measurement of $\dcp$ by a factor of two, while
a measurement of $\theta_{23}$ will remain robust. In addition, it is possible to improve the existing upper limits on some of the NSI parameters by a factors of two.

\acknowledgments

This work was supported by the G{\"o}ran Gustafsson Foundation (M.B.). 
S.C.~acknowledges support from the Neutrino Project under the XII plan of 
Harish-Chandra Research Institute and partial support from the European 
Union FP7 ITN INVISIBLES (Marie Curie Actions, PITN-GA-2011-289442).
We thank Pilar Coloma, Michal Malinsk{\'y}, Suprabh Prakash, Bed\v{r}ich Roskovec 
and Nikos Vassilopoulos for useful discussions. 


\begin{thebibliography}{34}
\expandafter\ifx\csname natexlab\endcsname\relax\def\natexlab#1{#1}\fi
\expandafter\ifx\csname bibnamefont\endcsname\relax
  \def\bibnamefont#1{#1}\fi
\expandafter\ifx\csname bibfnamefont\endcsname\relax
  \def\bibfnamefont#1{#1}\fi
\expandafter\ifx\csname citenamefont\endcsname\relax
  \def\citenamefont#1{#1}\fi
\expandafter\ifx\csname url\endcsname\relax
  \def\url#1{\texttt{#1}}\fi
\expandafter\ifx\csname urlprefix\endcsname\relax\def\urlprefix{URL }\fi
\providecommand{\bibinfo}[2]{#2}
\providecommand{\eprint}[2][]{\url{#2}}

\bibitem[{\citenamefont{Aad et~al.}(2012)}]{Aad:2012tfa}
\bibinfo{author}{\bibfnamefont{G.}~\bibnamefont{Aad}} \bibnamefont{et~al.}
  (\bibinfo{collaboration}{ATLAS}), \bibinfo{journal}{Phys. Lett.}
  \textbf{\bibinfo{volume}{B716}}, \bibinfo{pages}{1} (\bibinfo{year}{2012}),
  \eprint{1207.7214}.

\bibitem[{\citenamefont{Chatrchyan et~al.}(2012)}]{Chatrchyan:2012ufa}
\bibinfo{author}{\bibfnamefont{S.}~\bibnamefont{Chatrchyan}}
  \bibnamefont{et~al.} (\bibinfo{collaboration}{CMS}), \bibinfo{journal}{Phys.
  Lett.} \textbf{\bibinfo{volume}{B716}}, \bibinfo{pages}{30}
  (\bibinfo{year}{2012}), \eprint{1207.7235}.

\bibitem[{\citenamefont{Weinberg}(1979)}]{Weinberg:1979sa}
\bibinfo{author}{\bibfnamefont{S.}~\bibnamefont{Weinberg}},
  \bibinfo{journal}{Phys. Rev. Lett.} \textbf{\bibinfo{volume}{43}},
  \bibinfo{pages}{1566} (\bibinfo{year}{1979}).

\bibitem[{\citenamefont{Davis et~al.}(1968)\citenamefont{Davis, Harmer, and
  Hoffman}}]{Davis:1968cp}
\bibinfo{author}{\bibfnamefont{J.}~\bibnamefont{Davis},
  \bibfnamefont{Raymond}}, \bibinfo{author}{\bibfnamefont{D.~S.}
  \bibnamefont{Harmer}}, \bibnamefont{and}
  \bibinfo{author}{\bibfnamefont{K.~C.} \bibnamefont{Hoffman}},
  \bibinfo{journal}{Phys. Rev. Lett.} \textbf{\bibinfo{volume}{20}},
  \bibinfo{pages}{1205} (\bibinfo{year}{1968}).

\bibitem[{\citenamefont{Fukuda et~al.}(1998)}]{Fukuda:1998mi}
\bibinfo{author}{\bibfnamefont{Y.}~\bibnamefont{Fukuda}} \bibnamefont{et~al.}
  (\bibinfo{collaboration}{Super-Kamiokande}), \bibinfo{journal}{Phys. Rev.
  Lett.} \textbf{\bibinfo{volume}{81}}, \bibinfo{pages}{1562}
  (\bibinfo{year}{1998}), \eprint{hep-ex/9807003}.

\bibitem[{\citenamefont{Capozzi et~al.}(2014)\citenamefont{Capozzi, Fogli,
  Lisi, Marrone, Montanino et~al.}}]{Capozzi:2013csa}
\bibinfo{author}{\bibfnamefont{F.}~\bibnamefont{Capozzi}},
  \bibinfo{author}{\bibfnamefont{G.}~\bibnamefont{Fogli}},
  \bibinfo{author}{\bibfnamefont{E.}~\bibnamefont{Lisi}},
  \bibinfo{author}{\bibfnamefont{A.}~\bibnamefont{Marrone}},
  \bibinfo{author}{\bibfnamefont{D.}~\bibnamefont{Montanino}},
  \bibnamefont{et~al.}, \bibinfo{journal}{Phys. Rev.}
  \textbf{\bibinfo{volume}{D89}}, \bibinfo{pages}{093018}
  (\bibinfo{year}{2014}), \eprint{1312.2878}.

\bibitem[{\citenamefont{Forero et~al.}(2014)\citenamefont{Forero, Tortola, and
  Valle}}]{Forero:2014bxa}
\bibinfo{author}{\bibfnamefont{D.}~\bibnamefont{Forero}},
  \bibinfo{author}{\bibfnamefont{M.}~\bibnamefont{Tortola}}, \bibnamefont{and}
  \bibinfo{author}{\bibfnamefont{J.}~\bibnamefont{Valle}},
  \bibinfo{journal}{Phys. Rev.} \textbf{\bibinfo{volume}{D90}},
  \bibinfo{pages}{093006} (\bibinfo{year}{2014}), \eprint{1405.7540}.

\bibitem[{\citenamefont{Gonzalez-Garcia
  et~al.}(2014)\citenamefont{Gonzalez-Garcia, Maltoni, and
  Schwetz}}]{Gonzalez-Garcia:2014bfa}
\bibinfo{author}{\bibfnamefont{M.}~\bibnamefont{Gonzalez-Garcia}},
  \bibinfo{author}{\bibfnamefont{M.}~\bibnamefont{Maltoni}}, \bibnamefont{and}
  \bibinfo{author}{\bibfnamefont{T.}~\bibnamefont{Schwetz}},
  \bibinfo{journal}{JHEP} \textbf{\bibinfo{volume}{11}}, \bibinfo{pages}{052}
  (\bibinfo{year}{2014}), \eprint{1409.5439}.

\bibitem[{\citenamefont{Baussan et~al.}(2014)}]{Baussan:2013zcy}
\bibinfo{author}{\bibfnamefont{E.}~\bibnamefont{Baussan}} \bibnamefont{et~al.}
  (\bibinfo{collaboration}{ESSnuSB}), \bibinfo{journal}{Nucl. Phys.}
  \textbf{\bibinfo{volume}{B885}}, \bibinfo{pages}{127} (\bibinfo{year}{2014}),
  \eprint{1309.7022}.

\bibitem[{\citenamefont{Agarwalla et~al.}(2014)\citenamefont{Agarwalla,
  Choubey, and Prakash}}]{Agarwalla:2014tpa}
\bibinfo{author}{\bibfnamefont{S.~K.} \bibnamefont{Agarwalla}},
  \bibinfo{author}{\bibfnamefont{S.}~\bibnamefont{Choubey}}, \bibnamefont{and}
  \bibinfo{author}{\bibfnamefont{S.}~\bibnamefont{Prakash}},
  \bibinfo{journal}{JHEP} \textbf{\bibinfo{volume}{12}}, \bibinfo{pages}{020}
  (\bibinfo{year}{2014}), \eprint{1406.2219}.

\bibitem[{\citenamefont{Blennow et~al.}(2014)\citenamefont{Blennow, Coloma, and
  Fernandez-Martinez}}]{Blennow:2014fqa}
\bibinfo{author}{\bibfnamefont{M.}~\bibnamefont{Blennow}},
  \bibinfo{author}{\bibfnamefont{P.}~\bibnamefont{Coloma}}, \bibnamefont{and}
  \bibinfo{author}{\bibfnamefont{E.}~\bibnamefont{Fernandez-Martinez}},
  \bibinfo{journal}{JHEP} \textbf{\bibinfo{volume}{12}}, \bibinfo{pages}{120}
  (\bibinfo{year}{2014}), \eprint{1407.1317}.

\bibitem[{\citenamefont{Ohlsson}(2013)}]{Ohlsson:2012kf}
\bibinfo{author}{\bibfnamefont{T.}~\bibnamefont{Ohlsson}},
  \bibinfo{journal}{Rept. Prog. Phys.} \textbf{\bibinfo{volume}{76}},
  \bibinfo{pages}{044201} (\bibinfo{year}{2013}), \eprint{1209.2710}.

\bibitem[{\citenamefont{Miranda and Nunokawa}(2015)}]{Miranda:2015dra}
\bibinfo{author}{\bibfnamefont{O.}~\bibnamefont{Miranda}} \bibnamefont{and}
  \bibinfo{author}{\bibfnamefont{H.}~\bibnamefont{Nunokawa}}
  (\bibinfo{year}{2015}), \eprint{1505.06254}.

\bibitem[{\citenamefont{Antusch et~al.}(2009)\citenamefont{Antusch, Baumann,
  and Fernandez-Martinez}}]{Antusch:2008tz}
\bibinfo{author}{\bibfnamefont{S.}~\bibnamefont{Antusch}},
  \bibinfo{author}{\bibfnamefont{J.~P.} \bibnamefont{Baumann}},
  \bibnamefont{and}
  \bibinfo{author}{\bibfnamefont{E.}~\bibnamefont{Fernandez-Martinez}},
  \bibinfo{journal}{Nucl. Phys.} \textbf{\bibinfo{volume}{B810}},
  \bibinfo{pages}{369} (\bibinfo{year}{2009}), \eprint{0807.1003}.

\bibitem[{\citenamefont{Biggio et~al.}(2009{\natexlab{a}})\citenamefont{Biggio,
  Blennow, and Fernandez-Martinez}}]{Biggio:2009kv}
\bibinfo{author}{\bibfnamefont{C.}~\bibnamefont{Biggio}},
  \bibinfo{author}{\bibfnamefont{M.}~\bibnamefont{Blennow}}, \bibnamefont{and}
  \bibinfo{author}{\bibfnamefont{E.}~\bibnamefont{Fernandez-Martinez}},
  \bibinfo{journal}{JHEP} \textbf{\bibinfo{volume}{03}}, \bibinfo{pages}{139}
  (\bibinfo{year}{2009}{\natexlab{a}}), \eprint{0902.0607}.

\bibitem[{\citenamefont{Wolfenstein}(1978)}]{Wolfenstein:1977ue}
\bibinfo{author}{\bibfnamefont{L.}~\bibnamefont{Wolfenstein}},
  \bibinfo{journal}{Phys. Rev.} \textbf{\bibinfo{volume}{D17}},
  \bibinfo{pages}{2369} (\bibinfo{year}{1978}).

\bibitem[{\citenamefont{Barger et~al.}(1991)\citenamefont{Barger, Phillips, and
  Whisnant}}]{Barger:1991ae}
\bibinfo{author}{\bibfnamefont{V.~D.} \bibnamefont{Barger}},
  \bibinfo{author}{\bibfnamefont{R.}~\bibnamefont{Phillips}}, \bibnamefont{and}
  \bibinfo{author}{\bibfnamefont{K.}~\bibnamefont{Whisnant}},
  \bibinfo{journal}{Phys. Rev.} \textbf{\bibinfo{volume}{D44}},
  \bibinfo{pages}{1629} (\bibinfo{year}{1991}).

\bibitem[{\citenamefont{Mikheyev and Smirnov}(1985)}]{Mikheev:1986gs}
\bibinfo{author}{\bibfnamefont{S.}~\bibnamefont{Mikheyev}} \bibnamefont{and}
  \bibinfo{author}{\bibfnamefont{A.~Y.} \bibnamefont{Smirnov}},
  \bibinfo{journal}{Sov. J. Nucl. Phys.} \textbf{\bibinfo{volume}{42}},
  \bibinfo{pages}{913} (\bibinfo{year}{1985}).

\bibitem[{\citenamefont{Mikheyev and Smirnov}(1986)}]{Mikheev:1986wj}
\bibinfo{author}{\bibfnamefont{S.}~\bibnamefont{Mikheyev}} \bibnamefont{and}
  \bibinfo{author}{\bibfnamefont{A.~Y.} \bibnamefont{Smirnov}},
  \bibinfo{journal}{Nuovo Cim.} \textbf{\bibinfo{volume}{C9}},
  \bibinfo{pages}{17} (\bibinfo{year}{1986}).

\bibitem[{\citenamefont{Grossman}(1995)}]{Grossman:1995wx}
\bibinfo{author}{\bibfnamefont{Y.}~\bibnamefont{Grossman}},
  \bibinfo{journal}{Phys. Lett.} \textbf{\bibinfo{volume}{B359}},
  \bibinfo{pages}{141} (\bibinfo{year}{1995}), \eprint{hep-ph/9507344}.

\bibitem[{\citenamefont{Biggio et~al.}(2009{\natexlab{b}})\citenamefont{Biggio,
  Blennow, and Fernandez-Martinez}}]{Biggio:2009nt}
\bibinfo{author}{\bibfnamefont{C.}~\bibnamefont{Biggio}},
  \bibinfo{author}{\bibfnamefont{M.}~\bibnamefont{Blennow}}, \bibnamefont{and}
  \bibinfo{author}{\bibfnamefont{E.}~\bibnamefont{Fernandez-Martinez}},
  \bibinfo{journal}{JHEP} \textbf{\bibinfo{volume}{08}}, \bibinfo{pages}{090}
  (\bibinfo{year}{2009}{\natexlab{b}}), \eprint{0907.0097}.

\bibitem[{\citenamefont{Ohlsson and Zhang}(2009)}]{Ohlsson:2008gx}
\bibinfo{author}{\bibfnamefont{T.}~\bibnamefont{Ohlsson}} \bibnamefont{and}
  \bibinfo{author}{\bibfnamefont{H.}~\bibnamefont{Zhang}},
  \bibinfo{journal}{Phys. Lett.} \textbf{\bibinfo{volume}{B671}},
  \bibinfo{pages}{99} (\bibinfo{year}{2009}), \eprint{0809.4835}.

\bibitem[{\citenamefont{Agarwalla et~al.}(2015)\citenamefont{Agarwalla, Bagchi,
  Forero, and T\'ortola}}]{Agarwalla:2014bsa}
\bibinfo{author}{\bibfnamefont{S.~K.} \bibnamefont{Agarwalla}},
  \bibinfo{author}{\bibfnamefont{P.}~\bibnamefont{Bagchi}},
  \bibinfo{author}{\bibfnamefont{D.~V.} \bibnamefont{Forero}},
  \bibnamefont{and}
  \bibinfo{author}{\bibfnamefont{M.}~\bibnamefont{T\'ortola}},
  \bibinfo{journal}{JHEP} \textbf{\bibinfo{volume}{07}}, \bibinfo{pages}{060}
  (\bibinfo{year}{2015}), \eprint{1412.1064}.

\bibitem[{\citenamefont{Agostino et~al.}(2013)}]{Agostino:2012fd}
\bibinfo{author}{\bibfnamefont{L.}~\bibnamefont{Agostino}} \bibnamefont{et~al.}
  (\bibinfo{collaboration}{MEMPHYS}), \bibinfo{journal}{JCAP}
  \textbf{\bibinfo{volume}{01}}, \bibinfo{pages}{024} (\bibinfo{year}{2013}),
  \eprint{1206.6665}.

\bibitem[{\citenamefont{Huber et~al.}(2005)\citenamefont{Huber, Lindner, and
  Winter}}]{Huber:2004ka}
\bibinfo{author}{\bibfnamefont{P.}~\bibnamefont{Huber}},
  \bibinfo{author}{\bibfnamefont{M.}~\bibnamefont{Lindner}}, \bibnamefont{and}
  \bibinfo{author}{\bibfnamefont{W.}~\bibnamefont{Winter}},
  \bibinfo{journal}{Comput. Phys. Commun.} \textbf{\bibinfo{volume}{167}},
  \bibinfo{pages}{195} (\bibinfo{year}{2005}), \eprint{hep-ph/0407333}.

\bibitem[{\citenamefont{Huber et~al.}(2007)\citenamefont{Huber, Kopp, Lindner,
  Rolinec, and Winter}}]{Huber:2007ji}
\bibinfo{author}{\bibfnamefont{P.}~\bibnamefont{Huber}},
  \bibinfo{author}{\bibfnamefont{J.}~\bibnamefont{Kopp}},
  \bibinfo{author}{\bibfnamefont{M.}~\bibnamefont{Lindner}},
  \bibinfo{author}{\bibfnamefont{M.}~\bibnamefont{Rolinec}}, \bibnamefont{and}
  \bibinfo{author}{\bibfnamefont{W.}~\bibnamefont{Winter}},
  \bibinfo{journal}{Comput. Phys. Commun.} \textbf{\bibinfo{volume}{177}},
  \bibinfo{pages}{432} (\bibinfo{year}{2007}), \eprint{hep-ph/0701187}.

\bibitem[{\citenamefont{Blennow and Fernandez-Martinez}(2010)}]{Blennow:2009pk}
\bibinfo{author}{\bibfnamefont{M.}~\bibnamefont{Blennow}} \bibnamefont{and}
  \bibinfo{author}{\bibfnamefont{E.}~\bibnamefont{Fernandez-Martinez}},
  \bibinfo{journal}{Comput. Phys. Commun.} \textbf{\bibinfo{volume}{181}},
  \bibinfo{pages}{227} (\bibinfo{year}{2010}), \eprint{0903.3985}.

\bibitem[{\citenamefont{Gonzalez-Garcia
  et~al.}(2001)\citenamefont{Gonzalez-Garcia, Grossman, Gusso, and
  Nir}}]{GonzalezGarcia:2001mp}
\bibinfo{author}{\bibfnamefont{M.}~\bibnamefont{Gonzalez-Garcia}},
  \bibinfo{author}{\bibfnamefont{Y.}~\bibnamefont{Grossman}},
  \bibinfo{author}{\bibfnamefont{A.}~\bibnamefont{Gusso}}, \bibnamefont{and}
  \bibinfo{author}{\bibfnamefont{Y.}~\bibnamefont{Nir}},
  \bibinfo{journal}{Phys. Rev.} \textbf{\bibinfo{volume}{D64}},
  \bibinfo{pages}{096006} (\bibinfo{year}{2001}), \eprint{hep-ph/0105159}.

\bibitem[{\citenamefont{Bilenky and Giunti}(1993)}]{Bilenky:1992wv}
\bibinfo{author}{\bibfnamefont{S.~M.} \bibnamefont{Bilenky}} \bibnamefont{and}
  \bibinfo{author}{\bibfnamefont{C.}~\bibnamefont{Giunti}},
  \bibinfo{journal}{Phys. Lett.} \textbf{\bibinfo{volume}{B300}},
  \bibinfo{pages}{137} (\bibinfo{year}{1993}), \eprint{hep-ph/9211269}.

\bibitem[{\citenamefont{Meloni et~al.}(2010)\citenamefont{Meloni, Ohlsson,
  Winter, and Zhang}}]{Meloni:2009cg}
\bibinfo{author}{\bibfnamefont{D.}~\bibnamefont{Meloni}},
  \bibinfo{author}{\bibfnamefont{T.}~\bibnamefont{Ohlsson}},
  \bibinfo{author}{\bibfnamefont{W.}~\bibnamefont{Winter}}, \bibnamefont{and}
  \bibinfo{author}{\bibfnamefont{H.}~\bibnamefont{Zhang}},
  \bibinfo{journal}{JHEP} \textbf{\bibinfo{volume}{04}}, \bibinfo{pages}{041}
  (\bibinfo{year}{2010}), \eprint{0912.2735}.

\bibitem[{\citenamefont{Cervera et~al.}(2000)\citenamefont{Cervera, Donini,
  Gavela, Gomez~Cadenas, Hernandez et~al.}}]{Cervera:2000kp}
\bibinfo{author}{\bibfnamefont{A.}~\bibnamefont{Cervera}},
  \bibinfo{author}{\bibfnamefont{A.}~\bibnamefont{Donini}},
  \bibinfo{author}{\bibfnamefont{M.}~\bibnamefont{Gavela}},
  \bibinfo{author}{\bibfnamefont{J.}~\bibnamefont{Gomez~Cadenas}},
  \bibinfo{author}{\bibfnamefont{P.}~\bibnamefont{Hernandez}},
  \bibnamefont{et~al.}, \bibinfo{journal}{Nucl. Phys.}
  \textbf{\bibinfo{volume}{B579}}, \bibinfo{pages}{17} (\bibinfo{year}{2000}),
  \eprint{hep-ph/0002108}.

\bibitem[{\citenamefont{Freund}(2001)}]{Freund:2001pn}
\bibinfo{author}{\bibfnamefont{M.}~\bibnamefont{Freund}},
  \bibinfo{journal}{Phys. Rev.} \textbf{\bibinfo{volume}{D64}},
  \bibinfo{pages}{053003} (\bibinfo{year}{2001}), \eprint{hep-ph/0103300}.

\bibitem[{\citenamefont{Akhmedov et~al.}(2004)\citenamefont{Akhmedov,
  Johansson, Lindner, Ohlsson, and Schwetz}}]{Akhmedov:2004ny}
\bibinfo{author}{\bibfnamefont{E.~K.} \bibnamefont{Akhmedov}},
  \bibinfo{author}{\bibfnamefont{R.}~\bibnamefont{Johansson}},
  \bibinfo{author}{\bibfnamefont{M.}~\bibnamefont{Lindner}},
  \bibinfo{author}{\bibfnamefont{T.}~\bibnamefont{Ohlsson}}, \bibnamefont{and}
  \bibinfo{author}{\bibfnamefont{T.}~\bibnamefont{Schwetz}},
  \bibinfo{journal}{JHEP} \textbf{\bibinfo{volume}{04}}, \bibinfo{pages}{078}
  (\bibinfo{year}{2004}), \eprint{hep-ph/0402175}.

\bibitem[{\citenamefont{Kopp et~al.}(2008)\citenamefont{Kopp, Lindner, Ota, and
  Sato}}]{Kopp:2007ne}
\bibinfo{author}{\bibfnamefont{J.}~\bibnamefont{Kopp}},
  \bibinfo{author}{\bibfnamefont{M.}~\bibnamefont{Lindner}},
  \bibinfo{author}{\bibfnamefont{T.}~\bibnamefont{Ota}}, \bibnamefont{and}
  \bibinfo{author}{\bibfnamefont{J.}~\bibnamefont{Sato}},
  \bibinfo{journal}{Phys. Rev.} \textbf{\bibinfo{volume}{D77}},
  \bibinfo{pages}{013007} (\bibinfo{year}{2008}), \eprint{0708.0152}.

\end{thebibliography}

\end{document}